\documentclass[12pt]{iopart}
\usepackage{iopams}
\expandafter\let\csname equation*\endcsname\relax

\expandafter\let\csname endequation*\endcsname\relax

\usepackage{bm,bbm}
\usepackage{amsmath}
\usepackage{amsfonts}
\usepackage{amstext,amsthm}
\usepackage{amssymb}   
\usepackage{graphicx}
\usepackage{verbatim}
\usepackage{cite}
\usepackage{caption}
\usepackage{hyperref}
\usepackage{xcolor}

\begin{document}

\hyphenation{ge-ne-ra-li-zed}

\title[Second largest Eigenpair Statistics]{Second largest Eigenpair Statistics for Sparse Graphs}

\author{Vito A R Susca, Pierpaolo Vivo and Reimer K\"uhn}

\address{                 
King's College London, Department of Mathematics, Strand, London WC2R 2LS, United Kingdom
}

\begin{abstract}
We develop a formalism to compute the statistics of the second largest  eigenpair of weighted sparse graphs with $N\gg 1$ nodes, finite mean connectivity and bounded maximal degree, in cases where the top eigenpair statistics is known. The problem can be cast in terms of optimisation of a quadratic form on the sphere with a fictitious temperature, after a suitable deflation of the original matrix model. We use the cavity and replica methods to find the solution in terms of self-consistent equations for auxiliary probability density functions, which can be solved by an improved population dynamics algorithm enforcing eigenvector orthogonality on-the-fly. The analytical results are in perfect agreement with numerical diagonalisation of large (weighted) adjacency matrices, focussing on the cases of random regular and Erd\H{o}s-R\'enyi graphs. We further analyse the case of sparse Markov transition matrices for unbiased random walks, whose second largest eigenpair describes the non-equilibrium mode with the largest relaxation time. We also show that the population dynamics algorithm with population size $N_P$ does not actually capture the thermodynamic limit $N\to\infty$ as commonly assumed: the accuracy of the population dynamics algorithm has a strongly non-monotonic behaviour as a function of $N_P$, thus implying that an optimal size $N_P^\star=N_P^\star(N)$ must be chosen to best reproduce the results from numerical diagonalisation of graphs of finite size $N$. 
\end{abstract}


\maketitle

\tableofcontents

\section{Introduction \label{sec:intro}}
The second largest eigenvalue and the associated \emph{second} eigenvector of a $N\times N$ matrix $J$ is of great significance in many areas of science, with plenty of applications. In coding theory, the Hamming distance of a binary linear code can be expressed as a function of the second largest eigenvalue of the \emph{coset} graph associated to the code \cite{friedman2005generalized}. In biology, it has been shown in \cite{tomic2018evaluation} that the second largest eigenvalue of cancer metabolic networks describes the speed of cancer processes. In the context of clustering methods based on the adjacency matrix of a graph, the second eigenvector encodes inter-cluster connectivity, complementing the information about intra-cluster connectivity included in the top eigenvector \cite{lucinska2018clustering,Shi2000}. Moreover, in Principal Component Analysis, the second eigenvector of the covariance matrix of standardised data represents the direction that accounts for the second largest source of variability within the dataset \cite{shlens2014tutorial,jolliffe2016principal}.

The second largest eigenvalue plays a pivotal role in the study of complex systems and graph theory, representing topological features of the graphs \cite{cvetkovic1995second}. If the spectral gap, i.e. the distance between the largest and second largest eigenvalue, is large, then the graph has good connectivity and expansion properties \cite{Brouwer2011}. Therefore, many results have been derived about bounds for the second largest eigenvalue (see e.g. \cite{simic2015notes,kolotilina2013upper}). In particular, bipartite regular graphs with very wide spectral gaps are called \emph{expanders} (\emph{magnifiers} if not bipartite) and have been widely studied since the seminal work of Alon \cite{alon1986eigenvalues}. To shed light on the expansion properties of regular graphs, specific bounds have been derived for their second largest eigenvalue (see e.g. \cite{friedman2005generalized} and \cite{nilli1991second}).
 
The knowledge of the spectral gap is essential for random walks on undirected graphs, which are substantially equivalent to finite time-reversible Markov chains, as pointed out by Lovasz in his survey \cite{lovasz1993random}. Indeed, up to log-factors, the inverse spectral gap of the transition matrix represents the \emph{mixing} rate of the Markov chain, i.e. how fast the state probability vector of a Markov chain approaches the limiting stationary distribution \cite{lovasz2007eigenvalues}, given by the top right eigenvector of the transition matrix. The inverse of absolute value of second largest eigenvalue of the transition matrix denotes the largest relaxation time or \emph{mixing} time, and the corresponding eigenvector describes the non-equilibrium mode with the slowest decay rate.
The second largest eigenpair of Markov transition matrices also plays an important role in all processes that are described by means of random walks on graphs, such as out of equilibrium dynamics of glassy systems (see e.g. \cite{moretti2011complex,margiotta2019glassy}) and search algorithms such as Google PageRank
\cite{haveliwala2003second}.

In our analysis, we will be dealing with sparse symmetric random matrices, i.e. weighted adjacency matrices of undirected graphs. We focus on the case of high sparsity, i.e. when the probability of two nodes being connected is $p=c/N$, with $c$ being the constant mean degree of nodes. In this sparse limit, numerical studies have shown that most of the eigenvectors of a random regular graph, as well as almost-eigenvectors\footnote{An almost-eigenvector of a matrix $A$ with eigenvalue $\lambda$ is a normalised vector $\bm{v}$ that satisfies the eigenvector equation $(A-\lambda I)\bm v=\bm 0$ within some small tolerance $\epsilon$, i.e. $||A\bm v-\lambda \bm v||_2\leq\epsilon$.} \cite{Backhausz2016}, follow a Gaussian distribution \cite{Elon2008}, whereas Erd\H{o}s-R\'enyi eigenvectors are localised especially for low values of $c$. The statistics of the first eigenvector components for very sparse symmetric random matrices has been first considered in the seminal work by Kabashima and collaborators \cite{Kabashima2010} and subsequently in a more systematic way in our previous work \cite{Susca2019}. Localisation properties of eigenvectors of sparse non-Hermitian random matrices have been investigated in \cite{metz2020localization}.

Following the framework developed in \cite{Susca2019}, we look at the second largest eigenpair problem as the top eigenpair problem for a \emph{deflated} version of the original sparse matrix (see discussion in Section \ref{sec:Formulation}). We will be implementing a Statistical Mechanics formulation of the top eigenpair problem of the deflated matrix, using both the cavity (Section \ref{sec:cavity_formulation}) and replica (\ref{sec:replica_gen}) methods in a unified way. 

Both the replica and cavity methods from the physics of disordered systems have been employed in the realm of random matrix theory for a long time. The replica method, traditionally used in the physics of spin glasses \cite{Zamponi2010}, was first introduced in the context of random matrices by Edwards and Jones \cite{Edwards1976} to compute the average spectral density of random matrices defined in terms of the joint probability density function of their entries. Later on, the same approach proved useful to derive the spectral density of Erd\H{o}s-R\'enyi  adjacency matrices as the solution of an intractable integral equation in the seminal paper of Bray and Rodgers \cite{Rodgers1988}. Later, approximation schemes such as the single defect approximation (SDA) and the effective medium approximation (EMA) \cite{Biroli1999, Semerjian2002} were developed. An exact alternative approach was introduced in \cite{Kuehn2008}: starting from Bray-Rodgers replica-symmetric setup \cite{Rodgers1988}, the functional order parameters of the theory are expressed as continuous superpositions of Gaussians with fluctuating variances, as suggested by earlier solutions of models for finitely coordinated harmonically coupled systems \cite{Kuehn2007}. This formulation gives rise to non-linear integral equations for the probability densities of such variances, which can be efficiently solved by a population dynamics algorithm. We will follow a similar approach in \ref{sec:replica_gen}.
 
The cavity method \cite{Mezard1987} has been employed in the study of disordered systems and sparse random matrices as a more direct alternative to replicas. It is exact for highly sparse tree structures \cite{Bordenave2010}. As shown in \cite{Rogers2008}, one of the advantages of the cavity method is that it provides the spectral density for very large single instances of sparse random graphs.
Both methods,  known to lead to the same results for the spectral density \cite{Slanina2011}, recover the Kesten-McKay law for the spectra of random regular graphs \cite{Kesten1959, McKay1981}, the Mar\v{c}enko-Pastur law and the Wigner's semicircle law respectively for sparse covariance matrices, and for Erd\H{o}s-R\'enyi adjacency matrices in the large $c$ limit \cite{Kuehn2008,Rogers2008}. Likewise, the spectral density of sparse Markov matrices \cite{Kuehn2015, Kuehn2015a}, graphs with modular \cite{Erguen2009} and small-world \cite{Kuehn2011} structure, and with topological constraints \cite{Rogers2010} have been obtained. Also, the spectral density in the complex plane of sparse non-Hermitian matrices has been considered in  \cite{Rogers2009,Neri2012,Neri2016,Metz2018}.

As in \cite{Susca2019}, we will provide a cavity single-instance derivation for our problem. Generalising the single-instance results in the thermodynamic limit, we will show that even for the second eigenpair problem the cavity method leads to the same stochastic recursions obtained from the replica treatment. The crucial difference between the present work and \cite{Susca2019} is the presence of the orthogonality condition between the top and second eigenvectors in the set of final recursion equations. The population dynamics algorithm employed to solve these recursions, complemented by a wise implementation of the orthogonality constraint, allows us to characterise the distributions of the cavity fields in the thermodynamic limit, and to disentangle the individual contributions of different degrees to the second eigenvector's entries.
 
The plan of the paper is as follows. In Section \ref{sec:Formulation}, we will formulate the problem in terms of \emph{deflation} and provide the main starting point. In Section \ref{sec:cavity_formulation}, we will describe the cavity approach to the problem, first for the single instance case, and then in the thermodynamic limit, highlighting the role of the orthogonality constraint (in \ref{sec:cavity_td_orto}). To complement the cavity results, we offer an equivalent replica treatment in \ref{sec:replica_gen}. In Section \ref{sec:RRG_ansatz} we focus on the case of the random regular graph: we analytically show how the solution for the top eigenpair of the \emph{deflated} adjacency matrix gets modified as the deflation parameter is changed. In Section \ref{sec:Markov}, we specialise our results to the case of Markov transition matrices representing random walks on graphs. In Section \ref{sec:pd}, we provide the details of the population dynamics algorithm, focussing on how the extra orthogonality constraint is implemented. We also provide convincing evidence that -- at odds with what is commonly believed -- the algorithm with finite population size $N_P$ does not actually capture the thermodynamic limit $N\to\infty$, in that there is a non-trivial relation between the size $N$ of the adjacency matrix being diagonalised, and the size $N_P$ of the population one should ideally use to numerically compute its spectral properties. More precisely, the accuracy -- measured with different metrics -- with which the population dynamics algorithm reproduces numerical diagonalisation of matrices (graphs) of size $N$ has a strongly non-monotonic behaviour as a function of $N_P$, thus implying that an optimal size $N_P^\star=N_P^\star(N)$ must be chosen to best reproduce the diagonalisation results. Finally, in Section \ref{sec:conclusion} we offer a summary of results.

\section{Formulation of the problem} \label{sec:Formulation}

We consider a real sparse symmetric random matrix $J =(J_{ij})$ and assume that its top eigenpair $(\lambda_1, \bm u)$ is known. We define a \emph{deflated} matrix $\tilde J(x)= (\tilde J_{ij} (x))$ by
\begin{equation}
\tilde{J}_{ij}(x)=J_{ij}-\frac{x}{N}u_{i}u_{j}\ ,\label{eq:matrix_def}
\end{equation}
where $x$ represents the deflation parameter. The top eigenvector $\bm{u}$ of $J$ is normalised such that $|\bm{u}|^2=N$.\footnote{The same normalisation convention applies to all the other eigenvectors of $J$, $\bm{v}_\alpha$ with $\alpha=2,...,N$.} In what follows, the vector $\bm{u}$ will be also referred to as the \emph{probe} eigenvector. The dense matrix $\bm{u}\bm{u}^T/N$ represents the projector onto the top eigenspace of the original matrix $J$.  
The  $J_{ij}=c_{ij}K_{ij}$ are the i.i.d. entries of the original sparse symmetric random matrix $J$. They are defined in terms of the connectivity matrix $c_{ij}\in\{0,1\}$, i.e. the adjacency matrix of the underlying graph, and the random variables $K_{ij}$ encoding bond weights. Within our formalism, we are able to handle any kind of highly sparse degree connectivity - where the mean node degree $\langle k\rangle=c$ is a finite constant that does not scale with $N$ (entailing $c/N\rightarrow0$ as $N\rightarrow\infty$). We will typically consider bounded degree distributions: a candidate of interest can be represented by a bounded Poisson distribution
\begin{equation}
P(k_i=k)=\mathcal{N}^{-1}\mathrm{e}^{-\bar{c}}\bar{c}^k/k!\ ,\qquad k=0,\ldots,k_{\mathrm{max}}\ ,\label{eq:degree_pmf}
\end{equation}
with the mean degree $c\equiv\left\langle k\right\rangle$ and $\mathcal{N}=\sum_{k=0}^{k_\mathrm{max}}\mathrm{e}^{-\bar{c}}\bar{c}^k/k!$ for normalisation. The bond weights $K_{ij}$ will be i.i.d. random variables drawn from a parent pdf $p_{K}(K)$ with bounded support. This setting is sufficient to ensure that the largest eigenvalue $\lambda_1$ of $J$ will remain of $\sim\mathcal{O}(1)$ for $N\to\infty$.
 
The spectral theorem ensures that $\tilde{J}(x)$ can be diagonalised via an orthonormal basis of
eigenvectors $\bm{v}_{\alpha}(x)$ with corresponding real eigenvalues
$\tilde{\lambda}_{\alpha}(x)$ ($\alpha=1,\ldots,N$),

\begin{equation}
\tilde{J}\bm{v}_{\alpha}=\tilde{\lambda}_{\alpha}\bm{v}_{\alpha}\ ,
\end{equation}
for each eigenpair $\alpha=1,\ldots,N$, where to simplify notation we have omitted the $x$-dependence. Assume that there is no eigenvalue degeneracy, and that they are sorted $\tilde{\lambda}_1>\tilde{\lambda}_2>\ldots >\tilde{\lambda}_N$, and the same holds for the eigenvalues $\lambda_\alpha$ ($\alpha=1,\ldots,N$) of the original matrix $J$.

For any value of $x$, the matrices $J$ and $\tilde{J}(x)$ share the same set of eigenvectors (see Section 3.3.2 in \cite{Bjoerck2015}). The range of the deflation parameter $x$ is $[0,\lambda_1]$, where the boundaries of this range correspond respectively to no deflation ($x=0\Rightarrow J=\tilde J$) and \emph{full deflation} ($x=\lambda_1$). 
\begin{itemize}
\item When the value of $x$ is smaller than the spectral gap $g=\lambda_1-\lambda_2$, the top eigenvalue of $\tilde{J}(x)$ is given by $\lambda_1-x$ with corresponding eigenvector $\bm{u}$. Indeed:
\begin{equation}
\tilde{J}\bm{u}=\left(J-\frac{x}{N}\bm{u}\bm{u}^T\right)\bm{u}=(\lambda_1-x)\bm{u}
\label{eq:deflation}
\end{equation}
with $\lambda_1-x>\lambda_2$. We recall that $\bm{u}^T\bm{u}=N$.
\item Conversely, when $x>g$  then the second largest eigenvalue of $J$, $\lambda_2$, and the corresponding eigenvector $\bm{v}_2$ become the top eigenpair of the matrix $\tilde{J}$. Indeed, following \eqref{eq:deflation}, the top eigenvector of $J$, $\bm{u}$, is still an eigenvector of $\tilde{J}$ related to the eigenvalue $\lambda_1-x$ but now $\lambda_2>\lambda_1-x$. Clearly,
\begin{equation}
\tilde{J}\bm{v}_2=\left(J-\frac{x}{N}\bm{u}\bm{u}^T\right)\bm{v}_2=\lambda_2\bm{v}_2\ ,
\end{equation}
in view of the orthogonality between $\bm{u}=\bm{v}_1$ and $\bm{v}_2$.
\item In particular, when $x=\lambda_1$, i.e. for full deflation\footnote{In the thermodynamic limit, the value of $x$ such that full deflation is achieved is actually the average largest eigenvalue $\left\langle \lambda_1 \right\rangle_J$ of the matrix $J$.},  the top eigenvector of $J$, $\bm{u}$, is still an eigenvector of $\tilde{J}$, but corresponding to a zero eigenvalue. Indeed,
\begin{equation}
\tilde{J}\bm{u}=\left(J-\frac{\lambda_1}{N}\bm{u}\bm{u}^T\right)\bm{u}=(\lambda_1-\lambda_1)\bm{u}=0\bm{u}\ .
\end{equation}
\item All other eigenpairs are unchanged. 

\end{itemize}
 
By setting up a formalism based on the statistical mechanics of disordered systems, we aim to find the average (or typical) value $\langle \lambda_2 \rangle_J$ of the second largest eigenvalue $\lambda_2$ of $J$, and the density $\rho_{J,2}(v)=\Big\langle\frac{1}{N}\sum_{i=1}^N\delta(v-v_2^{(i)}) \Big\rangle_J$ of the corresponding \emph{second largest} eigenvector's components, $\bm v_2 =(v_2^{(1)},\ldots,v_2^{(N)})$. 
The second eigenpair statistics of the matrix $J$ is obtained by finding the \emph{top} eigenpair of the \emph{deflated} matrix $\tilde{J}(x)$ when $x=\lambda_1$. Thus, in order to obtain the desired quantities, we analyse the average largest eigenvalue $\langle \tilde{\lambda}_1\rangle_{\tilde{J}}$  and the density $\rho_{\tilde{J}}(v)=\Big\langle\frac{1}{N}\sum_{i=1}^N\delta(v-v_1^{(i)}) \Big\rangle_{\tilde{J}}$ of the top eigenvector's components, $\bm v_1 =(v_1^{(1)},\ldots,v_1^{(N)})$ of the deflated matrix $\tilde{J}$, where the average $\langle\cdot\rangle_{\tilde{J}}$ is taken over the distribution of the matrix $\tilde{J}$.

We provide:
\begin{itemize}
\item the second largest eigenpair statistics $\langle \lambda_2\rangle_J$ and $\rho_{J,2}(v)$ of  the matrix $J$, i.e. the solution corresponding to the maximum deflation for $\tilde{J}$, in the case of a generic connectivity $p(k)$ with bounded maximum degree, found via the cavity method (Section \ref{sec:cavity_formulation}). 
We also offer an equivalent replica derivation for the same problem (\ref{sec:replica_gen}). In this general case, the solution is available via population dynamics simulations (Section \ref{sec:pd});
\item an explicit analytical solution for $\langle \tilde{\lambda}_1\rangle_{\tilde{J}}$  and $\rho_{\tilde{J}}(v)$  in the specific case of the adjacency matrix of a random regular graph (RRG), showing that the solution requires that the deflation parameter $x$ exceed the spectral gap (Section \ref{sec:RRG_ansatz});
\item  the second largest eigenpair statistics of the unbiased random walk Markov transition matrix. In this case, the deflation parameter $x$ is set precisely to 1, i.e. equal to the largest eigenvalue of the Markov transition matrix. Also in this case, an analytical description is provided for the RRG connectivity case (Section \ref{sec:Markov}).
\end{itemize}

The equations \eqref{eq:cav_pi_last}, \eqref{eq:cav_norm_last}, \eqref{eq:cav_orto_last} and \eqref{eq:cav_evect_last} found below within the cavity framework (see Section \ref{sec:cavity_td_orto}) represent the solution of the second largest eigenpair problem in the thermodynamic limit, and constitute the main result of this paper. We notice that they are completely equivalent to the equations \eqref{eq:pi_last}, \eqref{eq:norm_last}, \eqref{eq:orto_last} and  \eqref{eq:evect_last} found within the replica framework (see \ref{sec:replica_evect}).

We will follow the same protocol used in \cite{Susca2019}. Focussing on the matrix  $\tilde{J}$, the problem can be formulated as the optimisation 
of a quadratic function $\hat{H}(\bm{v})$, according to which $\bm{v}_{1}$ is
the vector normalised to $N$ that realises the condition

\begin{equation}
N\tilde{\lambda}_1=\min_{| \bm{v}|^2 =N}\left[ \hat{H}(\bm{v})\right] =\min_{| \bm{v}|^2 =N}\left[-\frac{1}{2}\left(\bm{v},\tilde{J}\bm{v}\right)\right]\ ,\label{eq:minimum}
\end{equation}
as dictated by the Courant-Fischer definition of eigenvectors. The round
brackets $\left(\cdot,\cdot\right)$ indicate the dot product between
vectors in $\mathbb{R}^{N}$. It is easy to show that $\hat{H}\left(\bm{v}\right)$ is bounded
\begin{equation}
-\frac{1}{2}\tilde{\lambda}_{1}N\leq\hat{H}\left(\bm{v}\right)\leq-\frac{1}{2}\tilde{\lambda}_{N}N\,,
\end{equation}
and attains its minimum
when computed on the top eigenvector.

For a fixed matrix $\tilde{J}$, the minimum in \eqref{eq:minimum} can be computed by introducing a fictitious canonical ensemble of $N$-dimensional vectors $\bm v$ at inverse temperature $\beta$, whose Gibbs-Boltzmann distribution reads
\begin{equation}
P_{\beta,\tilde{J}}(\bm v)=\frac{1}{Z}\exp\left[\frac{\beta}{2}(\bm v,\tilde{J}\bm v)\right]\delta(|\bm v|^2-N)\ ,\label{eq:hard}
\end{equation}
where the delta function enforces normalisation. Clearly, in the low temperature limit $\beta\to\infty$, only one ``state'' remains populated, which corresponds to $\bm v=\bm v_1$, the top eigenvector of the matrix $\tilde{J}$.

\section{Full deflation: cavity method} \label{sec:cavity_formulation}

In this section, we present the cavity derivation of the single instance equation for the second largest eigenpair problem. The formalism shown here differs from that presented in \cite{Susca2019}: here we analyse the partition function of the Boltzmann distribution \eqref{eq:hard}, rather than a soft-constrained version of it. This allows us to include hard constraints within the cavity framework. 
The equations expressing the solution can be easily generalised to the thermodynamic limit case, reproducing the same equations that will be found by the replica formalism in \ref{sec:replica_gen}, which constitute the main results of this work.

\subsection{Top eigenpair of a single instance: generic deflation case \label{sec:general_deflation}}
Given a single instance matrix $\tilde{J}$, its largest eigenvalue $\tilde\lambda_1$  can be defined as
\begin{equation}
\tilde\lambda_1=\lim_{\beta\rightarrow\infty}\frac{2}{\beta N} \ln Z_N,\qquad Z_N=\int\mathrm{d}\bm{v}\exp\left[ \frac{\beta}{2}\left(\bm{v},\tilde{J}\bm{v}\right)\right] \delta\left(\left|\bm{v}\right|^{2}-N\right)\ .\label{eq:single_lambda2}
\end{equation}
The partition function explicitly reads
\begin{equation}
Z_N=\int \mathrm{d}\bm{v} \exp\left[ \frac{\beta}{2}\left(\bm{v},J\bm{v}\right)-\frac{\beta x}{2N}\left(\bm{u},\bm{v}\right)^2 \right]\delta\left(\left|\bm{v}\right|^{2}-N\right)\ .\label{eq:cavity_part_funct}
\end{equation}
The square in the exponent can be written as
\begin{equation}
\frac{1}{N}\left(\bm{u},\bm{v}\right)^2=N\left[\frac{1}{N}\left(\bm{u},\bm{v}\right)\right]^2 =Nq^2\ ,
\end{equation}
with the identification
\begin{equation}
q=\frac{1}{N}\left(\bm{u},\bm{v}\right)\ .\label{eq:q_def}
\end{equation}
The definition of the order parameter $q$ is enforced via the integral identity
\begin{equation}
1=\int N\beta \frac{\mathrm{d}q\mathrm{d}\hat{q}}{2\pi} \exp\left(\mathrm{i}N\beta q \hat{q}-\mathrm{i}\beta \hat{q} (\bm{u},\bm{v})\right)\ .\label{eq:integral_q}
\end{equation}
By also employing a Fourier representation of the Dirac delta enforcing the normalisation constraint and including all the pre-factors in $\mathcal{C}$, the partition function becomes
\begin{align}
Z_N&=\mathcal{C}\int \mathrm{d}q\mathrm{d}\hat{q}\mathrm{d}\lambda \exp \left[ \beta N S_N(q,\hat{q},\lambda) \right]\ ,\label{eq:cavity_part_funct_sp}
\end{align}
where
\begin{equation}
 S_N(q,\hat{q},\lambda) =\mathrm{i}q\hat{q}-\frac{x}{2}q^2+\mathrm{i}\frac{\lambda}{2}+\frac{1}{N\beta}\mathrm{Log} \tilde{Z}_N\left(\hat{q},\lambda\right)
 \label{eq:action_SN}
\end{equation}
defines the action with
\begin{align}
\nonumber \tilde{Z}_N\left(\hat{q},\lambda\right)&=\int \prod_{i=1}^N \mathrm{d}v_i \exp \left[ - \frac{\beta}{2}\left(\bm{v},A\bm{v}\right)-\mathrm{i}\hat{q}\beta \left(\bm{u},\bm{v}\right)\right]\\
&=\sqrt{\frac{(2\pi)^N}{\beta^N\det(A)}}\exp\left(-\frac{\beta}{2}\hat{q}^2 \bm{u}^T A^{-1} \bm{u} \right)\ .
\label{eq:cavity_reduced_part_funct}
\end{align}
The  matrix $A^{-1}=(\mathrm{i}\lambda\mathbbm{1}_N-J)^{-1}$ is related to the \emph{resolvent} of $J$. It has the same eigenvectors as $J$, thus using the spectral theorem it can be expressed as
\begin{equation}
A^{-1}=(\mathrm{i}\lambda\mathbbm{1}_N-J)^{-1}=\sum_{\alpha=1}^N \frac{1}{\mathrm{i}\lambda-\lambda_\alpha}\tilde{\bm{u}}_\alpha \tilde{\bm{u}}_\alpha^T\ ,
\end{equation}
where the $\lambda_\alpha$ are the eigenvalues of $J$ and the $\tilde{\bm{u}}_\alpha$ are their corresponding eigenvectors. Notice that the $\tilde{\bm{u}}_\alpha$ are normalised such that $|\tilde{\bm{u}}_\alpha|^2=1$. On the other hand, the vector $\bm{u}$ appearing in the exponent of $\tilde{Z}_N$ is the top eigenvector of $J$, $\bm{u}_1$, normalised such that $|\bm{u}|^2=|\bm{u}_1|^2=N$. Therefore, 
\begin{equation}
\bm{u}=\bm{u}_1=\sqrt{N}\tilde{\bm{u}}_1\ ,
\end{equation}
entailing that 
\begin{equation}
\tilde{Z}_N\left(\hat{q},\lambda\right)\propto\exp\left(-\frac{\beta N}{2}\frac{\hat{q}^2}{\mathrm{i}\lambda-\lambda_1}\right)\ .
\end{equation}
In turn, the action \eqref{eq:action_SN} becomes
\begin{equation}
S_N(q,\hat{q},\lambda)=\mathrm{i}q\hat{q}-\frac{\lambda_1}{2}q^2+\mathrm{i}\frac{\lambda}{2}-\frac{\hat{q}^2}{2(\mathrm{i}\lambda-\lambda_1)}\ .
\label{eq:action_SN_ok}
\end{equation}

Taking into account \eqref{eq:action_SN_ok}, Eq. \eqref{eq:cavity_part_funct_sp} can be evaluated with a saddle-point approximation for large $\beta$. The stationarity of \eqref{eq:action_SN_ok} w.r.t. to $\lambda$, $\hat{q}$ and $q$ implies that
\begin{align}
&1 =\frac{(\mathrm{i}\hat{q}^\star)^2}{(\mathrm{i}\lambda^\star-\lambda_1)^2}\ ,
\label{eq:stat_cond_lambda}\\
&q^\star=\frac{-\mathrm{i}\hat{q}^\star}{\mathrm{i}\lambda^\star-\lambda_1}\ ,
\label{eq:stat_cond_hatq}\\
&q^\star x=\mathrm{i}\hat{q}^\star\ .
\label{eq:stat_cond_q}
\end{align}
Using \eqref{eq:stat_cond_q} in \eqref{eq:stat_cond_hatq}, one finds
\begin{equation}
q^\star=\frac{-q^\star x}{\mathrm{i}\lambda^\star-\lambda_1}\ .
\label{eq:q_in_hatq}
\end{equation}

Two cases can be distinguished, depending on the value of $q^\star$.
\begin{enumerate}
\item Assuming $q^\star\neq 0$, Eq. \eqref{eq:q_in_hatq} yields $\mathrm{i}\lambda^\star=\lambda_1-x$, while from \eqref{eq:stat_cond_lambda} it follows that ${q^\star}^2=1$. Using these results to express the action $S_N$, one finds
\begin{equation}
S_N=\frac{\mathrm{i}\lambda^\star}{2}+\frac{x}{2}-\frac{x}{2}=\frac{\lambda_1-x}{2}\ ,
\label{eq:action_sp_q1}
\end{equation}
entailing for the largest eigenvalue $\tilde\lambda_1$ defined in  \eqref{eq:single_lambda2}
\begin{equation}
\tilde\lambda_1=\mathrm{i}\lambda^\star=\lambda_1-x\ .
\label{eq:top_eval_deflated}
\end{equation}
As stated in Section \ref{sec:Formulation}, this is the top eigenvalue of $\tilde{J}$ when $x<g$. Indeed, the value $q^\star=\pm1$ indicates that the the probe eigenvector $\bm{u}$ and the top eigenvector of $\tilde{J}$ corresponding to $\lambda_1-x$ coincide. Thus, when the deflation parameter $x$ is smaller that the spectral gap $g$, there is no need to use the cavity method to obtain the top eigenpair of the deflated matrix, which is simply given by $(\lambda_1-x,\bm{u})$.
\item Assuming $q^\star=0$, it follows from \eqref{eq:stat_cond_q} that $\hat{q}^\star=0$. Thus the action reduces to
\begin{equation}
S_N=\frac{\mathrm{i}\lambda^\star}{2}\Rightarrow \tilde\lambda_1=\mathrm{i}\lambda^\star\ .
\end{equation}
The case $q^\star=0$ provides the top eigenvalue of the deflated matrix $\tilde{J}$ in the case $x>g$, including the case of full deflation $x=\lambda_1$. Therefore, $q^\star=\hat{q}^\star=0$ represents the orthogonality condition between the solution $\bm{v}$ and the probe eigenvector $\bm{u}$. In this scenario, the top eigenvalue $\tilde\lambda_1$ of the matrix $\tilde{J}$ is the \emph{second largest} eigenvalue of the original matrix $J$, viz.
\begin{equation}
\tilde\lambda_1=\lambda_2=\mathrm{i}\lambda^\star\ .  
\label{eq:cavity_second_eval}
\end{equation}
However, the stationarity conditions \eqref{eq:stat_cond_lambda},\eqref{eq:stat_cond_hatq} and \eqref{eq:stat_cond_q} do not provide the actual (real) value $\mathrm{i}\lambda^\star\equiv\lambda$, nor the components of the corresponding eigenvector $\bm{v}$.
\end{enumerate}
To sum up, the top eigenvalue of the deflated matrix $\tilde{J}$ is always given by the value $\mathrm{i}\lambda^\star\equiv\lambda$, regardless the value of $x$. However, for $x>g$ this value needs to be determined via the cavity method, as detailed in the next subsection.

\subsection{Cavity derivation for a single instance in case of full deflation \label{sec:cavity_eval}}
We focus on the case of full deflation $x=\lambda_1$. This choice is not restrictive, since the solution does not depend on $x$, for any $x>g$. As shown before, in the range $x>g$ one has $q^\star=0$.
However, for the time being we proceed with a generic $q^\star$.  Its actual value will be made explicit in the final result.

One looks at \eqref{eq:cavity_part_funct_sp}, without performing explicitly the integration in \eqref{eq:cavity_reduced_part_funct}.
Considering the stationarity of the action $S_N(q,\hat{q},\lambda)$ w.r.t $q$, $\hat{q}$ and $\lambda$, the following conditions hold,
\begin{align}
\mathrm{i}\hat{q}^\star&=\lambda_1 q^\star\ \label{eq:sp1}, \\
q^\star&=\frac{1}{N}\sum_{i=1}^N u_i \left\langle v_i \right\rangle\  \label{eq:q_cav}\ , \\
1&=\frac{1}{N} \sum_{i=1}^N \left\langle v_i^2 \right\rangle  \label{eq:norm}\ ,
\end{align}
where the starred quantities indicate the saddle-point values of the parameters. The angular brackets indicate averaging w.r.t. the distribution
\begin{equation}
P_{\beta}(\bm{v}|\hat{q}^\star,\lambda^\star)=\frac{1}{\tilde{Z}_N(\hat{q}^\star,\lambda^\star)}\exp \left( -\mathrm{i}\lambda^\star \frac{\beta}{2}\sum_{i} v_i^2-\mathrm{i}\hat{q}^\star\beta \left(\bm{u},\bm{v}\right)+\frac{\beta}{2}\left(\bm{v},J\bm{v}\right)\right)\ .
\label{eq:cavity_p_first}
\end{equation} 
By looking at the saddle point condition \eqref{eq:sp1}, in what follows we can identify $\mathrm{i}\hat{q}^\star=\lambda_1 q^\star=\lambda_1 q$ (omitting the star for brevity) and define $\mathrm{i}\lambda^\star\equiv\lambda$, such that \eqref{eq:cavity_p_first} becomes
\begin{equation}
P_{\beta}(\bm{v}|\lambda_1q,\lambda)=\frac{1}{\tilde{Z}_N(\lambda_1q,\lambda)}\exp \left( -\lambda\frac{\beta}{2}\sum_{i} v_i^2-\lambda_1q\beta \left(\bm{u},\bm{v}\right)+\frac{\beta}{2}\left(\bm{v},J\bm{v}\right)\right)\ .
\label{eq:cavity_p}
\end{equation}

The components $v_i$ are found in the $\beta\to\infty$ limit by the cavity method applied to the distribution \eqref{eq:cavity_p} \footnote{It can be noticed that the distribution \eqref{eq:cavity_p} is substantially equivalent to the \emph{grand-canonical} distribution (Eq. (7) in \cite{Susca2019}) which we adopt as the starting point of the cavity treatment in \cite{Susca2019}.}.  Here we will follow the protocol detailed in Section 3.1 of \cite{Susca2019}, reporting the key steps to make this paper self-contained.

By making a tree-like assumption on the structure of the highly sparse graph encoded in the original matrix $J$ that we deflate, the marginal pdf w.r.t. a certain component $i$ is given by
\begin{equation}
P_i(v_i|\lambda_1q,\lambda)=\frac{1}{Z_i}\exp\left( -\frac{\beta}{2}\lambda v_i^2-\beta\lambda_1qu_iv_i\right)\prod_{j\in\partial i}\int \mathrm{d}v_j \exp\left(\beta v_i J_{ij} v_j \right) P_j^{(i)}(v_j|\lambda_1q,\lambda)\ ,
\end{equation}
where $\partial i$ denotes the immediate neighbourhood of $i$. The factorisation over the neighbouring nodes of $i$ is due to the fact that in a tree-like structure the nodes $j\in\partial i$ are connected with each other only through $i$. The distribution $P_j^{(i)}(v_j|\lambda_1q,\lambda)$ is called \emph{marginal cavity} distribution: it is the distribution of the components $v_j$ defined on the neighbouring nodes of $i$, in the network in which $i$ has been removed.

In the same way (see for instance Eq.~(11) in \cite{Susca2019}), for any $j \in\partial i$ the cavity marginal pdf satisfies the self-consistent equation
\begin{equation}
P_j^{(i)}(v_j|\lambda_1q,\lambda)=\frac{1}{Z_j^{(i)}}\exp\left( -\frac{\beta}{2}\lambda v_j^2-\beta\lambda_1qu_jv_j\right)\prod_{\ell\in\partial j\backslash i}\int \mathrm{d}v_{\ell} \exp\left(\beta v_j J_{j\ell} v_{\ell} \right) P_{\ell}^{(j)}(v_{\ell}|\lambda_1q,\lambda)\ ,\label{eq:cavity_pdf}
\end{equation}
where $\partial j\backslash i$ indicates the set of neighbours of the node $j$ with the exclusion of $i$.

A Gaussian ansatz provides the solution to the self consistent equation, viz.
\begin{equation}
P_j^{(i)}(v_j|\lambda_1q,\lambda)=\sqrt{\frac{\beta \omega_j^{(i)}}{2\pi}}\exp\left(-\frac{\beta h_j^{(i)}}{2 \omega_j^{(i)}}\right) \exp\left( -\frac{\beta}{2}\omega_j^{(i)}v_j^2+\beta h_j^{(i)}v_j\right)\ ,\label{eq:cavity_ansatz}
\end{equation}
where the parameters $\omega_j^{(i)}$ and $h_j^{(i)}$ are called \emph{cavity fields}. By inserting the ansatz in \eqref{eq:cavity_pdf} and performing the Gaussian integrals, the set of self-consistent equations represented by \eqref{eq:cavity_pdf} translates into a set of recursions for the cavity fields, 
\begin{equation}
\omega_j^{(i)}=\lambda- \sum_{\ell \in \partial j \backslash i} \frac{J_{j\ell}^2}{\omega_\ell^{(j)}}, \label{eq:cavity_omega}
\end{equation}
\begin{equation}
h_j^{(i)}=-\lambda_1qu_j+\sum_{\ell\in\partial j \backslash i} \frac{J_{j\ell}h_\ell^{(j)}}{\omega_\ell^{(j)}} \ . \label{eq:cavity_h}
\end{equation}
Likewise, by means of \eqref{eq:cavity_ansatz}, the marginal distribution $P_i(v_i|\lambda_1q,\lambda)$ can be written as
\begin{equation}
P_i(v_i|\lambda_1q,\lambda)=\frac{1}{Z_i}\exp\left(-\frac{\beta}{2}\omega_i v_i^2+\beta h_i v_i \right)\ ,
\label{eq:cavity_marginal_factorised}
\end{equation}
where
\begin{equation}
\omega_i=\lambda-\sum_{j\in\partial i} \frac{J_{ij}^2}{\omega_j^{(i)}}\ , \label{eq:cavity_omega_marg}
\end{equation}
\begin{equation}
h_i=-\lambda_1qu_i+\sum_{j\in\partial i} \frac{J_{ij}h_j^{(i)}}{\omega_j^{(i)}}\ . \label{eq:cavity_h_marg}
\end{equation}

Using the cavity factorisation in \eqref{eq:cavity_marginal_factorised} to express \eqref{eq:cavity_p}, we eventually obtain
\begin{equation}
P_{\beta}(\bm{v}|\lambda_1q,\lambda)=\prod_{i=1}^N\frac{1}{Z_i}\exp\left(-\frac{\beta}{2}\omega_i v_i^2+\beta h_i v_i \right)\ .
\label{eq:cavity_p_factorised}
\end{equation}
In the $\beta\to\infty$ limit,
\begin{equation}
P_{\beta}(\bm{v}|\lambda_1 q, \lambda)\to\prod_{i=1}^N \delta\left( v_i-\frac{h_i}{\omega_i}\right)\ ,\label{eq:cav_components}
\end{equation}
entailing that the components $v_i$ of the top eigenvector of the fully deflated matrix $\tilde{J}$, representing the ground state of the system with Boltzmann distribution \eqref{eq:hard}, are given by the ratios $h_i/\omega_i$. The $\omega_i$ and the $h_i$ are determined respectively by Eq.  \eqref{eq:cavity_omega_marg} and  \eqref{eq:cavity_h_marg}.  Because of the full deflation, $\bm{v}$ also represents the \emph{second largest} eigenvector of the original matrix $J$.

In terms of Eq. \eqref{eq:cav_components}, the conditions \eqref{eq:q_cav} and \eqref{eq:norm} read
\begin{align}
q=&\frac{1}{N}\sum_{i=1}^N u_i \frac{h_i}{\omega_i}\ ,\label{eq:q_cav_single}\\
1=&\frac{1}{N}\sum_{i=1}^N \left( \frac{h_i}{\omega_i} \right)^2\ .
\label{eq:lambda_cav_single}
\end{align}
At this point, we recall that for any $x>g$ (in particular $x=\lambda_1$), we have $q^\star\equiv q=0$. Therefore $q=0$ must be considered in Eq. \eqref{eq:cavity_h} and \eqref{eq:cavity_h_marg}, and the condition \eqref{eq:q_cav_single} becomes
\begin{equation}
0=\frac{1}{N}\sum_{i=1}^N u_i \frac{h_i}{\omega_i}\ . \label{eq:cavity_q0}
\end{equation}
As anticipated in Section \ref{sec:general_deflation}, Eq. \eqref{eq:cavity_q0} expresses the orthogonality condition between $\bm{u}$ and $\bm{v}$. The components $u_i$ and $v_i$ in \eqref{eq:cavity_q0} are naturally referring to the same node $i$ with degree $k_i$ of the network represented by $J$.

To summarise, in the single instance full-deflation case the solution is given by the cavity recursions \eqref{eq:cavity_omega} and \eqref{eq:cavity_h} along with the normalisation condition \eqref{eq:lambda_cav_single} and the orthogonality constraint  \eqref{eq:cavity_q0}.
 The value $\lambda=\lambda_2$ represents the second largest eigenvalue of the matrix $J$ (i.e. the top eigenvalue of the deflated matrix $\tilde{J}$), with corresponding eigenvalue $\bm{v}$ whose components are defined in Eq. \eqref{eq:cav_components}. According to the same mechanism explained in Appendix A of \cite{Susca2019}, it is the only value that satisfies the normalisation condition \eqref{eq:lambda_cav_single}.

\subsection{Cavity method: thermodynamic limit}\label{sec:cavity_td}
Following the reasoning of Section 3.2 in \cite{Susca2019}, in the limit $N\to\infty$ we can consider the \emph{joint probability density} of the cavity fields $\omega_j^{(i)}$ and $h_j^{(i)}$ taking values around respectively $\omega$ and $h$,
\begin{align}
\nonumber \pi\left(\omega,h\right)=&\sum_{k=1}^{k_{\mathrm{max}}}\frac{k}{c}p\left(k\right)\int\mathrm{d}u\rho_{J}(u|k)\!\!\int\left[\prod_{\ell=1}^{k-1}\mathrm{d}\pi\left(\omega_{\ell},h_{\ell}\right)\right]\\
&\times\left\langle \delta\left(\omega-\lambda+\sum_{\ell=1}^{k-1}\frac{K_{\ell}^{2}}{\omega_{\ell}}\right)\delta\left(h-\left(-qu\left\langle \lambda_1 \right\rangle_{J}+\sum_{\ell=1}^{k-1}\frac{h_{\ell}K_{\ell}}{\omega_{\ell}}\right)\right)\right\rangle _{{\{K\}}_{k-1}}\ ,\label{eq:cavity_pdf_td}
\end{align}
where $\mathrm{d}\pi\left(\omega_{\ell},h_{\ell}\right)\equiv\mathrm{d}\omega_\ell\mathrm{d}h_\ell \pi\left(\omega_{\ell},h_{\ell}\right)$, and the average $\langle\cdot\rangle_{{\{K\}}_{k-1}}$ is taken over $k-1$ independent realisations of the bond weights $K$. Here, $\rho_J(u|k)$ is the distribution of the top eigenvector's component of $J$ conditioned to the degree $k$. The distribution $\frac{k}{c}p(k)$ represents the probability that a randomly chosen link points to a node of degree $k$ and $c=\langle k\rangle$, and appears in \eqref{eq:cavity_pdf_td} as cavity fields are related to links. Eq. \eqref{eq:cavity_pdf_td} generalises in the thermodynamic limit the recursions \eqref{eq:cavity_omega} and \eqref{eq:cavity_h} in the case of full deflation ($x=\lambda_1$).

By using the law of large numbers, in the thermodynamic limit the normalisation condition \eqref{eq:lambda_cav_single} reads
\begin{equation}
1=\sum_{k=0}^{k_{\mathrm{max}}}p(k)\int\mathrm{d}u\rho_{J}(u|k)\!\!\int\left[\prod_{\ell=1}^{k}\mathrm{d}\pi\left(\omega_{\ell},h_{\ell}\right)\right]\left\langle 
\left(\frac{-qu\left\langle \lambda_1 \right\rangle_{J}+\sum_{\ell=1}^{k}\frac{h_{\ell}K_{\ell}}{\omega_{\ell}}}{\lambda-\sum_{\ell=1}^{k}\frac{K_{\ell}^{2}}{\omega_{\ell}}}\right  )^2\right\rangle _{{\{K\}}_{k}}\ ,\label{eq:cavity_norm_td}
\end{equation}
whereas the orthogonality constraint \eqref{eq:q_cav_single} becomes
\begin{equation}
q=\sum_{k=0}^{k_{\mathrm{max}}}p(k)\int\mathrm{d}u\rho_J(u|k)u\!\!\int\left[\prod_{\ell=1}^{k}\mathrm{d}\pi\left(\omega_{\ell},h_{\ell}\right)\right]\left\langle 
\frac{-qu\left\langle \lambda_1 \right\rangle_{J}+\sum_{\ell=1}^{k}\frac{h_{\ell}K_{\ell}}{\omega_{\ell}}}{\lambda-\sum_{\ell=1}^{k}\frac{K_{\ell}^{2}}{\omega_{\ell}}}\right\rangle _{{\{K\}}_{k}}\ .\label{eq:cavity_ort_td}
\end{equation}
Similarly, the distribution of the top eigenvector's components of the fully deflated matrix $\tilde{J}$, i.e. the second largest eigenvector of $J$, is obtained in terms of averages w.r.t. the distribution $\pi(\omega,h)$ as
\begin{equation}
\rho_{\tilde{J}}(v)=\rho_{J,2}(v)=\sum_{k=0}^{k_{\mathrm{max}}}p(k)\int\mathrm{d}u\rho_J(u|k)\!\!\int\left[\prod_{\ell=1}^{k}\mathrm{d}\pi\left(\omega_{\ell},h_{\ell}\right)\right] \left\langle \delta \left( v-\frac{-qu\left\langle \lambda_1 \right\rangle_{J}+ \sum_{\ell=1}^{k}\frac{h_{\ell}K_{\ell}}{\omega_{\ell}}}{{\lambda-\sum_{\ell=1}^{k}\frac{K_{\ell}^{2}}{\omega_{\ell}}}} \right)\right\rangle_{{\{K\}}_{k}}\ .\label{eq:cavity_evect_td}
\end{equation}
We notice that in the equations \eqref{eq:cavity_norm_td}, \eqref{eq:cavity_ort_td} and \eqref{eq:cavity_evect_td}, the degree distribution $p(k)$ naturally crops up, as they encode properties related to nodes, rather than links.  Moreover, Eq. \eqref{eq:cavity_second_eval} generalises to the thermodynamic limit case, as 
\begin{equation}
\langle\tilde{\lambda}_1\rangle_{\tilde{J}}=\langle\lambda_2\rangle_J=\lambda\ .\label{eq:cavity_second_eval_td}
\end{equation}
for any $x>g$. We anticipate that the latter result is equivalent to the average second largest eigenvalue explicitly found by the replica approach in Eq. \eqref{eq:lambda2_gen} in \ref{sec:replica_gen}. Finally, we remark that Eq. \eqref{eq:top_eval_deflated} generalises at the ensemble level too, entailing the condition
\begin{equation}
\langle \lambda_1 \rangle_{\tilde{J}}=\lambda=\langle \lambda_1 \rangle_J-x
\label{eq:cavity_eval_td_out}
\end{equation}
which is valid when $x<g$.

\subsubsection{Cavity method: the orthogonality condition }\label{sec:cavity_td_orto}
The condition $q=0$, valid whenever $x$ exceeds the spectral gap, holds at the ensemble level as well. Indeed, when considering $q=0$, Eq. \eqref{eq:cavity_ort_td} encodes the orthogonality-on-average condition between the \emph{probe} eigenvector $\bm{u}$ and the top eigenvector $\bm{v}$ of the deflated matrix $\tilde{J}$, corresponding to the second largest eigenvector of the original matrix $J$. The interpretation of Eq. \eqref{eq:cavity_ort_td} for $q=0$ is made clearer by simply considering the average orthogonality condition between $\bm{u}$ and $\bm{v}$, viz.
\begin{align}
\nonumber 0&=\int\mathrm{d}u\mathrm{d}vP_J(u,v)uv\\
\nonumber &=\sum_{k=0}^{k_{\mathrm{max}}}p(k)\int\mathrm{d}u\rho_{J}(u|k)\mathrm{d}v\rho_{J,2}(v|u,k)uv\\
&=\sum_{k=0}^{k_{\mathrm{max}}}p(k)\int\mathrm{d}u\rho_{J}(u|k)u\int\left[\prod_{\ell=1}^{k}\mathrm{d}\pi\left(\omega_{\ell},h_{\ell}\right)\right] \left\langle \left(\frac{-qu\left\langle \lambda_1 \right\rangle_{J}+\sum_{\ell=1}^k \frac{h_\ell K_\ell}{\omega_\ell}}{\lambda-\sum_{\ell=1}^k \frac{K_\ell^2}{\omega_\ell}}\right)\right\rangle_{\{K\}_{k}}\ ,
\label{eq:cavity_orto_gen}
\end{align}
where $P_J(u,v)$ indicates the joint probability density of the first and  second largest eigenvector's components of $J$, and the conditional pdf $\rho_{J,2}(v|u,k)$ is obtained from \eqref{eq:cavity_evect_td} erasing the $u$-integration and the $k$-sum. The conditional pdf $\rho_J(u|k)$ is given by omitting the $k$-sum in the expression for the density of the top eigenvector components $\rho_J(u)$ \eqref{eq:density_top}. Comparing Eq. \eqref{eq:cavity_orto_gen} with \eqref{eq:cavity_ort_td} for $q=0$, it follows that they are equivalent.

Taking into account the average orthogonality condition $q=0$, the equations \eqref{eq:cavity_pdf_td}, \eqref{eq:cavity_norm_td}, \eqref{eq:cavity_ort_td}, \eqref{eq:cavity_evect_td} and \eqref{eq:cavity_second_eval_td} simplify to
\begin{align}
\pi(\omega,h)&=\sum_{k=1}^{k_{\mathrm{max}}}p(k)\frac{k}{c}\int\{\mathrm{d}\pi\} _{k-1}\left\langle \delta\left(\omega-\left(\lambda-\sum_{\ell=1}^{k-1}\frac{K_{\ell}^2}{\omega_{\ell}}\right)\right)\delta\left(h-\left( \sum_{\ell=1}^{k-1}\frac{h_{\ell}K_{\ell}}{\omega_{\ell}}\right)\right)\right\rangle_{\{K\}_{k-1}}\ ,\label{eq:cav_pi_last} \\ 
1&=\sum_{k=0}^{k_{\mathrm{max}}}p(k)\int\{\mathrm{d}\pi\} _{k} \left\langle \left(\frac{\sum_{\ell=1}^k \frac{h_\ell K_\ell}{\omega_\ell}}{\lambda-\sum_{\ell=1}^k \frac{K_\ell^2}{\omega_\ell}}\right)^{2}\right\rangle_{\{K\}_{k}}\ ,\label{eq:cav_norm_last} \\ 
0&=\sum_{k=0}^{k_{\mathrm{max}}}p(k)\int\mathrm{d}u \rho_J(u|k)u \int \{\mathrm{d}\pi\} _{k} \left\langle \left(\frac {\sum_{\ell=1}^k \frac{h_\ell K_\ell}{\omega_\ell}}{\lambda-\sum_{\ell=1}^k \frac{K_\ell^2}{\omega_\ell}}\right)\right\rangle_{\{K\}_{k}}\ ,\label{eq:cav_orto_last} \\ 
\rho_{\tilde{J}}(v)&\equiv\rho_{J,2}(v)=\sum_{k=0}^{k_{\mathrm{max}}}p(k)\int\{\mathrm{d}\pi\} _{k}\left\langle \delta\left(v-\frac{\sum_{\ell=1}^k \frac{h_\ell K_\ell}{\omega_\ell}}{\lambda-\sum_{\ell=1}^k \frac{K_\ell^2}{\omega_\ell}}\right)\right\rangle_{\{K\}_{k}}\ ,\label{eq:cav_evect_last}\\
\left\langle \tilde{\lambda_1} \right\rangle_{\tilde{J}}&\equiv \left\langle \lambda_2 \right\rangle_J=\lambda \label{eq:cav_eval_last}\ ,
\end{align}
where we have used the shorthand $\{\mathrm{d}\pi\} _{k}=\prod_{\ell=1}^{k}\mathrm{d}\omega_{\ell}\mathrm{d}h_{\ell}\pi(\omega_{\ell},h_{\ell})$ .

Enforcing the orthogonality condition given by \eqref{eq:cav_orto_last} is crucial to find the correct solution. The conditional pdf $\rho_J(u|k)$ appearing in \eqref{eq:cav_orto_last} is given by omitting the $k$-sum in the expression for the density of the top eigenvector components $\rho_J(u)$ (see Eq. (111) in \cite{Susca2019}), reported here 
\begin{equation}
\rho_J(u)=\sum_{k=0}^{k_{\mathrm{max}}}p(k)\int\left\{\mathrm{d}\pi_1\right\}_{k}\left\langle\delta\left(u-\frac{\sum_{\ell=1}^{k}\frac{b_{\ell}K_{\ell}}{a_{\ell}}}{\langle\lambda_1\rangle_J-\sum_{\ell=1}^{k}\frac{K_{\ell}^2}{a_{\ell}}}\right)\right\rangle_{\{K\}_k}\ ,\label{eq:density_top}
\end{equation}
where $\pi_1(a,b)$ indicates the distribution of cavity fields of type $a$ and $b$ for the top eigenpair problem\footnote{In the context of the top eigenpair problem, the cavity field of type $a$  has the role of an inverse cavity variance (similarly to $\omega$ for the second largest eigenpair problem), whereas $b$ represents a cavity bias (similarly to $h$ here). See Section 3 in \cite{Susca2019}.}. The integration w.r.t. the conditional distribution $\rho_J(u|k)$ in \eqref{eq:cav_orto_last} generalises to the thermodynamic limit the fact that both the components $u_i$ and $v_i=\frac{h_i}{\omega_i}$ in \eqref{eq:cavity_q0} refer to the same node $i$ with degree $k_i$. Indeed, by comparing \eqref{eq:density_top} with \eqref{eq:cav_evect_last} and \eqref{eq:cav_orto_last}, we notice that the components of $\bm{u}$ are still coupled to those of $\bm{v}$ in \eqref{eq:cav_orto_last} through their structure, as they both refer to the same degree $k$ (see Section \ref{sec:pd} for more details). The replica derivation in \ref{sec:replica_gen} provides an independent proof of this result.

Therefore, in order to enforce the constraint \eqref{eq:cav_orto_last} correctly, we need to impose strict orthogonality on-the-fly, i.e. while the components of the top eigenvector $\bm{u}$ and the components of the second largest eigenvector $\bm{v}$ are being evaluated at the same time by averaging w.r.t. respectively $\pi_1$ and $\pi$, as prescribed  by \eqref{eq:density_top} and \eqref{eq:cav_pi_last}. The way strict orthogonality is imposed is via a correction to the components of $\bm{v}$: the details of this procedure and the corresponding algorithm are given in Section \ref{sec:pd}. 
We remark that the condition $q=0$ holds whenever $x$ exceeds the spectral gap.

To summarise, the equations \eqref{eq:cav_pi_last},  \eqref{eq:cav_norm_last}, \eqref{eq:cav_orto_last}, \eqref{eq:cav_evect_last} and \eqref{eq:cav_eval_last} represent the solution of the second largest eigenpair problem in the thermodynamic limit and constitute the main result of this paper. This set of equations must be generally solved by a population dynamics algorithm, as detailed in Section \ref{sec:pd}.
It is completely equivalent to the equations \eqref{eq:pi_last}, \eqref{eq:norm_last}, \eqref{eq:orto_last}, \eqref{eq:eval_last} and  \eqref{eq:evect_last}, respectively, found within the replica framework (See \ref{sec:replica_evect}).

\begin{figure}
\begin{centering}
\includegraphics[scale=0.3]{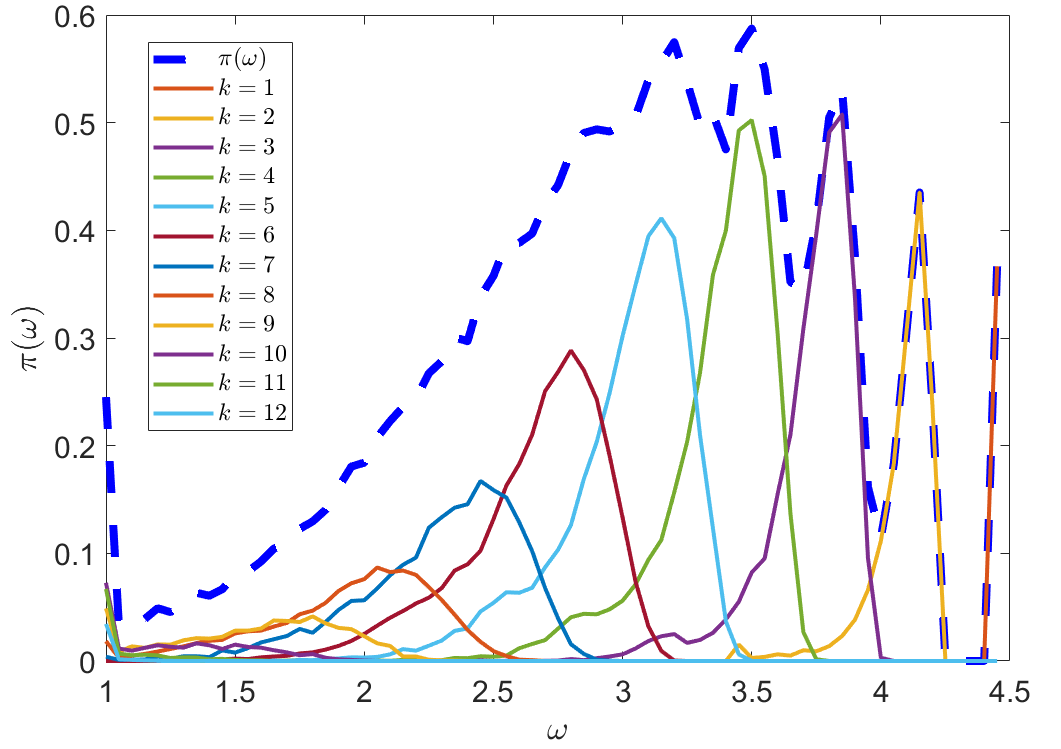}\includegraphics[scale=0.37]{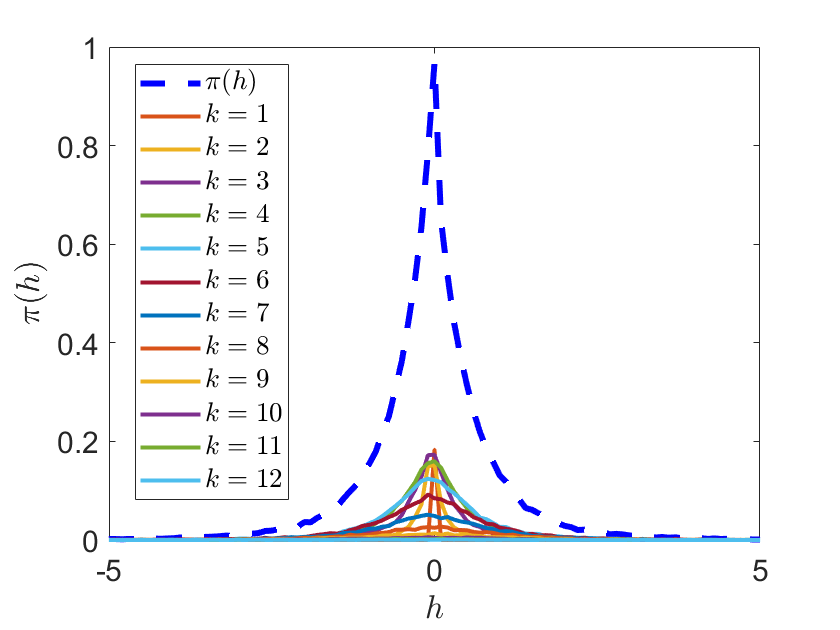}
\par\end{centering}
\begin{centering}
\includegraphics[scale=0.35]{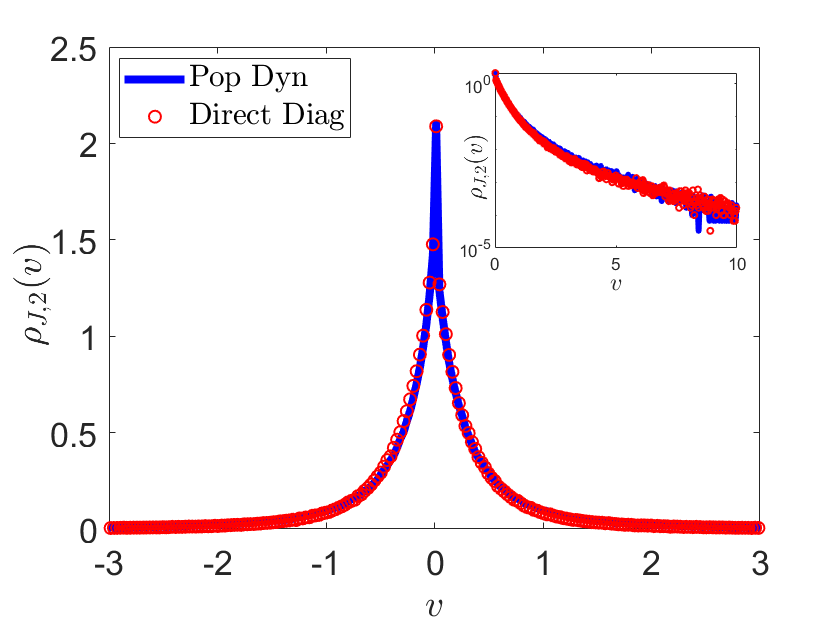}\includegraphics[scale=0.36]{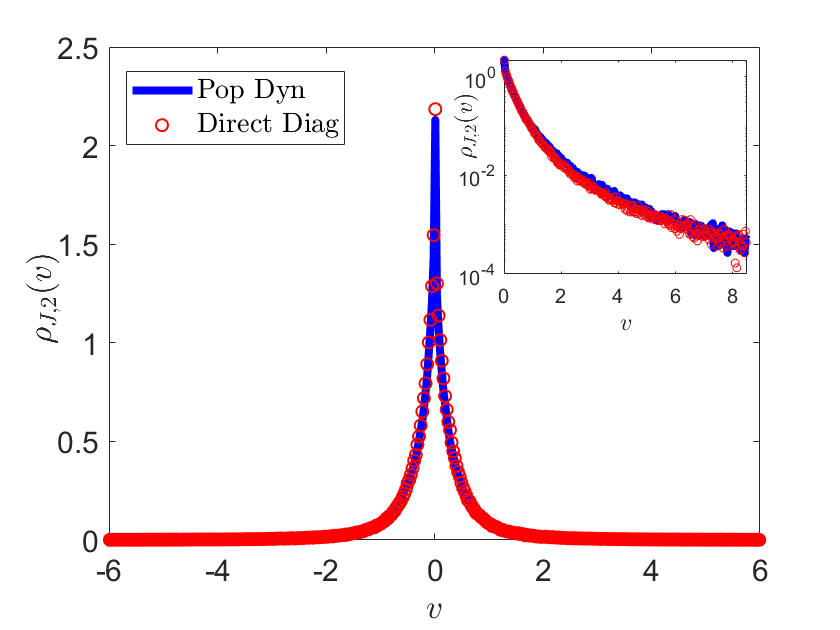}
\par\end{centering}
\begin{centering}
\caption{Second largest eigenpair of the ER adjacency matrix. {\bf Top left
panel}: marginal distribution of the inverse single site variances $\omega$. The thick dashed line represents the full pdf, the thinner curves underneath stand for the single degree contributions, from $k=1$ to $k=12$. The rightmost peak at $\omega=\lambda$ corresponds to $k=1$: the peaks are centered at lower $\omega$ as the degree $k$ increases. {\bf Top right panel}: marginal pdf
of the single-site bias fields $h$. Again, the thick dashed line represents the full pdf,
the thinner solid curves stand for the degree contributions from $k=1$ to $k=12$. Each curve corresponding to a degree $k$ is symmetric around $h=0$. As $k$ grows, their variance broadens and the curves flatten.
{\bf Bottom left panel}: pdf of the second largest eigenvector's components (see \eqref{eq:cav_evect_last}), obtained by population dynamics (solid blue) and by direct diagonalisation of $2000$ matrices of size $N=5000$ (red circles) showing excellent agreement. The population size is $N_P=10^5$. The inset shows the right tail of the pdf in log scale. {\bf Bottom right panel}: pdf of the second largest eigenvector's components in the case of ER weighted adjacency matrices, obtained by population dynamics (solid blue) and by direct diagonalisation of $2000$ matrices of size $N=5000$ (red circles) showing excellent agreement. Also in this case, the population size is $N_P=10^5$ and the inset shows the right tail of the pdf in log scale.\label{fig:er2_c4}}
\par\end{centering}
\centering{}
\end{figure}

Figure \ref{fig:er2_c4} shows the numerical results in the case of an Erd\H{o}s-R\'enyi (ER) adjacency matrix with $c=4$ and $k_{\mathrm{max}}=12$.  We find $\langle \lambda_2 \rangle_J=4.463$, within a 2\% error w.r.t. the value $\lambda_{2,\infty}=4.565$ obtained by extrapolation from the direct diagonalisation data. The bottom right panel of Figure \ref{fig:er2_c4} refers instead to the case of ER weighted adjacency matrix with $c=4$ and $k_{\mathrm{max}}=12$.  We consider the case of uniform distribution of bond weights, $p_{K}(K)=1/2$ for $K\in[1,3]$. In this case, we find $\langle \lambda_2 \rangle_J=9.5016$, within a 2.5\% error w.r.t. the reference value $\lambda_{2,\infty}=9.7452$ obtained by extrapolation from the direct diagonalisation data. In the plot, we compare the pdf of second largest eigenvector's components obtained via population dynamics with results from the direct diagonalisation of $2000$ matrices of size $N=5000$.

\begin{figure}

\begin{centering}
\includegraphics[scale=0.38]{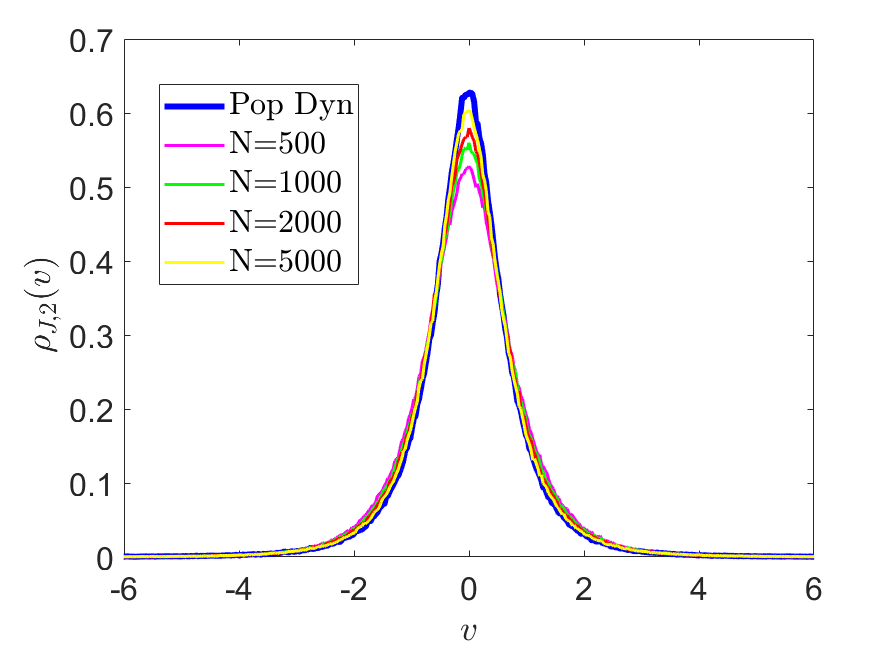}\includegraphics[scale=0.4]{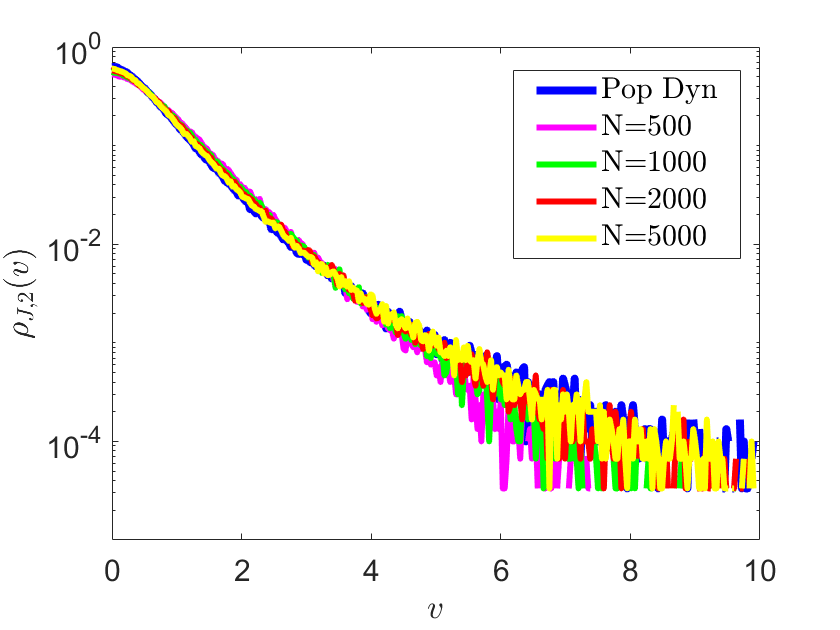}
\par\end{centering}
\begin{centering}
\includegraphics[scale=0.38]{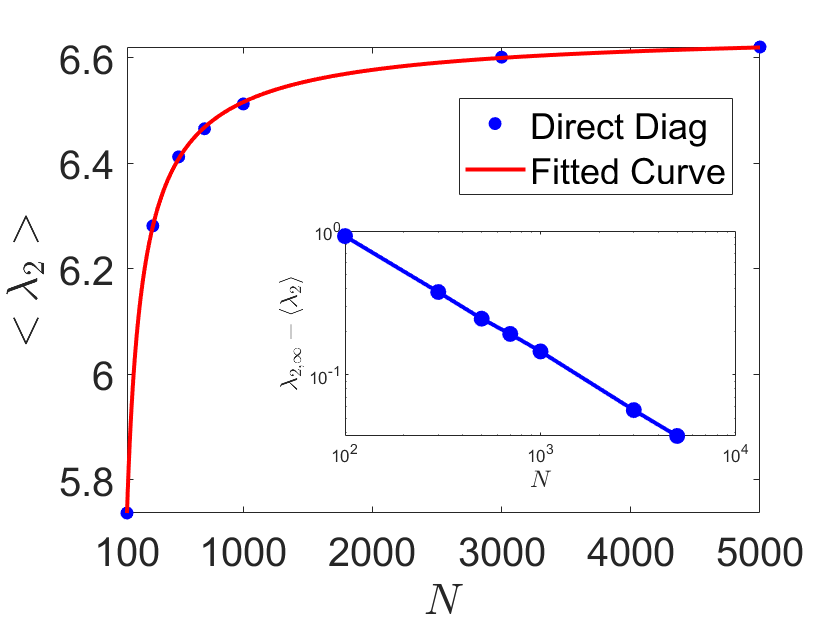}
\par\end{centering}
\begin{centering}
\caption{Pdf of second largest eigenvector's components in the ER adjacency matrix case, with $c=10$ and $k_{\mathrm{max}}=22$. {\bf Top left panel}: Results from population dynamics (blue thick curve) are compared with diagonalisation of matrices of size $N=500$ (light purple), $N=1000$ (green), $N=2000$ (red) and $N=5000$ (yellow). As $N$ increases, we notice that the direct diagonalisation curves approach the pdf generated by population dynamics with a fairly large population size, $N_P=10^5$. {\bf Top right panel}: the (right) tails of the distributions shown in the top left panel, shown in log scale. {\bf Bottom panel}: the average second largest eigenvalue $\langle \lambda_2 \rangle$ as a function of $N$, obtained with direct diagonalisation. The power law fit is superimposed in red. As discussed in the main text, the inset shows the plot of $\lambda_{2,\infty}-\langle\lambda_2\rangle$ vs $N$ in log scale. \label{fig:finite_size}}
\par\end{centering}
\centering{}
\end{figure}

Figure \ref{fig:finite_size} compares the theoretical results for the pdf of the second largest eigenvector's components with results of direct numerical diagonalisation for adjacency matrices of ER graphs  with $c=10$ and $k_{\mathrm{max}}=22$. In this case, we find $\langle \lambda_2 \rangle_J=6.656$, within a $0.4\%$ error w.r.t. the value $\lambda_{2,\infty}=6.658$ obtained by extrapolation from the direct diagonalisation data. We observe that there are finite size effects in the distribution of eigenvector components that are significantly stronger than those observed in the eigenvalue problem. The bottom panel of figure \ref{fig:finite_size} shows the average second largest eigenvalue $\langle \lambda_2 \rangle$ as a function of the matrix size $N$, obtained via direct diagonalisation of adjacency matrices of ER graphs with $c=10$ and $k_{\mathrm{max}}=22$. The data are fitted by a power law curve $\langle \lambda_2 \rangle=aN^{-b}+\lambda_{2,\infty}$, with $b\simeq0.8115$ for this type of network. The inset shows the plot of $\lambda_{2,\infty}-\langle\lambda_2\rangle$ against $N$ in log scale, confirming that the power law exponent $b$ is positive. The power law convergence is a common behaviour found in all ensembles analysed in this paper, though the value of the exponent $b$ depends on details of the systems.

\section{Random regular graphs\label{sec:RRG_ansatz}}
\begin{figure}
\begin{centering}
\includegraphics[scale=0.5]{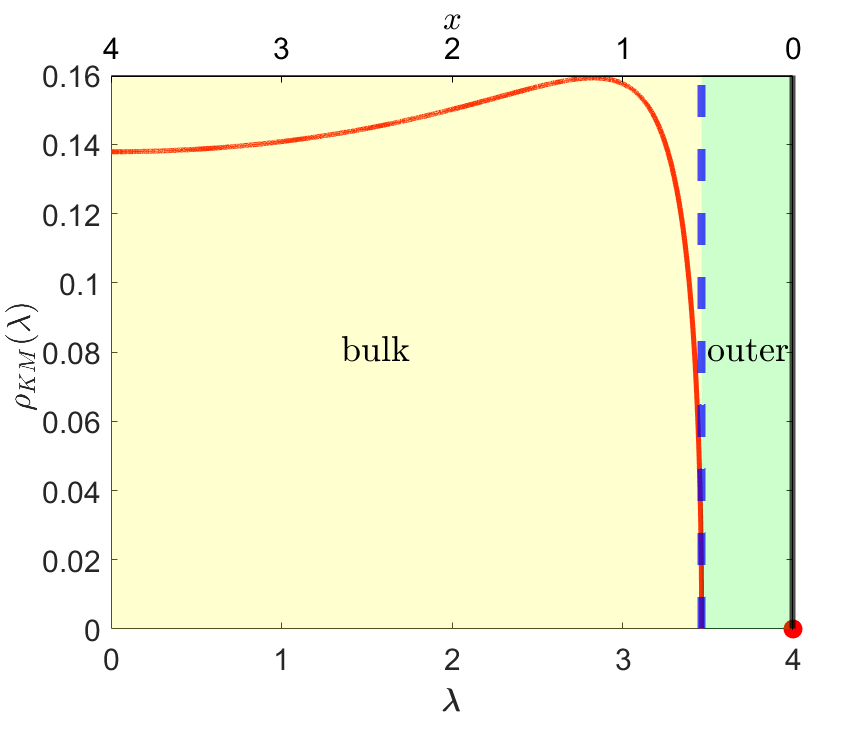}
\par\end{centering}
\begin{centering}
\caption{The positive branch of the Kesten-McKay distribution \eqref{eq:kmd_pdf} in solid red for $c=4$. The red dot at $\lambda=4$ represents the top eigenvalue $\lambda_1=c$, which is an outlier.
The dashed blue vertical line at $\lambda=2\sqrt{(c-1)}$ or equivalently $x=c-2\sqrt{(c-1)}$ separates the \emph{outer} regime (light green) from the \emph{bulk} regime (light yellow). \label{fig:regimes_rrg2} }
\par\end{centering}
\centering{}
\end{figure}

For non-weighted adjacency matrices of RRGs, the degree distribution is simply $p(s)=\delta_{s,c}$, and the bond weights distribution is trivially $p_{K}(K)=\delta(K-1)$, resulting in a constant \emph{probe} top eigenvector $\bm{u}$, i.e. $\rho_J(u)=\rho_J(u|c)=\delta(u-1)$. The largest eigenvalue $\lambda_1$ is non-random and pinned to the value $\lambda_1=c$. The spectral density is given by the Kesten-McKay distribution (See Figure \ref{fig:regimes_rrg2}),
\begin{equation}
\rho_{KM}(\lambda)=\frac{c \sqrt{4(c-1)-\lambda^2}}{2\pi (c^2-\lambda^2)}\ ,\qquad |\lambda|\leq2\sqrt{c-1}\ . \label{eq:kmd_pdf}
\end{equation} 
In this section we look at the behaviour of the solution for a generic value of the deflation parameter $x$ in the range $[0,c]$. Therefore, the value of $q$ is in principle non-zero. We remark that $q=0$ holds surely in the case of full deflation, as in Section \ref{sec:cavity_formulation} (and \ref{sec:replica_gen}).
For a general value of the deflation parameter $x$, the equation \eqref{eq:cavity_pdf_td} for $\pi(\omega,h)$, along with the conditions \eqref{eq:cavity_norm_td} and \eqref{eq:cavity_ort_td} become respectively
\begin{align}
\pi(\omega,h)&=\int\{\mathrm{d}\pi\} _{c-1}\delta\left(\omega-\left(\lambda-\sum_{\ell=1}^{c-1}\frac{1}{\omega_{\ell}}\right)\right)\delta\left(h-\left(-qx+ \sum_{\ell=1}^{c-1}\frac{h_{\ell}}{\omega_{\ell}}\right)\right)\ , \label{eq:pi_rrg_x} \\ 
1&= \int\{\mathrm{d}\pi\} _{c} \left(\frac{-qx+\sum_{\ell=1}^c\frac{h_\ell}{\omega_\ell}}{\lambda-\sum_{\ell=1}^c \frac{1}{\omega_\ell}}\right)^{2}\ ,\label{eq:lambda_rrg_x}\\ 
q&=\int\{\mathrm{d}\pi\} _{c} \left(\frac {-qx+\sum_{\ell=1}^c \frac{h_\ell}{\omega_\ell}}{\lambda-\sum_{\ell=1}^c \frac{1}{\omega_\ell}}\right)\ ,\label{eq:q_rrg_x}
\end{align}
and the density of the top eigenvector's components of the deflated matrix $\tilde{J}$ \eqref{eq:cavity_evect_td} is given for general $x$ by
\begin{align}
\rho_{\tilde{J}}(v)&=\int\{\mathrm{d}\pi\} _{c}~ \delta\left(v-\frac{-qx +\sum_{\ell=1}^c \frac{h_\ell}{\omega_\ell}}{\lambda-\sum_{\ell=1}^c \frac{1}{\omega_\ell}}\right)\ .
\label{eq:evect_rrg_x}
\end{align}
We will show that the solution of the self-consistency equation \eqref{eq:pi_rrg_x} along with \eqref{eq:lambda_rrg_x}, \eqref{eq:q_rrg_x} and \eqref{eq:evect_rrg_x} crucially depends on the value of the deflation parameter $x$. We recall  here that the range of $x$ is $[0,c]$, where the boundaries of this range correspond respectively to no deflation ($x=0$) and full deflation ($x=c$).

We anticipate that in the \emph{outer} regime $0\leq x<c-2\sqrt{c-1}$ (see Figure \ref{fig:regimes_rrg2}), the \emph{probe} eigenvector $\bm{u}=\{1,1,\ldots,1\}$, i.e. the top eigenvector of the original matrix $J$, is also the top eigenvector of the deflated matrix $\tilde{J}$, with corresponding largest eigenvalue $c-x$ lying outside the bulk of the Kesten-McKay spectrum \cite{Kesten1959,McKay1981}. Conversely, in the \emph{bulk} regime i.e. when $x>c-2\sqrt{c-1}$,  the top eigenvector's components density is a standard normal distribution, with corresponding largest eigenvalue $2\sqrt{c-1}$. The \emph{probe} all-one eigenvector $\bm{u}$ is still an eigenvector of $\tilde{J}$ but refers to an eigenvalue $c-x<2\sqrt{c-1}$. In other words, we show that the second largest eigenpair of the RRG adjacency matrix is given by $\langle \lambda_2\rangle_J=2\sqrt{c-1}$ and $\rho_{J,2}(v)=\mathcal{N}(0,1)$. Figure \ref{fig:regimes_rrg2} explains graphically the \emph{outer} and \emph{bulk} regimes.

The abrupt change of the solution (from constant $\bm{u}$ to normally distributed when $x$ hits the value $c-2\sqrt{c-1}$) reflects the fact that the usual peaked ansatz for the RRG case (see \cite{Susca2019}) is not valid in the \emph{bulk} regime $c-2\sqrt{c-1}<x\leq c$. Therefore, in order to solve the self-consistency equation  \eqref{eq:pi_rrg_x}, we choose a ``mixed'' ansatz of the form 
\begin{equation}
\pi(\omega,h)=\delta(\omega-\bar\omega)\sqrt{\frac{1}{2\pi\sigma^2}}\exp \left[-\frac{(h-\bar{h})^2}{2\sigma^2} \right]\ ,
\label{eq:gauss-ansatz}
\end{equation}
for real $\bar{\omega}$ and $\bar{h}$.

We further show that in the range $0\leq x<c-2\sqrt{c-1}$, the solution reduces to a peaked ansatz, i.e. $\sigma^2=0$ - just like in the case of the largest eigenpair of the original matrix $J$ - whereas in the range $c-2\sqrt{c-1}\leq x<c$, the variance $\sigma^2$ must be finite. 

Indeed, by inserting \eqref{eq:gauss-ansatz} into \eqref{eq:pi_rrg_x} and performing the r.h.s. integrals, we find 
\begin{equation}
\pi(\omega,h)=\delta \left(\omega-\left(\lambda-\frac{c-1}{\bar{\omega}}\right)\right)\sqrt{\frac{\bar{\omega}^2}{2\pi\sigma^2 (c-1)}}\exp \left[-\frac{\left(h-(-qx+\frac{\bar{h}}{\bar{\omega}}(c-1))\right)^2}{2\sigma^2(c-1)/\bar{\omega}} \right]\ .
\label{eq:ansatz_resolved}
\end{equation}
Comparing \eqref{eq:ansatz_resolved} with the ansatz \eqref{eq:gauss-ansatz}, we find that the following relations must be satisfied
\begin{align}
\bar{\omega}&=\lambda-\frac{c-1}{\bar{\omega}}\ ,\label{eq:baromega}\\
\bar{h} &= -qx+\frac{\bar{h}}{\bar{\omega}}(c-1)\ ,\label{eq:barh}\\
\sigma^2 & =\sigma^2\frac{c-1}{\bar{\omega}^2}\ .\label{eq:sigma-1}
\end{align}
From the last condition \eqref{eq:sigma-1}, we can infer that if $\sigma^2>0$, then $\bar\omega=\sqrt{c-1}$, i.e. a finite variance of the distribution of components pins $\bar\omega$ to a specific value. Only if $\sigma^2=0$ , then $\bar\omega$ can assume values other than $\sqrt{c-1}$, according to Eq. \eqref{eq:baromega}.

Inserting the ansatz \eqref{eq:gauss-ansatz} in the normalisation condition \eqref{eq:lambda_rrg_x} and in the condition \eqref{eq:q_rrg_x}, we find two extra conditions to fix respectively $\sigma^2$ and $q$,
\begin{align}
&\sigma^2  = \frac{\bar{\omega}^2}{c}\left[ \left(\lambda-\frac{c}{\bar\omega}\right)^2-\left(c\frac{c}{\bar\omega}-qx\right)^2\right]\ ,\label{eq:sigma-2}\\
&q\left(\lambda-\frac{c}{\bar\omega}\right) =\left(c\frac{\bar{h}}{\bar\omega}-qx\right)\ .\label{eq:z-2}
\end{align}

By combining \eqref{eq:baromega}, \eqref{eq:barh} and \eqref{eq:z-2}, we find an expression for $q$ in terms of $\bar\omega$ and $\bar{h}$,
\begin{equation}
q=\frac{\bar{h}}{\bar\omega-1}\ ,
\label{eq:z-h-omega}
\end{equation}
which in turn can be inserted into Eq. \eqref{eq:barh} to give
\begin{equation}
\bar{h}\left(1+\frac{x}{\bar\omega-1}-\frac{c-1}{\bar\omega}\right)=0\ .\label{eq:new-barh}
\end{equation}
Comparing eq. \eqref{eq:baromega} rewritten as
\begin{equation}
\bar{\omega}^2-\lambda\bar\omega+c-1=0\ 
\label{eq:baromega_1}
\end{equation}
with a slight rewriting of the condition that the expression in the round brackets of \eqref{eq:new-barh} be zero, viz.
\begin{equation}
\bar{\omega}^2-(c-x)\bar\omega+c-1=0\ ,
\label{eq:omega-with-x}
\end{equation}
we notice that \eqref{eq:baromega_1} and \eqref{eq:omega-with-x} can be compatible only if the coefficient of $\bar{\omega}$ is the same, entailing $\lambda=c-x$.
Moreover, by solving \eqref{eq:omega-with-x} for $\bar\omega$ we also find the explicit dependence of $\bar\omega$ on $x$. Indeed,   we get
\begin{equation}
\bar\omega(x)_{1,2}=\frac{c-x\pm\sqrt{(c-x)^2-4(c-1)}}{2}\ .
\label{eq:omega-with-x-2}
\end{equation}
By imposing that the radicand be positive in order to get a real solution, we find that eq. \eqref{eq:omega-with-x-2} yields a $x$-dependent real solution only for $0\leq x<c-2\sqrt{c-1}$. Only in this regime, $\bar{\omega}=\bar{\omega}(x)$ can assume values other than $\sqrt{c-1}$, entailing from \eqref{eq:sigma-1} a peaked solution for $\pi$.\footnote{We remark that in this regime a finite variance solution for $\pi$ that pins $\bar{\omega}$ to $\sqrt{c-1}$  is still possible, but yields a higher ground state free energy $\langle F \rangle_{\tilde{J}}$ than the peaked solution. Indeed, $\langle F \rangle_{\tilde{J}}=-\frac{N}{2}\langle \lambda_1 \rangle_{\tilde{J}}$. See Sections \ref{sec:RRG_peaked_eval} and \ref{sec:RRG_gauss_eval}.}

Conversely, for any $x> c-2\sqrt{c-1}$, Eq. \eqref{eq:omega-with-x-2} would produce a $x$-dependent \emph{complex} solution $\bar\omega(x)$, which is not acceptable for this problem (recall that $\omega$ and $h$ must be real), thus implying
\begin{equation}
\sigma^2>0\Leftrightarrow\bar\omega(x)=\sqrt{c-1}\:\;\; \forall x \in [c-2\sqrt{c-1},c]\ .
\label{eq:omega-gauss-fixed}
\end{equation}

\subsection{RRG-deflated top eigenvalue: \emph{outer} regime  \label{sec:RRG_peaked_eval}}
From \eqref{eq:omega-gauss-fixed}, it follows that $\sigma^2=0$ in the \emph{outer} regime. From \eqref{eq:sigma-2} and \eqref{eq:z-2}, we thus find

\begin{equation}
 \begin{cases}
\left(\lambda-\frac{c}{\omega}\right)^2=\left(c\frac{\bar{h}}{\bar\omega}-qx\right)^2\\
q\left(\lambda-\frac{c}{\omega}\right)=\left(c\frac{\bar{h}}{\bar\omega}-qx \right) \\
 \end{cases} \Rightarrow q=\pm1\ .
 \label{eq:z_peaked}
\end{equation}
When solving \eqref{eq:z_peaked}, we must discard the other possible solution $q=0$, since it would not satisfy the normalisation constraint \eqref{eq:lambda_rrg_x}.

Equipped with this information and also taking into account \eqref{eq:gauss-ansatz}, \eqref{eq:omega-with-x} and the identity $\bar{h}=\bar\omega-1$, which follows from \eqref{eq:z-h-omega}, we find for the average of the largest eigenvalue of $\tilde{J}$ the formula
\begin{equation}
\left\langle \tilde{\lambda}_{1}\right\rangle _{\tilde{J}}=c-x\ ,
\end{equation}
which is exactly equal to $\lambda$ as expected (see \eqref{eq:cavity_eval_td_out}). Details of the replica computation that leads to this result can be found in \ref{sec:RRG_appendix}.

Therefore, the deflation with a parameter $x$ in the regime $0\leq x<c-2\sqrt{c-1}$ has the effect of decreasing the top eigenvalue $c$ of the original RRG adjacency matrix $J$ by a quantity $x$, as long as it lies outside the spectral bulk of the Kesten-McKay distribution. This confirms the mechanism explained in Section \ref{sec:Formulation}. In the next subsection, we will show that the corresponding eigenvector is still the top eigenvector of $J$.

\subsection{RRG-deflated density of top eigenvector components: \emph{outer} regime 
\label{sec:RRG_peaked_evect}}
As found at the beginning of this Section, within the range $0\leq x<c-2\sqrt{c-1}$ , the ansatz for $\pi$ is delta-peaked, since $\sigma^2=0$. We show that a peaked ansatz of this sort corresponds to the top eigenvector of the matrix $\tilde{J}$ being all-ones: this means that for $0\leq x<c-2\sqrt{c-1}$ the top eigenvector of  $\tilde{J}$ is exactly the \emph{probe} eigenvector $\bm{u}$.

Indeed, by inserting the ansatz \eqref{eq:gauss-ansatz} in \eqref{eq:evect_rrg_x}  and taking into account \eqref{eq:omega-with-x} and \eqref{eq:z_peaked}, we find
\begin{equation}
\rho_{\tilde{J}}(v)=\delta\left(v-\frac{c\frac{\bar{h}}{\bar\omega}-qx}{\lambda-\frac{c}{\bar\omega}}\right)\ ,
\end{equation}
but, from \eqref{eq:z_peaked},
\begin{equation}
\left| c\frac{\bar{h}}{\bar\omega}-q x \right|=\left| \lambda-\frac{c}{\bar\omega}\right|\ ,
\end{equation}
implying
\begin{equation}
\rho_{\tilde{J}}(v)=\delta\left(v-1\right)\Rightarrow \bm{v}=\bm{u} \ ,
\end{equation}
where the choice of the ``$+$'' sign solution is not restrictive.

In conclusion, as long as the largest eigenvalue $c-x$ of the deflated matrix $\tilde{J}$ lies outside the spectral bulk (i.e. for $0\leq x<c-2\sqrt{c-1}$), the corresponding top eigenvector $\bm{v}$ is equal to the \emph{probe} eigenvector $\bm{u}=(1,...,1)^T$, i.e. the top eigenvector of $J$.

\subsection{RRG top eigenvalue: \emph{bulk} regime \label{sec:RRG_gauss_eval}}
In this range, we have shown in \eqref{eq:omega-gauss-fixed} that the variance $\sigma^2$ is positive, giving rise to a mixed ``delta-Gaussian'' ansatz for $\pi$. The parameter $\sigma^2$ being positive implies that $\bar\omega$ must be pinned to the value $\sqrt{c-1}$.
From \eqref{eq:baromega}, it follows that $\lambda=2\sqrt{c-1}$.
The values of $q$ and $\bar{h}$ are determined by the normalisation \eqref{eq:lambda_rrg_x} and orthogonality \eqref{eq:q_rrg_x} conditions. Indeed, the change in the ansatz corresponds to a change in the structure of the largest eigenvector $\bm{v}$ of $\tilde{J}$. As shown in Section \ref{sec:cavity_td}, the orthogonality condition reads
\begin{align}
\nonumber 0=&\int\mathrm{d}u\mathrm{d}v\rho_{\tilde{J}}(u|c)\rho_{\tilde{J}}(v|u,c)uv\\
=&\int\{\mathrm{d}\pi\}_c \frac{\sum_{\ell=1}^{c}\frac{h_{\ell}}{\omega_{\ell}}-qx}{\lambda-\sum_{\ell=1}^{c}\frac{1}{\omega_{\ell}}}\ ,
\label{eq:orthogonality_rrg}
\end{align}
where $\rho_{\tilde{J}}(u|c)=\delta(u-1)$ is the conditional distribution of the probe eigenvector's entries and \eqref{eq:evect_rrg_x} has been used.
Comparing \eqref{eq:orthogonality_rrg} with \eqref{eq:q_rrg_x} we infer that $q=0$. Moreover,  inserting $q=0$ in \eqref{eq:sigma-2} and \eqref{eq:z-2}, we can respectively infer that 
\begin{align}
\sigma^2 &=\frac{{\bar\omega}^2}{c}\left(\lambda-\frac{c}{\bar\omega}\right)^2=\frac{(c-2)^2}{c}\ ,\label{eq:sigma-gauss}\\
\bar{h}&=0\label{eq:barh-gauss}\ .
\end{align}

Equipped with this information and also by taking into account \eqref{eq:gauss-ansatz}, we find that the average of the largest eigenvalue of $\tilde{J}$ is
\begin{equation}
\left\langle \tilde{\lambda}_{1}\right\rangle _{\tilde{J}}=2\sqrt{c-1}\ ,
\end{equation}
corresponding to the upper edge of the Kesten-McKay distribution, and once again exactly equal to $\lambda$ (see \eqref{eq:cav_eval_last}). Also in this case, the details of the replica calculation are in \ref{sec:RRG_appendix}.

As expected, the eigenvalue does not depend on the normalisation of the corresponding eigenvector, encoded in $\sigma^2$.
Since this result holds for any $x$ in $c-2\sqrt{c-1}<x\leq c$, including the case of full deflation when $x=\langle\lambda_1\rangle_J=c$ and the first eigenmode $\bm{u}$ of the original matrix $J$ is associated to a zero eigenvalue, we conclude that the average second largest eigenvalue of the matrix $J$ is
\begin{equation}
\left\langle \lambda_{2}\right\rangle _{J}=\left\langle  \tilde{\lambda}_{1}\right\rangle _{\tilde{J}(x=c)}=\lambda=2\sqrt{c-1}\ .
\end{equation}
Also in this case, we find agreement with the general deflation framework described in Section \ref{sec:Formulation}.

\subsection{RRG density of top eigenvector components: \emph{bulk} regime
\label{sec:RRG_gauss_evect}}
In this range of values for $x$, we show that the ``delta-Gaussian" ansatz for $\pi(\omega,h)$ leads to a Gaussian-distributed top eigenvector of the matrix $\tilde{J}$. Since this result is valid also in case of full deflation, i.e. $x=c$, we can conclude that the eigenvector corresponding to the \emph{second largest} eigenvalue of a random regular graph adjacency matrix $J$ is normally distributed\footnote{We remark that our method cannot provide the eigenvector statistic for $x=c-2\sqrt{c-1}$. Indeed, for this specific value of $x$, the \emph{probe} eigenvector $\bm{u}$ is forced to correspond to the eigenvalue $2\sqrt{c-1}$, which retains its own eigenvector, thus artificially creating a degeneracy. Our method is based on the assumption of non-degeneracy of eigenvalues, so we are not able to give a result about eigenvectors in this marginal case.}.
We then identify in $x=c-2\sqrt{c-1}\iff\lambda=2\sqrt{c-1}$ a transition point for the structure of the distribution of the top eigenvector's components of $\tilde{J}(x)$, at which the parameter $q$ changes discontinuously from $q=\pm1$ to $0$.

\begin{figure}
\begin{centering}
\includegraphics[scale=0.5]{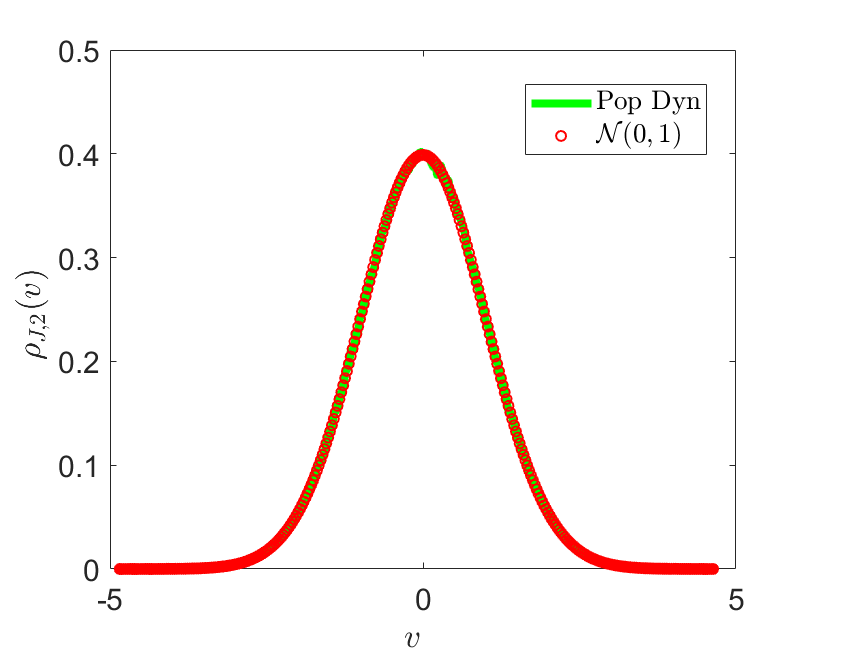}
\par\end{centering}
\begin{centering}
\caption{In green, the profile of the distribution of the second largest eigenvector's components \eqref{eq:rrg_gauss_evect} obtained via population dynamics, with population size $N_P=10^6$. As a reference, we plot the standard normal distribution (red circles), showing perfect matching.\label{fig:rrg2_normpdf} }
\par\end{centering}
\centering{}
\end{figure}

We now evaluate the density of the top eigenvector components in the range $c-2\sqrt{c-1}<x\leq c$. Inserting the ansatz \eqref{eq:gauss-ansatz} in \eqref{eq:evect_rrg_x}  and taking into account \eqref{eq:omega-gauss-fixed}, \eqref{eq:orthogonality_rrg}, \eqref{eq:sigma-gauss} and \eqref{eq:barh-gauss}, we find
\begin{equation}
\rho_{\tilde{J}}(v)\equiv\rho_{J,2}(v)=\frac{\exp(-v^2/2)}{\sqrt{2\pi}}\ . \label{eq:rrg_gauss_evect}
\end{equation}
We remark that this analytical result is in excellent agreement with the statistics of the second largest eigenvector components of the RRG adjacency matrices found by population dynamics, as shown in Figure \ref{fig:rrg2_normpdf}. Moreover, it is compatible with previous known results about eigenvectors of random regular graphs \cite{Elon2008,Backhausz2016}.

\section{Sparse random Markov transition matrices \label{sec:Markov}}
In this section, we apply the deflation formalism to an ensemble of transition matrices $W$ for discrete Markov chains in a $N$-dimensional state space, in order to characterise the statistics of the second largest eigenpair. This kind of Markov chain represents a random walk on a graph. We remark here that the second largest eigenpair encodes non-equilibrium properties of a Markov process. Indeed, the inverse of the (absolute value) of the second largest eigenvalue represents the slowest relaxation time, whereas the associated \emph{second} eigenvector is the non-equilibrium mode with the largest relaxation time.  

We will then employ a full deflation, by setting $x=\lambda_1(W)=1$. The evolution equation for the Markov chain states probability vector at time $t$, $\bm{p}(t)$, is given in terms of the matrix $W$ by
\begin{equation}
\bm{p}(t+1)=W\bm{p}(t)\ .
\end{equation}
The transition matrix $W$ is such that $W_{ij}\geq0\ \forall(i,j)$ and $\sum_{i}W_{ij}=1\ \forall j$. For an irreducible chain, the top right eigenvector of the matrix $W$  corresponding to the Perron-Frobenius eigenvalue $\lambda_1=1$ represents the unique equilibrium distribution, i.e. $\bm{v}^{(1)}=\bm{p}^{\mathrm{eq}}$.
The matrix $W$ is in general not symmetric. However, if the Markov process satisfies a detailed balance condition, i.e. $W_{ij}p_j^{\mathrm{eq}}=W_{ji}p_i^{\mathrm{eq}}\;\;\forall(i,j)$, it can be symmetrised via a similarity transformation, yielding
\begin{equation}
W_{ij}^S=(p_i^{\mathrm{eq}})^{-1/2}W_{ij}(p_j^{\mathrm{eq}})^{1/2}\ .
\end{equation}

The symmetrised matrix $W^S$ and its deflated version $\tilde{W}^S$ will be the target of our analysis: even though $W^S$  is not itself a Markov matrix since the columns normalisation constraint is lost, in view of the detailed balance condition $W^S$ has the same (real) spectrum as $W$, and its top eigenvector $\bm{u}$  is given in terms of the top right eigenvector of $W$, $\bm{p}_{\mathrm{eq}}$, as
\begin{equation}
u_i=(N p_i^{\mathrm{eq}})^{1/2}\ .
\end{equation}
It is actually well-known that the relation between the eigenvectors of $W$ and those of $W^S$  holds in general and is not limited to the case of the top one.

We will consider  the case of an unbiased random walk: the matrix $W$ is then defined as
\begin{equation}
W_{ij}=
\begin{cases}
\frac{c_{ij}}{k_j}, & i\neq j \\
1, & i=j\ \mathrm{and}\ k_j=0\ ,
\end{cases}
\end{equation}
where $c_{ij}$ represents the connectivity matrix and $k_j=\sum_i c_{ij}$ is the degree of  node $j$. In this case, the top right eigenvector of $W$ is proportional to the vector expressing the degree sequence: for our purposes, we choose  the inverse of the mean degree as proportionality constant, i.e. $p_i^{\mathrm{eq}}=k_i/(Nc)$. The symmetrised matrix $W^S$ is expressed as
\[
W_{ij}^S=
\begin{cases}
\frac{c_{ij}}{\sqrt{k_i k_j}}, & i\neq j \\
1, & i=j\ \mathrm{and}\ k_j=0\ ,
\end{cases}
\]
with its top eigenvector being $u_i^{(1)}=\sqrt{k_i/c}$. Thus, we have 
\begin{equation}
\rho_{W^S}(u)=\sum_{k\geq k_{\mathrm{min}}} p(k) \delta\left( u-\sqrt{\frac{k}{c}} \right)\ , \label{eq:density_markov_theory}
\end{equation}
where $p(k)$ is the degree distribution of the connectivity matrix $\{c_{ij}\}$.

In order to avoid isolated nodes and isolated clusters of nodes, we consider degree distributions with $k_{\mathrm{min}}\geq2$ and finite mean degree\footnote{A suitable candidate could be a shifted Poissonian degree distribution with $k_{\mathrm{min}}=2$, i.e. 
\begin{equation}
p(k)=\frac{\mathrm{e}^{-\bar{c}}\bar{c}^{k-2}}{\left( k-2\right)!}\mathbbm{1}_{k\geq2}\ ,
\end{equation}
with mean degree $c=\bar{c}+2$. }. We will provide a treatment for a generic distribution $p(k)$ with the aforementioned properties and the analytical solution for the random regular connectivity case with degree distribution $p(k)=\delta_{k,c}$. 

\subsection{Second largest eigenpair of Markov transition matrices\label{sec:eval_m}}
We focus on the fully deflated symmetrised version of the Markov matrix $W$, that is
\begin{equation}
\tilde{W}^S_{ij}=W^S_{ij}-\frac{1}{N}u_i u_j\ ,
\end{equation}
where $W^S_{ij}=\frac{c_{ij}}{\sqrt{k_i k_j}}$ and $\bm{u}$ represents the top eigenvector of $W^S$, normalised to $N$, i.e. $u_i=\sqrt{\frac{k_i}{c}}$. Here, $c$ represents the mean degree, $c=\langle k\rangle$. Our aim is to find the typical largest eigenvalue of $\tilde{W}^S$, which corresponds to the typical second largest eigenvalue of $W^S$. In the next subsection, we will characterise the distribution of the top eigenvector of $\tilde{W}^S$, equivalent to the \emph{second} eigenvector of $W^S$.

We follow the same formalism illustrated in Section \ref{sec:cavity_td}. An alternative replica derivation can be found in \ref{sec:markov_appendix}.  Here, we will just report the final equations, corresponding to \eqref{eq:cav_pi_last} along with \eqref{eq:cav_norm_last}, \eqref{eq:cav_orto_last} and \eqref{eq:cav_eval_last}. By taking into account \eqref{eq:density_markov_theory} and the existence of $k_{\mathrm{min}}=2$, we find
\begin{align}
\pi(\omega,h)&=\sum_{k=k_{\mathrm{min}}}^{k_{\mathrm{max}}}p(k)\frac{k}{c}\int\{\mathrm{d}\pi\} _{k-1} \delta\left(\omega-\left(\lambda k-\sum_{\ell=1}^{k-1}\frac{1}{\omega_{\ell}}\right)\right)\delta\left(h-\left( \sum_{\ell=1}^{k-1}\frac{h_{\ell}}{\omega_{\ell}}\right)\right)\ ,\label{eq:pi_last_m} \\ 
1&=\sum_{k=k_{\mathrm{min}}}^{k_{\mathrm{max}}}p(k)k\int\{\mathrm{d}\pi\} _{k}  \left(\frac{\sum_{\ell=1}^k \frac{h_\ell}{\omega_\ell}}{\lambda k-\sum_{\ell=1}^k \frac{1}{\omega_\ell}}\right)^{2}\ ,\label{eq:norm_last_m} \\ 
0&=\sum_{k=k_{\mathrm{min}}}^{k_{\mathrm{max}}}p(k)\frac{k}{\sqrt{c}}\int\{\mathrm{d}\pi\} _{k} \left(\frac {\sum_{\ell=1}^k \frac{h_\ell}{\omega_\ell}}{\lambda k-\sum_{\ell=1}^k \frac{1}{\omega_\ell}}\right)\ , \label{eq:orto_last_m} \\ 
\left\langle \tilde{\lambda_1} \right\rangle_{\tilde{J}}&\equiv \left\langle \lambda_2 \right\rangle_{W^S}=\lambda\ .
\end{align}

We remark that in the Markov case a bounded largest degree is not strictly necessary as the spectrum is always bounded. However, we will consider a $k_\mathrm{max}$ for practical purposes.
The self-consistency equation \eqref{eq:pi_last_m} along with the normalisation condition \eqref{eq:norm_last_m} and the orthogonality constraint \eqref{eq:orto_last_m} is solved by a population dynamics algorithm (See Section \ref{sec:pd}). The RRG connectivity case is analytically tractable, as shown in Section \ref{sec:markov_rrg}.

In analogy to Eq.\eqref{eq:cav_evect_last}, the density of the top eigenvector's component of the matrix ${\tilde{W}}^S$, corresponding to the second largest eigenvector of $W^S$, is given by

\begin{equation}
\rho_{\tilde{W}^S}(v)\equiv\rho_{W^S,2}(v) =\sum_{k= k_{\mathrm{min}}}^{k_{\mathrm{max}}}p(k)\int\left\{\mathrm{d}\pi\right\} _{k}\delta\left(v-\frac{\sum_{j=1}^k\frac{h_j}{\omega_j}}{\lambda k-\sum_{j=1}^k\frac{1}{\omega_j}}\sqrt{k}\right)\ ,\label{eq:density_with_pi_m}
\end{equation}
where $\pi(\omega,h)$ satisfies the self-consistency equation \eqref{eq:pi_last_m}, supplemented by the normalisation condition \eqref{eq:norm_last_m} and the orthogonality condition \eqref{eq:orto_last_m}.

\begin{figure}
\begin{centering}
\includegraphics[scale=0.37]{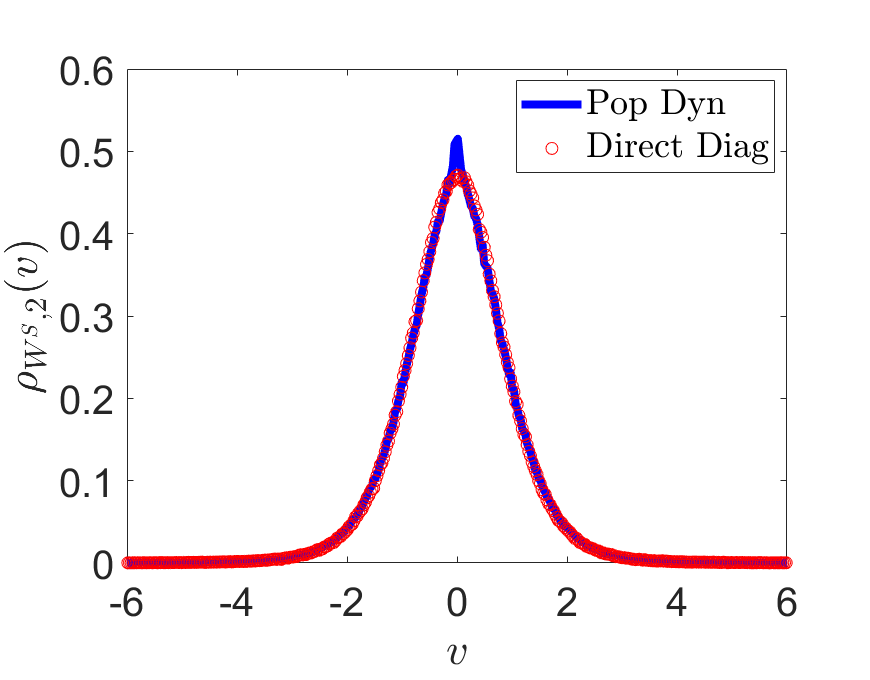}\includegraphics[scale=0.37]{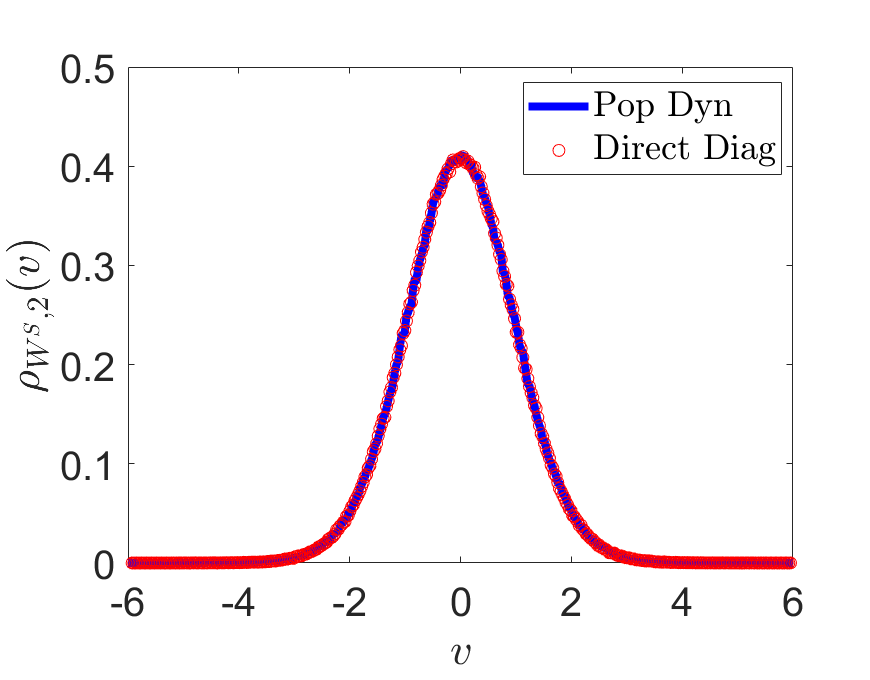}
\par\end{centering}

\begin{centering}
\caption{Pdf of the components of the second largest eigenvector for the unbiased random walk Markov matrix case (see \eqref{eq:density_with_pi_m}), with shifted Poisson degree distribution ($k_{\mathrm{min}}=2$). {\bf Left
panel}: mean degree $c\simeq6$ and $k_{\mathrm{max}}=12$.  Results from population dynamics with $N_P=5000$ (solid blue) compared with the direct diagonalisation of 4000 matrices of size $N=1000$ (red circles) finding a good agreement. {\bf Right panel}: mean degree $c\simeq12$ and $k_{\mathrm{max}}=22$.  Results from population dynamics with $N_P=1500$  (solid blue) compared with the direct diagonalisation of 2000 matrices of size $N=1000$ (red circles), with excellent agreement. In both cases, the size of the population used is $N_P^\star$, the optimal value corresponding to the finite size $N$ of the matrices being diagonalised (see Section \ref{sec:finite_pd}).}\label{fig5}
\par\end{centering}
\centering{}
\end{figure}

Figure \ref{fig5} compares the pdf of the second largest eigenvector's components obtained via population dynamics with results obtained via direct diagonalisation, for the unbiased random walk Markov matrix case with shifted Poisson degree distribution ($k_{\mathrm{min}}=2$). We study both a low ($c\simeq6$, left panel) and a high ($c\simeq12$, right panel) connectivity case. In the $c\simeq6$ case with $k_{\mathrm{max}}=12$, we find $\langle \lambda_2 \rangle_{W^S}=0.7456$, within a 0.7\% error w.r.t. the value $\lambda_{2,\infty}=0.7504$ obtained by extrapolation from the direct diagonalisation data. In the $c\simeq12$ case with $k_{\mathrm{max}}=22$, we find $\langle \lambda_2 \rangle_{W^S}=0.5530$, within a 0.1\% error w.r.t. the value $\lambda_{2,\infty}=0.5524$ obtained by extrapolation from the direct diagonalisation data. As a reference point, the average value of the second largest eigenvalue in the RRG case with the same $c$ is $\lambda_{2}(W^S)_{RRG}=0.5528$. We notice that the agreement near the peak of the distribution is slightly worse for the low connectivity case: this is in agreement with the finding that finite-size effects are generally more pronounced for lower $c$ (see also discussion in section \ref{sec:finite_pd}).

\subsection{Unbiased random walk on a RRG: second largest eigenpair statistics\label{sec:markov_rrg}}
For a random regular graph, for which $p(k)=\delta_{k,c}$, we note that the matrix $W^S$ reduces to 
\begin{equation}
W^S_{ij}=\frac{c_{ij}}{c}\ ,
\end{equation}
implying that all results about the RRG adjacency matrix case stated in Sections \ref{sec:RRG_gauss_eval} and \ref{sec:RRG_gauss_evect} carry over to this case too, but with all eigenvalues rescaled by $1/c$. As expected, $\lambda_1(W^S)_{RRG}=1$, and the second largest eigenvalue corresponding to a $\mathcal{N}(0,1)$-distributed eigenvector is $\lambda_2(W^S)_{RRG}=\frac{2\sqrt{c-1}}{c}$. The spectral gap for this kind of Markov matrices as a function of $c$ is then $g(c)=1-\frac{2\sqrt{c-1}}{c}$.

\section{Population Dynamics \label{sec:pd}}

\setcounter{footnote}{0}

\subsection{The orthogonality challenge \label{sec:pd_orto}}
With the exception of the unweighted adjacency matrix of a RRG, Eq. \eqref{eq:cav_pi_last} -- supplemented with the conditions \eqref{eq:cav_norm_last} and \eqref{eq:cav_orto_last} -- must be generally solved via a Population Dynamics algorithm, a Monte Carlo technique deeply rooted in the statistical mechanics of spin glasses \cite{Mezard2001,Krzakala2016}.

The algorithm we use bears some similarity with the one employed in \cite{Susca2019}. Here, we will highlight the main differences that stem from the presence of the orthogonality condition \eqref{eq:cav_orto_last}. We recall that the Eqs. \eqref{eq:cav_pi_last}, \eqref{eq:cav_norm_last} and \eqref{eq:cav_orto_last} refer to the case of full deflation, where we look at the top eigenpair of the deflated matrix $\tilde{J}$ (the second largest eigenpair of the matrix $J$).

Some observations are in order before sketching the algorithm.
As we stated in \cite{Susca2019}, within the population dynamics algorithm the definition of the $h$ variables in Eq. \eqref{eq:cav_pi_last} is effectively converted into a stochastic linear update of $h$ values. Its stability can only be achieved for  $\lambda=\left\langle\lambda_1 \right\rangle_J$. For any $\lambda>\left\langle\lambda_1 \right\rangle_J$, the variables of type $h$ will shrink to zero, whereas for $\lambda<\left\langle\lambda_1 \right\rangle_J$ they will explode in norm. In our scenario, where we consider  $\lambda<\left\langle\lambda_1 \right\rangle_J$, the recursion is thus \emph{a priori} unstable, unless it is otherwise constrained. Therefore, if unconstrained, the population will never spontaneously evolve towards a stable regime, which would at the same time satisfy the conditions  \eqref{eq:cav_norm_last} and \eqref{eq:cav_orto_last}.

As anticipated in Section \ref{sec:cavity_td}, this observation entails that the orthogonality condition \eqref{eq:cav_orto_last} must be strictly enforced on-the-fly -- by imposing a correction to the fields $h$, which once again have no fixed scale given by their update equation. Enforcing the constraint  \eqref{eq:cav_orto_last} is equivalent to looking for a self-consistent solution of \eqref{eq:cav_pi_last} in a smaller, constrained space.
Only once the condition \eqref{eq:cav_orto_last} has been enforced, a new stable non-trivial fixed point arises, and the behaviour of the $h$-variables is similar to that in the top eigenvector case: for any value $\lambda>\left\langle\lambda_2 \right\rangle_J$, the variables $h$ under iteration of the modified population dynamics algorithm shrink to zero, whereas for $\lambda<\left\langle\lambda_2 \right\rangle_J$ they will explode in norm. Hence, Eq. \eqref{eq:cav_pi_last} -- taken together with the condition \eqref{eq:cav_orto_last} -- admits a stable, hence normalisable solution, such that Eq. \eqref{eq:cav_norm_last} is naturally satisfied only for $\lambda=\left\langle\lambda_2 \right\rangle_J$: after the orthogonality correction has been enforced, the procedure we follow is then exactly identical to that used in \cite{Susca2019}.

\subsection{The algorithm \label{sec:pd_algo}}
Taking into account the observations made in section \ref{sec:pd_orto}, we briefly sketch the algorithm in the case of full deflation.

Two pairs of (coupled) populations with $N_{P}$ members each  $\left\{ \left(a_{i},b_{i}\right)\right\} _{1\leq i\leq N_{P}}$  and $\left\{ \left(\omega_{i},h_{i}\right)\right\} _{1\leq i\leq N_{P}}$
are randomly initialised, taking into account that both $a_i$ and $\omega_{i}$ must be larger than $\zeta$, the upper edge of the support of the bond pdf $p_{K}(K)$. We typically choose $N_P=10^5$ or larger.
In what follows, the parameter $\lambda$ is the candidate second largest eigenvalue of $J$, whereas $\langle \lambda_1 \rangle_J$ is the average top largest eigenvalue of $J$. The first population is employed to solve the top eigenpair problem, and the other to solve the second eigenpair problem; the latter is  constrained by results of the former due to the orthogonality constraint.

\begin{figure}
\begin{centering}
\includegraphics[scale=0.8]{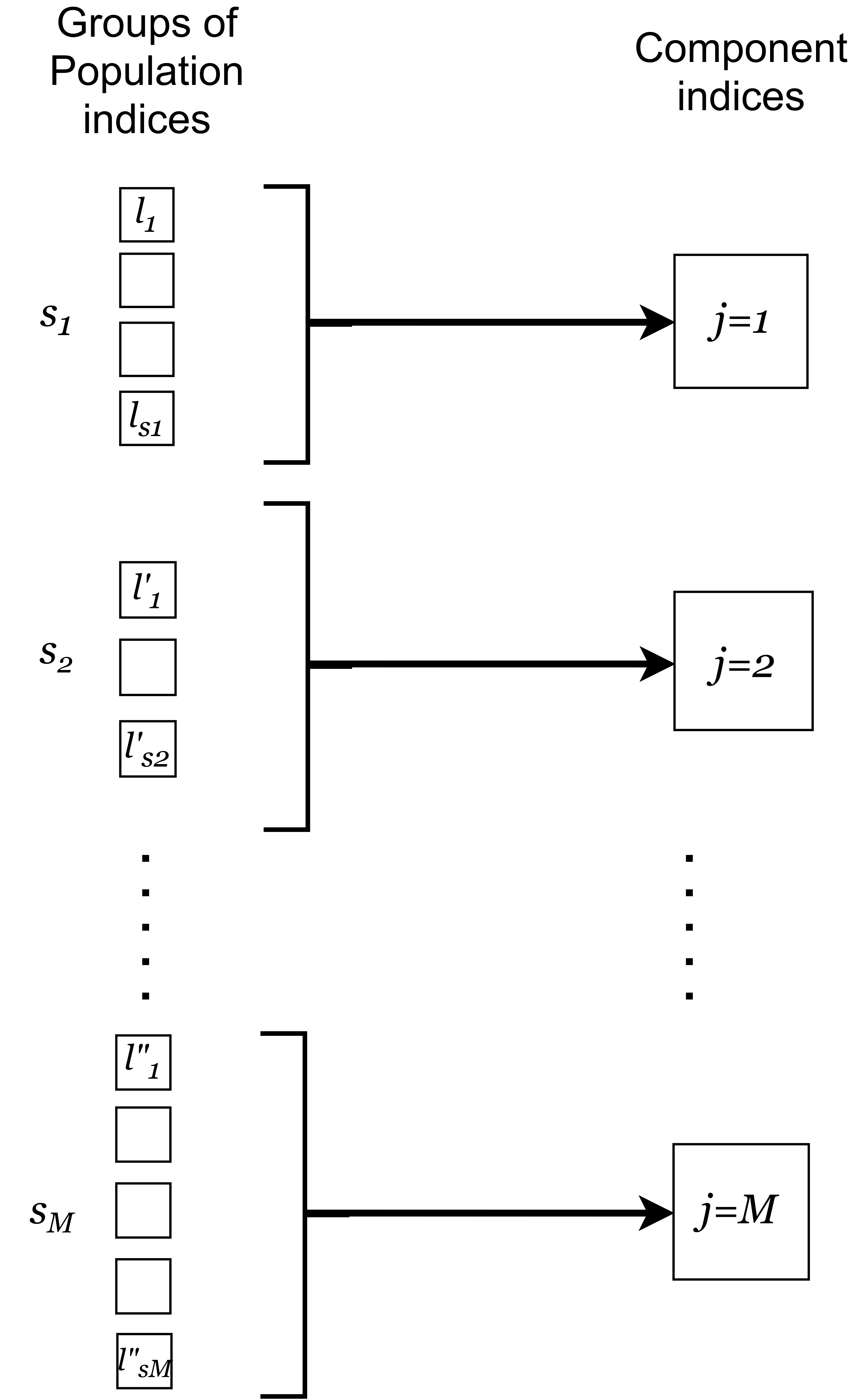}
\par\end{centering}
\begin{centering}
\caption{A schematic representation of the rigid matching between between each set of $s$ population indices and each component $j$ as illustrated in step (iv)(c) of the Population Dynamics algorithm in Section \ref{sec:pd_algo}. The labels $s_j$ with $j=1,...,M$ denote the number of population indices contributing to each component $j$, i.e. the size of each set $S_j(s=s_j)$. \label{fig:algo_scheme}}
\par\end{centering}
\centering{}
\end{figure}

We therefore first run a short population dynamics simulation following Section 6 in \cite{Susca2019} involving only the population $\left\{ \left(a_{i},b_{i}\right)\right\} _{1\leq i\leq N_{P}}$ to find the solution for the first eigenpair problem and the value $\langle \lambda_1 \rangle_J$. This first simulation acts as an equilibration phase for the fields contributing to the largest eigenpair. Then, for any suitable value of $\lambda\in\mathbb{R}<\langle \lambda_1 \rangle_J$, the following steps are iterated until stable populations are obtained:

\begin{enumerate}
\item Generate a random $s\sim\frac{s}{c}p_{c}\left(s\right)$, where $c=\left\langle s \right\rangle$
\item Generate $s-1$ i.i.d. random variables $K_{\ell}$ from the bond weights pdf $p_{K}(K)$
\item Select $s-1$ pairs $\left(a_{\ell},b_{\ell}\right)$ and  $\left(\omega_{\ell},h_{\ell}\right)$ from both populations at random, where the set of $s-1$ population indices for the two randomly selected samples is the same for both samples; compute
\begin{align}
a^{(new)}&=\langle \lambda_1 \rangle_J-\sum_{\ell=1}^{s-1}\frac{K_{\ell}^2}{a_{\ell}}\ ,\\
b^{(new)} & =  \sum_{\ell=1}^{s-1}\frac{b_{\ell}K_{\ell}}{a_{\ell}}\ ,\\
\omega^{(new)} & =  \lambda-\sum_{\ell=1}^{s-1}\frac{K_{\ell}^2}{\omega_{\ell}}\ ,\\
h^{(new)} & =  \sum_{\ell=1}^{s-1}\frac{h_{\ell}K_{\ell}}{\omega_{\ell}}\ ,
\end{align}
and replace two randomly selected pairs $\left(a_{i},b_{i}\right)$ and $\left(\omega_{i},h_{i}\right)$
where $i\in\{1,...,N_{P}\}$ with the pairs $\left(a^{(new)},b^{(new)}\right)$ and $\left(\omega^{(new)},h^{(new)}\right)$. 
\item  Compute the components of the top eigenvector $\bm{u}$ and the candidate second largest eigenvector $\bm{v}$. In order to create a sample estimate of the eigenvectors statistics corresponding to the two top eigenvalues, we initialise two empty vectors, respectively $\bm{u}=\{u_j\}_{1\leq j\leq M}$ and $\bm{v}=\{v_j\}_{1\leq j\leq M}$ of size $M$, where $M=[N_P/c]$ (typically $M=\mathcal{O}(10^4)$ if $N_P=\mathcal{O}(10^5)$). The square brackets indicate the integer part. Then for any $j=1,...,M$:
\begin{enumerate}
\item Generate $s\sim p(s)$
\item Generate $s$ i.i.d. random variables $K_\ell$ from the weights pdf $p_{K}(K)$
\item Randomly select a subset of $s$ indices from the population indices between $1$ and $N_P$. This subset is denoted by $S_j(s)$. Then, for any $\ell \in S_j(s)$ select $s$ pairs $\left(a_{\ell},b_{\ell}\right)$ and  $\left(\omega_{\ell},h_{\ell}\right)$ from both populations; compute 
\begin{align}
u_j&=\frac{\sum_{\ell\in S_j(s)}\frac{b_{\ell}K_{\ell}}{a_{\ell}}}{\langle \lambda_1 \rangle_J-\sum_{\ell\in S_j(s)}\frac{K_{\ell}^2}{a_{\ell}}}\ ,\label{eq:u_pd}\\
v_j& = \frac{\sum_{\ell\in S_j(s)}\frac{h_{\ell}K_{\ell}}{\omega_{\ell}}}{\lambda -\sum_{\ell\in S_j(s)}\frac{K_{\ell}^2}{\omega_{\ell}}}\ . \label{eq:v_pd}
\end{align} 

Each set $S_j(s)$ of $s$ population indices labelled by $\ell$ contributes uniquely to a single component $j$ of the vectors $\bm{u}$ and $\bm{v}$. There is a unique matching between each set of $s$ population indices and each component $j$ (see scheme in Figure \ref{fig:algo_scheme}): in other words, each group of $s$ pairs  $\left(a_{\ell},b_{\ell}\right)$ and  $\left(\omega_{\ell},h_{\ell}\right)$ takes part in the definition of just one component $j$, respectively $u_j$ and $v_j$. Each set $S_j(s)$ of $s$ population indices corresponding to a specific component $j$ is then saved, along with the set of $s$ weights $\{ K_\ell\}$.
\end{enumerate}
\item Compute  $q=\frac{(\bm{u},\bm{v})}{|\bm{u}|^2}$, where $(\cdot,\cdot)$ indicates the dot product. In order to enforce the condition $q=0$, for any component $j=1,...,M$ apply the correction
\begin{equation}
v_j\leftarrow v_j-q u_j\;\ .\label{eq:gs_orto}
\end{equation}
In view of the rigid connection between the population indices labelling the fields and every specific component of $\bm{u}$ and $\bm{v}$, the orthogonalisation in \eqref{eq:gs_orto} is practically achieved by correcting each field $h_\ell$ participating in the definition of every specific component $v_j$. The values of the indices $\ell$ here are those saved in each subset $S_j(s)$ in step (iv)(c), along with the corresponding weights $K_\ell$. For any  $j=1,...,M$ and for any $\ell\in S_j(s)$ contributing to the single component $j$ of both $\bm{u}$ and $\bm{v}$ we have
\begin{equation}
h_\ell\leftarrow h_\ell-qu_j\left(\frac{\lambda \omega_\ell}{K_\ell s}-K_\ell \right)\ ,
\end{equation}
where $s=k_j$ is exactly the ``degree'' drawn from $p(s)$ in step (iv)(a) and used to build each component $v_j$ in step (iv)(c).
\item Return to (i).
\end{enumerate}

A \emph{sweep} is completed when all the $N_P$ pairs $(a_i,b_i)$ and $(\omega_i,h_i)$  have been updated at least once according to the steps above. The update of the pairs $(a,b)$ is stable, thanks to the prior equilibration phase. The convergence is assessed by looking only at the first moments of the two vectors formed by the $N_P$ samples of the pairs $(\omega,h)$. The parameter $\lambda$ is varied according to the behaviour illustrated in Section \ref{sec:pd_orto}: starting from an initial ``large'' value $\lambda<\langle \lambda_1 \rangle_J$, it is then progressively decreased until a non trivial distribution for the $h$ is achieved, in correspondence of the value $\lambda=\langle \lambda_2 \rangle_J$. Indeed, we observe that for any $\lambda> \langle \lambda_2 \rangle_J$, the $h$ shrink to zero, whereas for any $\lambda< \langle \lambda_2 \rangle_J$, they blow up in norm.

Some comments are in order:
\begin{itemize}
\item the condition expressed in \eqref{eq:gs_orto} is a Gram-Schmidt orthogonalisation, taking place after every microscopic update of the fields;
\item the correction does not take place for components $v_j$ related to $s=0$, as both $v_j$ and $u_j$ are zero;
\item in step (iv)(c), we can clearly see that the components $u_j$ and $v_j$ are coupled through their degree and the set of bond weights, as anticipated in Section \ref{sec:cavity_td}. Indeed, for any $j$, the $s$ i.i.d. realisations of the weights $\{K\}_s$ and the ``local neighbourhood" $S_j(s)$ that we dynamically create at every step (c) must be exactly the same for both $u_j$ and $v_j$. In other words, both $u_j$ and $v_j$ must have the same update history.
\end{itemize}

\subsection{Potential for simplifications in special cases}
The steps (iv) and (v) of the algorithm are computationally heavy. We are able in some cases to simplify them.
\begin{itemize}
\item For adjacency matrices of RRGs, where the variables $a,b$ and $\omega$ are constant, the correction \eqref{eq:gs_orto} translates to forcing the mean of the $h$ to be zero after every update. Both steps (iv)-(v) are then replaced by
\begin{equation}
h_i\leftarrow h_i- \bar{h} \;\;\;\;\;\;\;\;\; \forall i=1,...,N_P\ ,\label{eq:corr_rrg}
\end{equation}
where $\bar{h}$ indicates the sample mean of the $h$ population.
\item In the ER case (both weighted and non-weighted), we take advantage of the fact that in the thermodynamic limit there is no statistical distinction between the cavity fields $\omega$ and $h$ (respectively $a$ and $b$) and the denominator and numerator in \eqref{eq:v_pd}, (respectively in \eqref{eq:u_pd}), even in presence of the truncation of the Poissonian degree distribution\footnote{Provided that the largest degree is reasonably large. The only difference between the distribution $\pi(\omega,h)$ and the distribution of the denominator and numerator of \eqref{eq:cav_evect_last} can be observed because of the contribution coming  from the largest degree, whose probability to occur is negligible.}. Hence, we can consider just one couple of fields per species to represent a component, so we identify $M$=$N_P$. Steps (iv) and (v) are then replaced by
\begin{enumerate}
\setcounter{enumi}{3}
\item Compute eigenvectors $\bm{u}$ and $\bm{v}$  as
\begin{align}
u_i&=\frac{b_i}{a_i}\ , \\ 
v_i&=\frac{h_i}{\omega_i}\;\;\; \forall i=1,...,N_P\ . 
\end{align}
\item Compute the correction as
\begin{equation}
h_i\leftarrow h_i-u_i\frac{(\bm{u},\bm{v})}{|\bm{u}|^2}\omega_i\;\;\;\forall i=1,...,N_P\ .\label{eq:corr_er}
\end{equation}
\end{enumerate}
\end{itemize}

\subsection{Population dynamics algorithm describes finite-size systems. \label{sec:finite_pd}}
When no simplification can be used, as in the case of Markov matrices, the population dynamics algorithm can be relatively slow, due to the number of nested updates it requires. In these cases, we have therefore been often forced to consider a population size $N_P$ smaller than the values we would have typically wished ($N_P=\mathcal{O}(10^5)$ or more).

However, what may appear as a limitation at first sight turned out to be a blessing, in that it made us aware of an interesting interplay between the size $N_P$ of the population dynamics, and the size $N$ of the graph whose spectral properties were to be reproduced.

Indeed, we have collected convincing evidence that population dynamics at finite $N_P$ does not really capture the thermodynamic limit $N\to\infty$: for a given graph size $N\gg 1$, there is an optimal size of the population $N_P^\star=N_P^\star(N)$ that best captures the spectral properties of that finite-size graph, and the degree of agreement between ``theory" and numerical diagonalisation has a strongly non-monotonic behaviour as a function of $N_P$. Similarly, a population of given size $N_P$ reproduces well spectral properties of graphs around a certain optimal size $N^\star$, but its accuracy rapidly deteriorates if the graph size $N$ is markedly different from $N^\star$. Of course, the higher $N_P$ (e.g. in cases where it is possible to employ $N_P=\mathcal{O}(10^5)$ or larger), the better the large $N$ limit is captured (see e.g. the case in Fig. \ref{fig:finite_size}). 

This intriguing phenomenon may be related to the existence of loops, which seem to be more relevant in the eigenvector problem than the spectral problem. Indeed, whatever $N_P$ is, the cavity fields of type $\omega$ and $h$ will have common predecessors within their own species after $\sim\ln(N_P)/\ln(c-1)$ updates. This implies the presence of loops in the population dynamics update history, which lead to correlations between different members of the population. Therefore, the assumption of population elements \emph{independently} drawn from an ensemble, which underlies \eqref{eq:cav_pi_last} (or equivalently \eqref{eq:pi_last}) is violated. That assumption in turn implements the notion that loops \emph{in the underlying graph} that is being described will diverge in the thermodynamic limit.

To quantify this effect, we compare the cumulative distribution function (CDF) of the second eigenvector's components of Markov matrices with Poissonian shifted degree distribution, obtained via population dynamics at various $N_P$, with the result from direct diagonalisation of matrices from the same ensemble at a given size $N=1000$ -- for both low and high mean degree.

\begin{figure}
\begin{centering}
\includegraphics[scale=0.35]{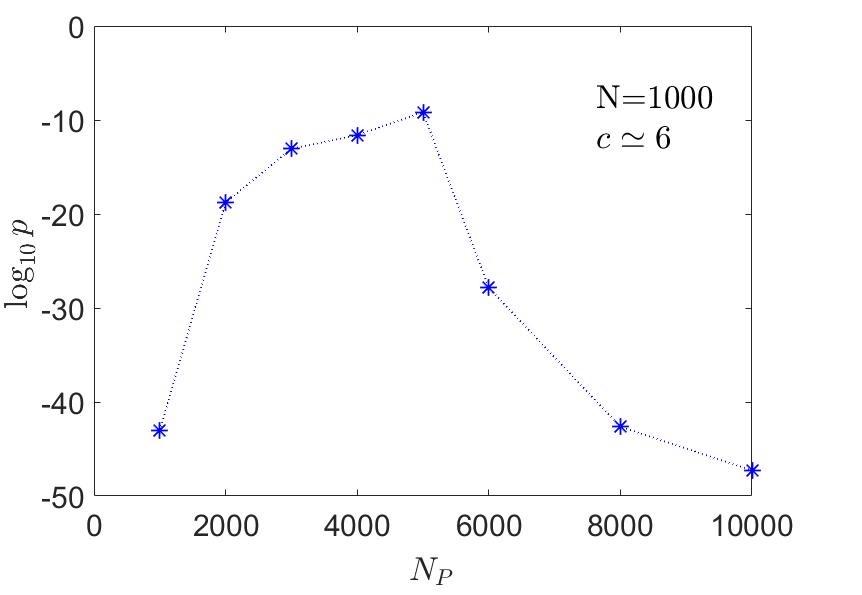}\includegraphics[scale=0.35]{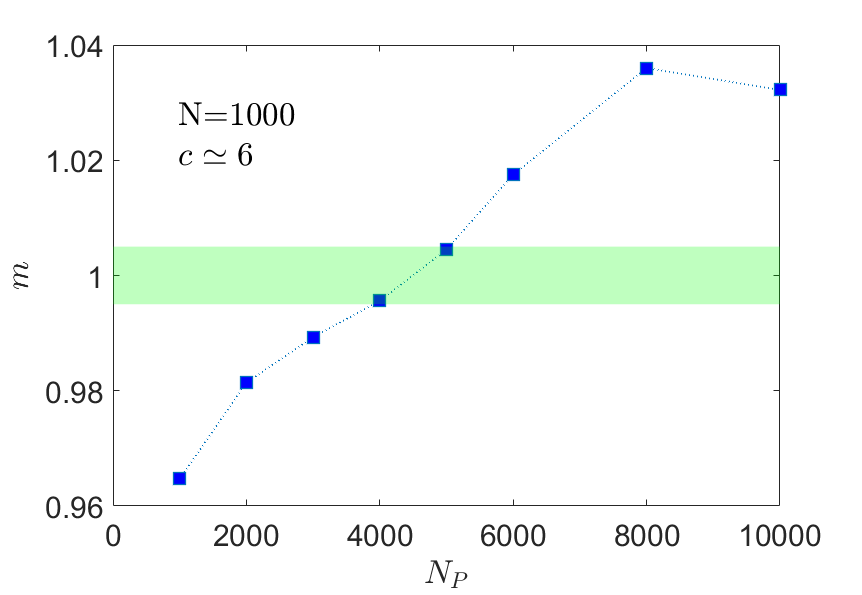}
\par\end{centering}
\begin{centering}
\includegraphics[scale=0.35]{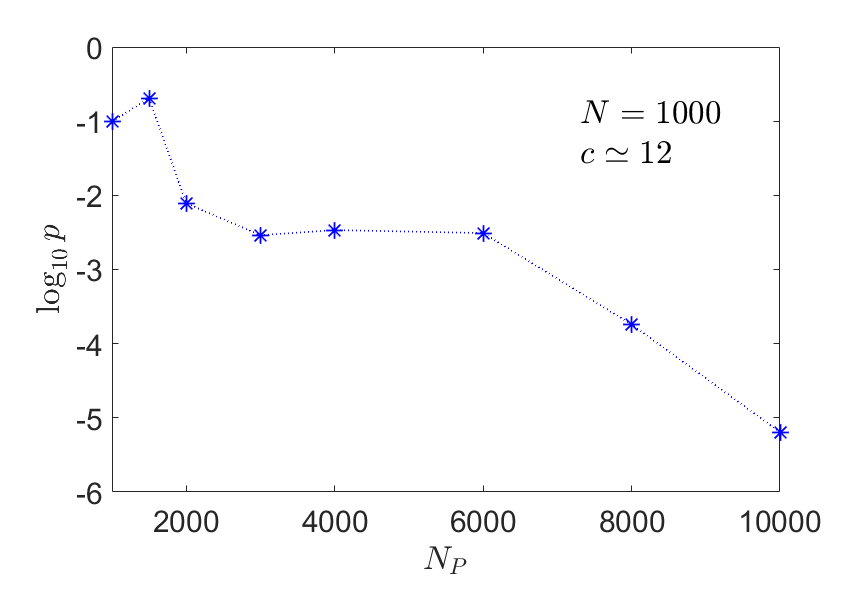}\includegraphics[scale=0.35]{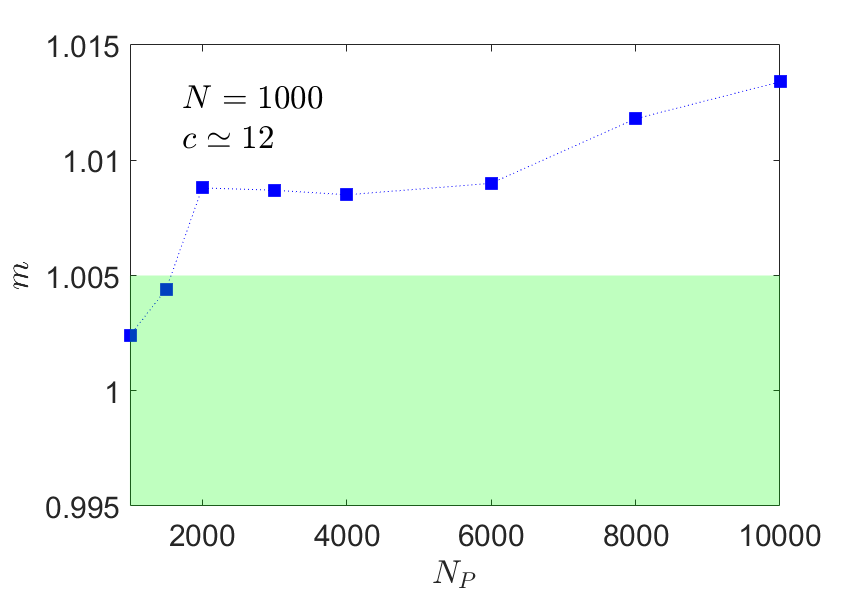}
\par\end{centering}
\begin{centering}
\caption{{\bf Top 
panels}: low mean degree case, $c\simeq6$,  reference matrix size $N=1000$. The left panel shows the base-$10$ logarithm of the $p$-value of the KS two-sample test comparing the two empirical cdfs corresponding to different population sizes. We notice that the $p$-values are all rather low, yet there is a clear maximum value at $N_P^\star\simeq 5000$, and the non-monotonic behaviour is quite pronounced. The right panel shows the slope $m$ of the best-fit regression line of the Q-Q plot between the $25\%$ and 75\% quantiles, for various population sizes. The closer $m$ is to 1, the better the agreement. The plot confirms again that the best agreement with our reference distribution is obtained with $N_P^\star\simeq 5000$. 
{\bf Bottom panels}: high mean degree case, $c\simeq12$, reference matrix size $N=1000$. On the left, we show the $p$-value of the KS two-sample test against $N_P$ in linear $y$-scale. The curve is much flatter than the low-$c$ case, and the $p$-values are all significant, suggesting a high level of similarity between the two distributions throughout the full range of $N_P$. On the right, we plot the slope $m$ of the best fit regression line of the Q-Q plot between the $25\%$ and 75\% quantiles, for various population sizes. For this figure of merit, we again observe a rather flat value of the slope between $N_P\simeq 2000$ and $N_P\simeq 6000$, where $m\simeq 1$ (within a $0.2\%$ error). At high $c$, we indeed observe negligible finite size effects in the direct diagonalisation samples at different sizes $N$, and this phenomenon seems to be present also in the population dynamics simulations. \label{fig:fit}}
\par\end{centering}
\centering{}
\end{figure}

In Figure \ref{fig:fit}, we assess the similarity of the two distributions using two figures of merit. The first (left) is the $p$-value of a 2-sample Kolmogorov-Smirnoff (KS) test: the larger the $p$-value, the strongest the evidence in favor of the hypothesis that the two distributions are the same.
The second (right) is based on the analysis of a so-called \emph{quantile-quantile plot} (Q-Q plot), which is the scatter plot of the quantiles of the two sets of data. Precisely, we focus on the slope $m$ of the best fit regression line $y=mx+b$ of the Q-Q plot, considered between the first and third quartile (respectively, the 0.25 and 0.75 quantiles), to limit spurious effects coming from the under-sampling of the tails. The slope $m$ is directly proportional to the correlation coefficient between the quantiles of the two distributions, and $m= 1$ for identical distributions.

The existence of an optimal population size $N_P^\star$ for a given graph size $N$ -- and the non-monotonic behaviour of the accuracy with $N_P$ -- is quite evident in the left panels. The optimal value of $N_P^\star(N)$ is consistently identified by both figures of merit. However, the effect is more pronounced in the case of low connectivity (top row of Figure \ref{fig:fit}) -- where finite size effects are indeed stronger -- than in the case of high connectivity (bottom row of Figure \ref{fig:fit}).

\section{Conclusions \label{sec:conclusion}}
In summary, we have developed a formalism to compute the statistics of the second largest eigenvalue and of the components of the corresponding eigenvector for some ensembles of sparse symmetric matrices, i.e. weighted adjacency matrices of graphs with finite mean connectivity. By assuming that the top eigenpair is known, we show that for a given matrix, computing the second largest eigenpair is equivalent to computing the top eigenpair of a \emph{deflated} matrix, obtained by subtracting from the original matrix the dense matrix representing a rank-one perturbation proportional to the projector onto its first eigenstate. As in \cite{Susca2019}, the search for the top eigenpair of the deflated matrix is then transformed into the optimisation of a quadratic Hamiltonian on a sphere: introducing the associated Gibbs-Boltzmann distribution and a fictitious inverse temperature $\beta$, the top eigenvector represents the ground state of the system, reached in the limit $\beta\to\infty$. In order to extract this limit, we have employed two Statistical Mechanics methods, cavity and replicas.  We started analysing the case of a single-instance matrix within the cavity framework, introducing a new cavity formulation that allows for the inclusion of hard constraints. 

The single-instance cavity method easily leads to recursion equations, which represent the essential ingredient to obtain the solution of the problem in the thermodynamic limit. We also obtain the exact same equations using replicas as an alternative approach, confirming the equivalence of the two methods in the thermodynamic limit. We employed an improved population dynamics algorithm to solve the stochastic recursion \eqref{eq:cav_pi_last} complemented by the conditions  \eqref{eq:cav_norm_last} and \eqref{eq:cav_orto_last}, (or equivalently \eqref{eq:pi_last} along with \eqref{eq:norm_last} and \eqref{eq:orto_last}) that enforce normalisation and orthogonality of eigenvectors corresponding to different eigenvalues. We found that the convergence of the algorithm is driven not only by the largest eigenvalue of the deflated matrix (i.e. the second largest eigenvalue of the original matrix) but, most essentially, by the fact that the orthogonality condition \eqref{eq:cav_orto_last} (or equivalently \eqref{eq:orto_last}) be correctly enforced. Some ensembles permit simplifications of the algorithm used to enforce orthogonality, which we exploited to speed up convergence.

We remark that from the theoretical point of view our method is applicable no matter the size of the spectral gap. However, if the gap is very narrow, numerical precision limit may not allow for a sufficiently accurate determination of $\lambda=\langle \lambda_2 \rangle_J$.

The simulations show excellent agreement between the theory and the direct diagonalisation of large matrices, and allow us to unpack the contributions to the average density of the \emph{second} eigenvector's components coming from nodes of different degrees.

Our study clearly demonstrates that --- in contrast to beliefs commonly held in the community --- population dynamics at finite $N_P$ is fundamentally incapable of analysing properties representing the thermodynamic limit behaviour. This discovery is in some sense due to the fact that finite size effects are much stronger for eigenvectors than for eigenvalues (in particular for matrices without random edge weights). That finite population size effects are quantitatively related to finite size effects is, in retrospect, not really surprising, given the clear analogy existing between the emergence of correlations in population values -- through loops of common ancestors of population updates -- and common ancestors created through loops in random graphs of finite size, in which the scaling of loop lengths with population and graph size follows basically the same logarithmic law.

In the case of the RRG adjacency matrix, we also analytically studied the pdf of the components of the top eigenvector of the deflated matrix as the deflation parameter is continuously changed, showing the abrupt change of the solution as soon as the deflation parameter becomes larger than the spectral gap of the Kesten-McKay distribution. 

Lastly, we applied our formalism to sparse Markov matrices representing unbiased random walks on a network, for which the second largest eigenpair plays an important role encoding non-equilibrium properties.

\ack{}{}

The authors acknowledge funding by the Engineering and Physical Sciences Research
Council (EPSRC) through the Centre for Doctoral Training in Cross Disciplinary Approaches to Non-Equilibrium Systems (CANES, Grant Nr. EP/L015854/1). 
VARS gratefully acknowledges insightful discussions with Sirio Belga Fedeli and Gioia Boschi.

\section*{\textemdash \textemdash \textemdash \textemdash \textemdash \textendash{}}

\appendix
\section{\\Full deflation: replica derivation \label{sec:replica_gen} }

In this section, we evaluate  the average (or typical) value of the largest eigenvalue and the density of top eigenvectors' components of the matrix $\tilde{J}$ within the replica framework. Our derivation applies to any graph with degree distribution $p(k)$ having finite mean. For weighted adjacency matrices with a Poissonian distribution, we also ask that its support be bounded to ensure that their average largest eigenvalue is finite in the thermodynamic limit.

\subsection{\\Typical largest eigenvalue}\label{sec:replica_eigenvalue}

Consider a $N\times N$ deflated symmetric matrix $\tilde{J}_{ij}(x)=c_{ij}K_{ij}-\frac{x}{N}u_i u_j$. We recall that the $u_i$ represents the $i$-th component of  the \emph{probe} eigenvector $\bm{u}$, i.e. the top eigenvector of the original matrix $J$ (normalised such that $|\bm{u}|^2=N$) which we assume to be known.
Within the framework of the configuration model \cite{Kuehn2011}, the joint distribution of the matrix entries $J_{ij}$ is

\begin{equation}
P\left(\left\{ J_{ij}\right\} \middle | \left\{k_i\right\} \right)=P\left(\left\{c_{ij}\right\} \middle | \left\{k_i\right\} \right)\prod_{i<j}\delta_{K_{ij},K_{ji}}p\left(K_{ij}\right)\ ,\label{eq:joint_true}
\end{equation}
where the distribution $P\left(\left\{c_{ij}\right\} \middle | \left\{k_i\right\} \right)$ of connectivities $\left\{c_{ij}\right\}$ compatible with a given degree sequence $\left\{k_i\right\}$ is given by
\begin{equation}
P\left(\left\{ c_{ij}\right\} \middle | \left\{k_i\right\} \right)=\frac{1}{\mathcal{M}}\prod_{i<j}\delta_{c_{ij},c_{ji}}\left(\frac{c}{N}\delta_{c_{ij},1}+\left(1-\frac{c}{N}\right)\delta_{c_{ij},0}\right)\prod_{i=1}^{N}\delta_{\sum_{j}c_{ij},k_{i}}\ ,\label{eq:true_connectivity}
\end{equation}
and the pdf $p_{K}\left(K_{ij}\right)$ of bond weights (over a compact support whose upper edge is denoted by $\zeta$) can be kept unspecified until the very end. Our derivation will follow the procedure presented in Appendix B in \cite{Susca2019}.

Here we fix $x=\left\langle \lambda_1 \right \rangle_J$: in this setting, the second largest eigenvalue of $J$ is given in terms of the largest eigenvalue of $\tilde{J}$. This can be computed as the formal limit
\begin{equation}
\left\langle\lambda_2\right\rangle_J=\left\langle \tilde{\lambda}_{1}\right\rangle_{\tilde{J}}=\lim_{\beta\rightarrow\infty}\frac{2}{\beta N}\left\langle \ln Z\right\rangle _{\tilde{J}},\qquad Z=\int\mathrm{d}\bm{v}\exp\left[ \frac{\beta}{2}\left(\bm{v},\tilde{J}\bm{v}\right)\right] \delta\left(\left|\bm{v}\right|^{2}-N\right)\ ,\label{eq:average_lambda1summary}
\end{equation}
in terms of the quenched free energy of the model defined in \eqref{eq:hard}. We recall that the round
brackets $\left(\cdot,\cdot\right)$ indicate the dot product between
vectors in $\mathbb{R}^{N}$.

The partition function explicitly reads
\begin{equation}
Z=\int \mathrm{d}\bm{v} \exp\left[ \frac{\beta}{2}\left(\bm{v},J\bm{v}\right)-\frac{\beta \langle \lambda_1 \rangle_J}{2N}\left(\bm{u},\bm{v}\right)^2 \right]\delta\left(\left|\bm{v}\right|^{2}-N\right)\ .\label{eq:part_funct}
\end{equation}

By calling $q=\frac{1}{N}\left(\bm{u},\bm{v}\right) $, we can linearise the square in the exponent of \eqref{eq:part_funct} by means of a Hubbard-Stratonovich identity as follows, 

\begin{equation}
\exp\left(-\frac{\beta \langle \lambda_1 \rangle_J Nq^2}{2}\right)=\sqrt{\frac{\beta \langle \lambda_1 \rangle_J N}{2\pi}}\int\mathrm{d}z \exp \left(-\frac{\beta \langle \lambda_1 \rangle_J N}{2}z^2+\beta\mathrm{i}z \langle \lambda_1 \rangle_J Nq\right)\ ,\label{eq:HS}
\end{equation}
and therefore the partition function reads

\begin{equation}
Z=\sqrt{\frac{\beta \langle \lambda_1 \rangle_J N}{2\pi}}\int \mathrm{d}\bm{v}\mathrm{d}z \exp \left(-\frac{\beta \langle \lambda_1 \rangle_J N}{2}z^2+\mathrm{i}\beta \langle \lambda_1 \rangle_J z \left(\bm{u},\bm{v}\right)+ \frac{\beta}{2} \left(\bm{v},J\bm{v}\right) \right)\delta\left(\left|\bm{v}\right|^{2}-N\right)\ .\label{eq:part_functHS}
\end{equation}

The average over $\tilde{J}$ then reduces to computing the average over $J$. It is computed using the replica trick as follows

\begin{equation}
\left\langle \tilde{\lambda}_{1}\right\rangle _{\tilde{J}}=\lim_{\beta\rightarrow\infty}\frac{2}{\beta N}\lim_{n\rightarrow0}\frac{1}{n}\ln\left\langle Z^{n}\right\rangle _{J}\ ,\label{eq:formula_largest}
\end{equation}
where $n$ is initially taken as an integer, and then analytically continued to real values in the vicinity of $n=0$. The replicated partition function is

\begin{align}
\nonumber\left\langle Z^{n}\right\rangle _{J} & =\left(\frac{\beta \langle \lambda_1 \rangle_J N}{2\pi}\right)^{\frac{n}{2}}\int\left(\prod_{a=1}^{n}\mathrm{d}\bm{v}_{a}\right)\left\langle \exp\left( \frac{\beta}{2}\sum_{a=1}^{n}\sum_{i,j}^{N}v_{ia}J_{ij}v_{ja}\right) \right\rangle _{J}\ \prod_{a=1}^{n}\delta\left(\left|\bm{v}_{a}\right|^{2}-N\right)\\
& \times\int\left(\prod_{a=1}^{n}\mathrm{d}z_{a}\right)\exp \left( -\frac{\beta \langle \lambda_1 \rangle_J N}{2}\sum_{a=1}^n z_{a}^2+\mathrm{i}\beta \langle \lambda_1 \rangle_J \sum_{a=1}^n\sum_{i=1}^N z_{a}v_{ia}u_i\right)\ .
\end{align}
Since the components of $\bm{u}$ are assumed to be known and fixed, they are not affected by the ensemble average. Taking the average w.r.t. the joint distribution \eqref{eq:true_connectivity} of matrix entries yields \cite{Susca2019,Kuehn2011}

\begin{align}
\left\langle \exp\left( \frac{\beta}{2}\sum_{a=1}^{n}\sum_{i,j}^{N}v_{ia}J_{ij}v_{ja}\right) \right\rangle _{J}&=\frac{1}{\mathcal{M}}\int_{-\pi}^{\pi}\left( \prod_{i=1}^{N}\frac{\mathrm{d}\phi_i}{2\pi}\right)\exp\left( -\mathrm{i}\sum_i\phi_i k_i\right)\nonumber \\
&\times\exp\left[ \frac{c}{2N}\sum_{i,j=1}^N\left(\left\langle \mathrm{e}^{\beta K\sum_{a}v_{ia}v_{ja}+\mathrm{i}(\phi_i+\phi_j)}\right\rangle _{K}-1\right)\right]\ ,\label{eq:ensemble_average}
\end{align}
where the average $\left\langle\cdot \right\rangle _{K}$ is taken w.r.t. the pdf of the bond weights $p_{K}(K)$. A Fourier representation of the Kronecker deltas expressing the degree constraints in \eqref{eq:true_connectivity} has been employed. Employing a Fourier representation of the Dirac delta enforcing the normalisation constraint, the replicated partition function thus becomes

\begin{align}
\nonumber &\left\langle Z^{n}\right\rangle _{J}  \propto\frac{1}{\mathcal{M}}\int\left(\prod_{a=1}^{n}\mathrm{d}\bm{v}_{a}\mathrm{d}\lambda_a\mathrm{d}z_{a}\right)\exp \left( -\frac{\beta \langle \lambda_1 \rangle_J N}{2}\sum_{a=1}^n z_{a}^2+\mathrm{i}\beta \langle \lambda_1 \rangle_J \sum_{a=1}^n\sum_{i=1}^N z_{a}v_{ia}u_i\right)\\
\nonumber& \times \exp\left( \mathrm{i}\frac{\beta}{2}N\sum_{a=1}^n\lambda_{a}\right) \exp\left( -\mathrm{i}\frac{\beta}{2}\sum_{a=1}^n\sum_{i=1}^N\lambda_{a}v_{ia}^{2}\right)\int_{-\pi}^{\pi}\left( \prod_{i=1}^{N}\frac{\mathrm{d}\phi_i}{2\pi}\right)\exp\left( -\mathrm{i}\sum_{i=1}^N\phi_i k_i\right)\\
&\times\exp\left[ \frac{c}{2N}\sum_{i,j=1}^N\left(\left\langle \mathrm{e}^{\beta K\sum_{a}v_{ia}v_{ja}+\mathrm{i}(\phi_i+\phi_j)}\right\rangle _{K}-1\right)\right]\ ,\label{replicatedeig}
\end{align}
where we omit irrelevant proportionality constants.

In order to decouple sites, we introduce the functional order parameter

\begin{equation}
\psi\left(\vec{v},\phi\right)=\frac{1}{N}\sum_{i=1}^{N}\delta\left(\phi-\phi_i\right)\prod_{a=1}^{n}\delta\left(v_{a}-v_{ia}\right)\,,
\end{equation}
where the symbol $\vec{v}$ denotes a $n$-dimensional
vector in replica space. We then consider its integrated version \cite{Susca2019,Kuehn2011}
\begin{equation}
\psi\left(\vec{v}\right)=\int \mathrm{d}\phi~ \mathrm{e}^{\mathrm{i}\phi}\psi\left(\vec{v},\phi\right) =\frac{1}{N}\sum_{i=1}^{N}\mathrm{e}^{\mathrm{i}\phi_i}\prod_{a=1}^{n}\delta\left(v_{a}-v_{ia}\right)\ ,\label{eq:integrated_rho}
\end{equation}
and enforce the latter definition using the integral identity

\begin{equation}
1=\int N\mathcal{D}\psi\mathcal{D}\hat{\psi}\exp\left\{ -\mathrm{i}\int\mathrm{d}\vec{v}\ \hat{\psi}\left(\vec{v}\right)\left[N\psi\left(\vec{v}\right)-\sum_{i=1}^N\mathrm{e}^{\mathrm{i}\phi_i}\prod_{a=1}^{n}\delta\left(v_{a}-v_{ia}\right)\right]\right\} \ .
\end{equation}
In terms of the integrated order parameter \eqref{eq:integrated_rho} and its conjugate, the replicated partition function can be written as

\begin{align}
&\nonumber \left\langle Z^{n}\right\rangle _{J} \propto\frac{1}{\mathcal{M}}\int\mathcal{D}\psi\mathcal{D}\hat{\psi}\mathrm{d}\vec{\lambda}\mathrm{d}\vec{z}\exp\left( -\mathrm{i}N\int\mathrm{d}\vec{v}\hat{\psi}\left(\vec{v}\right)\psi\left(\vec{v}\right)\right)\\
\nonumber & \times\exp\left[\frac{Nc}{2}\left(\int\mathrm{d}\vec{v}\mathrm{d}\vec{v^\prime}\psi(\vec{v})\psi(\vec{v^\prime})\left\langle \mathrm{e}^{\beta K \sum_{a}v_{a}v_{a}^{'}}\right\rangle_K-1\right)\right]\exp\left({ \mathrm{i}\frac{\beta}{2}N\sum_{a=1}^n\lambda_{a}} -\frac{\beta \langle \lambda_1 \rangle_J N}{2}\sum_{a=1}^n z_{a}^2\right)\\
\nonumber&\times\int_{-\pi}^{\pi}\left( \prod_{i=1}^{N}\frac{\mathrm{d}\phi_i}{2\pi}\right)\exp\left( -\mathrm{i}\sum_{i=1}^N\phi_i k_i\right)\int\prod_{a=1}^{n}\mathrm{d}\bm{v}_{a}\\
& \times\exp\left( -\mathrm{i}\frac{\beta}{2}\sum_{a=1}^n\sum_{i=1}^N\lambda_{a}v_{ia}^{2}+\mathrm{i}\beta \langle \lambda_1 \rangle_J \sum_{a=1}^n\sum_{i=1}^N z_{a}v_{ia}u_i+ \mathrm{i}\sum_{i=1}^N\mathrm{e}^{\mathrm{i}\phi_i}\int\mathrm{d}\vec{v}\hat{\psi}\left(\vec{v}\right)\prod_{a=1}^{n}\delta\left(v_{a}-v_{ia}\right)\right) \ .
\end{align}
The multiple integral in the last two lines is the product of $N$ $n$-dimensional integrals, each related to both $k_i$ and $u_i$, i.e. the degree and the eigenvector component of the node $i$. It can be expressed by means of the law of large numbers in the following way:

\begin{align}
\nonumber I=& \prod_{i=1}^N\int_{-\pi}^{\pi}\frac{\mathrm{d}\phi_i}{2\pi}\int\mathrm{d}\vec{v}_i\exp\left( -\mathrm{i}\phi_i k_i -\mathrm{i}\frac{\beta}{2}\sum_{a=1}^n\lambda_{a}v_{ia}^{2}+\mathrm{i}\beta\langle\lambda_1\rangle_J\sum_{a=1}^n z_{a}v_{ia}u_i+ \mathrm{i}\hat{\psi}(\vec{v}_i)\mathrm{e}^{\mathrm{i}\phi_i}\right)\\
=& \exp\left[ \sum_{i=1}^N \mathrm{Log} \int\mathrm{d}\vec{v}_i \exp\left(-\mathrm{i}\frac{\beta}{2}\sum_{a=1}^n\lambda_{a}v_{ia}^{2}+\mathrm{i}\beta\langle\lambda_1\rangle_J\sum_{a=1}^n z_{a}v_{ia}u_i \right)I[k_i,\vec{v}_i]\right] \ ,\label{eq:single_site_1}
\end{align}
where $\mathrm{Log}$ denotes the principal branch of the complex logarithm, and
\begin{equation}
I[k_i,\vec{v}_i]=\int_{-\pi}^{\pi}\frac{\mathrm{d}\phi_i}{2\pi}\exp\left(  -\mathrm{i}\phi_i k_i+\mathrm{i}\hat{\psi}(\vec{v}_i)\mathrm{e}^{\mathrm{i}\phi_i}\right)\ .
\end{equation}
Each of the $\phi_i$ integrals can be performed by rewriting the last exponential factor as a power series, viz.
\begin{align}
I[k_i,\vec{v}_i]=&\int_{-\pi}^{\pi}\frac{\mathrm{d}\phi_i}{2\pi}\mathrm{e}^{-\mathrm{i}\phi_i k_i}\sum_{s=0}^{\infty} \frac{\left(\mathrm{i}\hat\psi(\vec{v}_i)\right)^s}{s!}\mathrm{e}^{\mathrm{i}s\phi_i}
=\sum_{s=0}^{\infty} \frac{\left(\mathrm{i}\hat\psi(\vec{v}_i)\right)^s}{s!}\delta_{s,k_i}
=\frac{\left(\mathrm{i}\hat\psi(\vec{v}_i)\right)^{k_i}}{k_i!}\;\;\;\;\; \forall k_i\ ,
\end{align}
with $i=1,\ldots,N$. Therefore, by invoking the Law of Large Numbers, the single site integral $I$ \eqref{eq:single_site_1} can be expressed as
\begin{align}
I=&\exp\left[ \sum_{i=1}^N \mathrm{Log} \int\mathrm{d}\vec{v}_i \exp\left(-\mathrm{i}\frac{\beta}{2}\sum_{a=1}^n\lambda_{a}v_{ia}^{2}+\mathrm{i}\beta\langle\lambda_1\rangle_J\sum_{a=1}^n z_{a}v_{ia}u_i \right)\frac{\left(\mathrm{i}\hat\psi(\vec{v}_i)\right)^{k_i}}{k_i!}\right]\nonumber\\
\nonumber =&\exp \bigg\{ N\sum_{k=k_{\mathrm{min}}}^{k_{\mathrm{max}}} p(k)\bigg[\int\mathrm{d}u~\rho_J(u|k)\mathrm{Log}\int\mathrm{d}\vec{v}\exp\left(-\mathrm{i}\frac{\beta}{2}\sum_{a=1}^n\lambda_a v_a^2+\mathrm{i}\beta\langle\lambda_1\rangle_J u \sum_{a=1}^n z_{a}v_{a}\right)\\
&\times (\mathrm{i}\hat\psi(\vec{v}))^k-\mathrm{Log}(k!)\bigg]\bigg\}\ ,\label{eq:single_site_2}
\end{align}
where we have used 
\begin{equation}
\frac{1}{N}\sum_{i=1}^{N}\mathrm{Log}f(k_i,u_i)\simeq\sum_{k=k_{\mathrm{min}}}^{k_{\mathrm{max}}} p(k)\int\mathrm{d}u\rho_J(u|k) \mathrm{Log}f(k,u)\ .
\end{equation}
Here, $p(k)$ is the actual degree distribution of the graph and $\rho_J(u|k)$ represents the distribution of the top eigenvector's components of the original matrix $J$ conditioned on the degree $k$. As shown in \cite{Susca2019}, the variables $u_i$ are strongly correlated with the $k_i$ so their dependence on the $k_i$ must be taken into account.

Therefore, the replicated partition function takes a form amenable to a saddle point evaluation in the large $N$ limit (assuming we can safely exchange the limits $n\to 0$ and $N\to\infty$)

\begin{equation}
\left\langle Z^{n}\right\rangle _{J}\propto\int\mathcal{D}\psi\mathcal{D}\hat{\psi}\mathrm{d}\vec{\lambda}\mathrm{d}\vec{z}\exp\left(NS_{n}[\psi,\hat{\psi},\vec{\lambda},\vec{z}]\right)\ ,
\end{equation}
where
\begin{equation}
S_{n}[\psi,\hat{\psi},\vec{\lambda},\vec{z}]=S_{1}[\psi,\hat{\psi}]+S_{2}\left[\psi\right]+S_{3}(\vec{\lambda})+S_{4}(\vec{z})+S_{5}[\hat{\psi},\vec{\lambda},\vec{z}]\ ,\label{eq:action}
\end{equation}
and

\begin{align}
S_{1}[\psi,\hat{\psi}] & =-\mathrm{i}\int\mathrm{d}\vec{v}\hat{\psi}(\vec{v})\psi(\vec{v})\ ,\label{eq:S1}\\
S_{2}[\psi] & =\frac{c}{2}\left(\int\mathrm{d}\vec{v}\mathrm{\mathrm{d}}\vec{v^\prime}\psi(\vec{v})\psi(\vec{v^\prime})\left\langle  \mathrm{e}^{\beta K \sum_{a}v_{a}v_{a}^{'}}\right\rangle_K-1\right)\ ,\label{eq:S2}\\
S_{3}(\vec{\lambda}) & =\mathrm{i}\frac{\beta}{2}\sum_{a=1}^n\lambda_{a}\ ,\label{eq:S3}\\
S_{4}(\vec{z}) & =-\frac{\beta \langle \lambda_1 \rangle_J}{2}\sum_{a=1}^n z_{a}^2\ ,\label{eq:S4}\\
S_{5}[\hat{\psi},\vec{\lambda},\vec{z}] & =\sum_{k=k_{\mathrm{min}}}^{k_{\mathrm{max}}} p(k)\bigg[\int\mathrm{d}u~\rho_J(u|k)\mathrm{Log}\int\mathrm{d}\vec{v}\exp\left(-\mathrm{i}\frac{\beta}{2}\sum_{a=1}^n\lambda_a v_a^2+\mathrm{i}\beta\langle\lambda_1\rangle_J u \sum_{a=1}^n z_{a}v_{a}\right)\nonumber\\
&\;\;\;\;\;\;\;\;\;\;\;\;\;\;\;\;\;\;\;\;\;\;\times(\mathrm{i}\hat\psi(\vec{v}))^k-\mathrm{Log}(k!)\bigg]\ \ ,\label{eq:S5}
\end{align}
where we consider $k_{\mathrm{min}}=0$ henceforth.

The stationarity of the action $S_{n}$ w.r.t. variations of $\psi$
and $\hat{\psi}$ requires that the order parameter at the saddle point $\psi^\star$ and its
conjugate $\hat{\psi}^\star$ satisfy the following coupled equations

\begin{align}
\mathrm{i}\hat{\psi}^\star(\vec{v}) &=c\int\mathrm{d}\vec{v^\prime}\psi^\star(\vec{v^\prime}) \left\langle  \mathrm{e}^{\beta K \sum_{a}v_{a}v_{a}^{'}}\right\rangle_K\ ,\label{eq:rho_stat}\\
\psi^\star(\vec{v}) &=\sum_{k=1}^{k_{\mathrm{max}}} p(k)k\int \mathrm{d}u\rho_J(u|k) \frac{\exp\left[ -\mathrm{i}\frac{\beta}{2}\sum_{a}\lambda_{a}v_{a}^{2}+\mathrm{i}\beta \langle\lambda_1 \rangle_J u \sum_{a} z_{a}v_{a}\right]\left(\mathrm{i}\hat{\psi}^\star(\vec{v})\right)^{k-1} }{\int\mathrm{d}\vec{v^\prime}\exp\left[ -\mathrm{i}\frac{\beta}{2}\sum_{a}\lambda_{a}v_{a}^{\prime 2}+\mathrm{i}\beta  \langle\lambda_1 \rangle_J u \sum_{a} z_{a}v_{a}^{\prime}\right]\left(\mathrm{i}\hat{\psi}^\star(\vec{v^\prime})\right)^k}\ ,\label{eq:hat_rho_stat}
\end{align}
which have to be solved together with the stationarity conditions w.r.t. each component $\lambda_{\bar{a}}$ of $\vec{\lambda}$ and $z_{\bar{a}}$ of $\vec{z}$ (for $\bar{a}=1,\ldots,n$),

\begin{equation}
1=\sum_{k=0}^{k_{\mathrm{max}}} p(k)\int \mathrm{d}u\rho_J(u|k)\frac{\int\mathrm{d}\vec{v}\exp\left[ -\mathrm{i}\frac{\beta}{2}\sum_{a}\lambda_{a}v_{a}^{2}+\mathrm{i}\beta  \langle\lambda_1 \rangle_J u\sum_{a} z_{a}v_{a}\right]\left(\mathrm{i}\hat{\psi}^\star(\vec{v}) \right)^k v_{\bar{a}}^{2}}{\int\mathrm{d}\vec{v^\prime}\exp\left[ -\mathrm{i}\frac{\beta}{2}\sum_{a}\lambda_{a}v_{a}^{\prime2}+\mathrm{i}\beta  \langle\lambda_1 \rangle_J u\sum_{a} z_{a}v_{a}^\prime\right]\left(\mathrm{i}\hat{\psi}^\star(\vec{v^\prime}) \right)^k }\ ,\label{eq:lambda_stat}
\end{equation}

\begin{equation}
z_{\bar{a}}=\mathrm{i}\sum_{k=0}^{k_{\mathrm{max}}} p(k)\int \mathrm{d}u\rho_J(u|k)u\frac{\int\mathrm{d}\vec{v}\exp\left[ -\mathrm{i}\frac{\beta}{2}\sum_{a}\lambda_{a}v_{a}^{2}+\mathrm{i}\beta  \langle\lambda_1 \rangle_J u\sum_{a} z_{a}v_{a}\right]\left(\mathrm{i}\hat{\psi}^\star(\vec{v}) \right)^k v_{\bar{a}}}{\int\mathrm{d}\vec{v^\prime}\exp\left[ -\mathrm{i}\frac{\beta}{2}\sum_{a}\lambda_{a}v_{a}^{\prime2}+\mathrm{i}\beta  \langle\lambda_1 \rangle_J u\sum_{a} z_{a}v_{a}^\prime\right]\left(\mathrm{i}\hat{\psi}^\star(\vec{v^\prime}) \right)^k }\ .\label{eq:lambda_stat}
\end{equation}
Apart from the extra averages w.r.t. $p(k)$ and $\rho_J(u|k)$, the equations \eqref{eq:rho_stat} and \eqref{eq:hat_rho_stat} share some similarities with the saddle-point equations leading to the spectral density of sparse random graphs \cite{Rodgers1988,Kuehn2008}  and to those leading to the top eigenpair statistics of sparse symmetric matrices \cite{Susca2019}: similarly to the latter case, the harmonic ``Hamiltonian"  of this problem is real-valued and includes the inverse temperature $\beta$. Following \cite{Susca2019,Kuehn2008}, we will now search for replica-symmetric solutions written in the form of uncountably infinite superpositions of  Gaussians with a non-zero mean. As in the case for the top eigenvector, our ansatz will be preserving permutational symmetry between replicas, but not the rotational invariance in replica space, since this symmetry would not produce a physically meaningful result for this problem.

 \begin{align}
\lambda_{\bar{a}} &=\lambda\qquad\forall \bar{a}=1,\ldots,n\ ,\\
z_{\bar{a}}&=z\qquad\forall \bar{a}=1,\ldots,n\ ,\\
\psi^\star(\vec{v}) &=\psi_0\int\mathrm{d}\omega\mathrm{d}h\ \pi\left(\omega,h\right)\prod_{a=1}^{n}\frac{1}{Z_\beta(\omega,h)}\exp\left[ -\frac{\beta}{2}\omega v_{a}^{2}+\beta hv_{a}\right] \ ,\label{eq:ansatz_rep_1_1}\\
\hat{\psi}^\star(\vec{v}) &=\hat{\psi}_0\int\mathrm{d}\hat{\omega}\mathrm{d}\hat{h}\ \hat{\pi}(\hat{\omega},\hat{h})\prod_{a=1}^{n}\exp\left[ \frac{\beta}{2}\hat{\omega}v_{a}^{2}+\beta\hat{h}v_{a}\right] \ ,\label{eq:ansatz_rep_2_2}
\end{align}
where 

\begin{equation}
Z_\beta(x,y)=\sqrt{\frac{2\pi}{\beta x}}\exp\left(\frac{\beta y^{2}}{2 x}\right)\ .\label{eq:zomega_er}
\end{equation}

We remark that our replica symmetry assumption has proved to be generally exact in the random matrix context and specifically for the spectral problem of sparse random matrices \cite{Edwards1976,Rodgers1988,Kuehn2008,Khorunzhy2004}. Moreover, the representation of the order parameter as a superposition of Gaussian pdfs leads to the correct solution for harmonically coupled systems  \cite{Kuehn2007}, such as the one described in the present work.

The calculation follows the same path traced in Appendix B of \cite{Susca2019}. 
In \eqref{eq:ansatz_rep_1_1} and \eqref{eq:ansatz_rep_2_2}, $\pi$ and $\hat\pi$ are auxiliary normalised joint pdfs of the parameters appearing in the Gaussian distributions. The $\psi_0$ and $\hat\psi_0$ are determined such that the distributions $\pi(\omega,h)$ and $\hat\pi(\hat\omega,\hat{h})$ are normalised.

Expressing the order parameter in this form allows us to perform explicitly the $\vec{v}$-integrals in the action $S_n$, eventually leading to simpler coupled stationarity equations for $\pi$ and $\hat\pi$. The convergence of the $\vec{v}$-integrals requires $\omega>\hat\omega$ and $\omega >\zeta$ (where $\zeta$ is the upper edge of the support of the pdf $p_{K}(K)$ of bond weights).

Rewriting the action in terms of $\pi$ and $\hat\pi$, after performing the $\vec{v}$-integrations, and extracting the leading $n\to 0$ contribution the full action now reads
\begin{equation}
S_n=S_{1}[\pi,\hat{\pi}]+S_{2}[\pi]+S_{3}(\lambda)+S_{4}(z)+S_{5}[\hat{\pi},\lambda,z]\ ,\label{eq:action_pihatpi}
\end{equation}
with
\begin{align}
S_{1}[\pi,\hat{\pi}] & =-\mathrm{i}\psi_0\hat\psi_0-\mathrm{i}\psi_0\hat\psi_0n\int\mathrm{d}\pi(\omega,h)\mathrm{d}\hat{\pi}(\hat{\omega},\hat{h})\ln\frac{Z_\beta(\omega-\hat{\omega},h+\hat{h})}{Z_\beta(\omega,h)}\ ,\label{eq:S1_pi} & {}\\
S_{2}[\pi] & =\frac{c}{2}\left(\psi_0^2-1 \right)+\frac{c}{2}\psi_0^2 n \int\mathrm{d}\pi(\omega,h)\mathrm{d}\pi(\omega',h')\left\langle \ln\frac{Z^{(2)}_\beta\left(\omega,\omega',h,h',K\right)}{Z_\beta\left(\omega,h\right)Z_\beta\left(\omega',h'\right)}\right\rangle_K\ ,\label{eq:S2_pi}\\
S_{3}(\lambda) & =\mathrm{i}\frac{\beta}{2}n\lambda\ ,\label{eq:S3_pi}\\
S_{4}(z) &=-n\frac{\beta}{2}\langle \lambda_1 \rangle_J z^2\ ,\\
S_{5}[\hat{\pi},\lambda,z] & =c~\mathrm{Log}(\mathrm{i}\hat\psi_0)-\sum_{k=0}^{k_{\mathrm{max}}} p(k)\mathrm{Log}(k!)+n\sum_{k=0}^{k_{\mathrm{max}}} p(k)\int\mathrm{d}u\rho_J(u|k)\nonumber\\
&\times\int\{ \mathrm{d}\hat{\pi}\} _{k}~\mathrm{Log}~Z_\beta\left(\mathrm{i}\lambda-\{ \hat{\omega}\} _{k},\mathrm{i}z\langle \lambda_1 \rangle_J u+\{ \hat{h}\} _{k}\right)\label{S5beforesaddle}\ ,
\end{align}
where we have introduced the shorthands 
\begin{equation}
Z^{(2)}_\beta(\omega,\omega',h,h',K)=Z_\beta(\omega',h')Z_\beta\left(\omega-\frac{K^2}{\omega'},h+\frac{h'K}{\omega'}\right)
\end{equation}
and $\{\mathrm{d}\hat{\pi}\} _{k}=\prod_{\ell=1}^{k}\mathrm{d}\hat{\omega}_{\ell}\mathrm{d}\hat{h}_{\ell}\hat{\pi}(\hat{\omega}_{\ell},\hat{h}_{\ell})$,
along with $\{ \hat{\omega}\} _{s}=\sum_{\ell=1}^{s}\hat{\omega}_{\ell}$
and $\{ \hat{h}\} _{s}=\sum_{\ell=1}^{s}\hat{h}_{\ell}$. 

The action contains $\mathcal{O}(1)$ and $\mathcal{O}(n)$ terms as $n\to 0$: the $\mathcal{O}(1)$ terms are cancelled by the $\mathcal{O}(1)$ terms arising from the evaluation of the normalisation constant $\mathcal{M}$ at the saddle-point so that the action \eqref{eq:action_pihatpi} is $\mathcal{O}(n)$ as expected. We refer to Appendix B of \cite{Susca2019} for the evaluation of $\mathcal{M}$.

The stationarity condition w.r.t. $\lambda$ entails

\begin{equation}
\frac{\partial S}{\partial\lambda}\Big|_{\lambda=\lambda^\star}=0\Rightarrow 1=\sum_{k=0}^{k_{\mathrm{max}}} p(k)\int\mathrm{d}u\rho_J(u|k)\int\{\mathrm{d}\hat{\pi}\} _{k}\langle v^{2}\rangle_{\bar P} \ ,\label{eq:stat_k}
\end{equation}
where the average $\langle \cdot \rangle_{\bar{P}} $ is taken with respect to
the Gaussian measure
\begin{equation}
\bar{P}(v) = \sqrt{\frac{\beta\left(\mathrm{i}\lambda^\star-\{ \hat{\omega}\} _{k}\right)}{2\pi}}\exp\left[ -\frac{\beta}{2}\left(\mathrm{i}\lambda^\star-\{ \hat{\omega}\} _{k}\right)\left(v-\frac{\mathrm{i}z^{\star}\langle \lambda_1 \rangle_J u+\{ \hat{h}\} _{k}}{\mathrm{i}\lambda^\star-\{ \hat{\omega}\} _{k}}\right)^{2}\right]\ .\label{eq:p_bar}
\end{equation}
More explicitly, \eqref{eq:stat_k} reads
\begin{equation}
1=\sum_{k=0}^{k_{\mathrm{max}}} p(k)\int\mathrm{d}u\rho_J(u|k)\int\{\mathrm{d}\hat{\pi}\} _{k}\left[\frac{1}{\beta(\mathrm{i}\lambda^\star-\{ \hat{\omega}\} _{k})}+\left(\frac{\mathrm{i}z^{\star}\langle \lambda_1 \rangle_J u+\{ \hat{h}\} _{k}}{\mathrm{i}\lambda^\star-\{ \hat{\omega}\} _{k}}\right)^{2}\right].\label{eq:lambdastarcondition}
\end{equation}
We note that the $\beta$-dependent term vanishes as $\beta\rightarrow\infty$.

The stationarity condition w.r.t. $z$ entails
\begin{equation}
\frac{\partial S}{\partial z}\Big|_{z=z^\star}=0\Rightarrow z^{\star}=\mathrm{i}\sum_{k=0}^{k_{\mathrm{max}}} p(k)\int\mathrm{d}u\rho_J(u|k)u\int\{\mathrm{d}\hat{\pi}\} _{k}\langle v\rangle_{\bar P} \ ,\label{eq:stat_z}
\end{equation}
where the average $\langle \cdot \rangle_{\bar{P}} $ is taken with respect to
the Gaussian measure \eqref{eq:p_bar}. More explicitly,
\begin{equation}
z^{\star}=\mathrm{i}\sum_{k=0}^{k_{\mathrm{max}}} p(k)\int\mathrm{d}u\rho_J(u|k)u\int\{\mathrm{d}\hat{\pi}\} _{k}\left(\frac{\mathrm{i}z^{\star}\langle \lambda_1 \rangle_J u+\{ \hat{h}\} _{k}}{\mathrm{i}\lambda^\star-\{ \hat{\omega}\} _{k}}\right).\label{eq:zstarcondition}
\end{equation}

The stationarity condition with respect to variations of $\pi$, 
$
\frac{\delta S}{\delta\pi}  =  0
$, is
\begin{equation}
\hat{\pi}(\hat{\omega},\hat{h})=\int\mathrm{d}\omega\mathrm{d}h~\pi(\omega,h) \left\langle \delta\left(\hat{\omega}-\frac{K^2}{\omega}\right)\delta\left(\hat{h}-\frac{hK}{\omega}\right)\right\rangle_K\ .\label{eq:pi_hat}
\end{equation}

Similarly, the stationarity condition with respect to variations of $\hat{\pi}$, 
$
\frac{\delta S}{\delta\hat{\pi}}=0,
$
produces the condition
\begin{equation}
\pi(\omega,h)=\sum_{k=1}^{k_{\mathrm{max}}} p(k)\frac{k}{c}\int\mathrm{d}u\rho_J(u|k)\int\{ \mathrm{d}\hat{\pi}\} _{k-1}\delta\left(\omega-(\mathrm{i}\lambda^\star-\{ \hat{\omega}\} _{k-1})\right)\delta\left(h-(\mathrm{i}z^{\star}\langle\lambda_1\rangle_J u+\{ \hat{h}\} _{k-1})\right)\ .\label{eq:pi_firstform}
\end{equation}
Inserting \eqref{eq:pi_hat} into \eqref{eq:pi_firstform} yields
\begin{align}
\nonumber\pi(\omega,h)&=\sum_{k=1}^{k_{\mathrm{max}}} p(k)\frac{k}{c}\int\mathrm{d}u\rho_J(u|k)\int \{\mathrm{d}\pi\}_{k-1}\\
&\times\left\langle \delta\left(\omega-(\mathrm{i}\lambda^\star-\sum_{\ell=1}^{k-1}\frac{K^2_\ell}{\omega_{\ell}})\right)\delta\left(h-\left(\mathrm{i}z^{\star}\langle\lambda_1\rangle_J u+\sum_{\ell=1}^{k-1}\frac{h_{\ell}K_\ell}{\omega_{\ell}}\right)\right)\right\rangle_{\{K\}_{k-1}}\ ,\label{eq:pi}
\end{align}
where the brackets $\langle\cdot\rangle_{\{K\}_{k-1}}$ denote averaging with respect to a collection of $k-1$ i.i.d.
random variables $K$, each drawn from the bond weight pdf $p_{K}(K)$.
We recall that $p(k)$ appearing in  \eqref{eq:pi} is already the actual degree distribution of the graph with finite mean $c$ and bounded maximal degree.

Following \cite{Susca2019}, we relabel the constant terms $\lambda\equiv \mathrm{i}\lambda^\star$ and $q\equiv-\mathrm{i}z^{\star}$ since they both turn out to be real-valued. We eventually find 
\begin{align}
\nonumber \pi(\omega,h)&=\sum_{k=1}^{k_{\mathrm{max}}}p(k)\frac{k}{c}\int\mathrm{d}u\rho_{J}(u|k)\int\{\mathrm{d}\pi\} _{k-1}\\
&\times\left\langle \delta\left(\omega-\left(\lambda-\sum_{\ell=1}^{k-1}\frac{K_{\ell}^2}{\omega_{\ell}}\right)\right)\delta\left(h-\left(-qu\left\langle \lambda_1 \right\rangle_{J} +\sum_{\ell=1}^{k-1}\frac{h_{\ell}K_{\ell}}{\omega_{\ell}}\right)\right)\right\rangle_{\{K\}_{k-1}}\ .
\label{eq:pi_gen}
\end{align}

The parameter $\lambda$ must be tuned as to enforce the supplementary condition \eqref{eq:lambdastarcondition} as $\beta\to\infty$, which reads 

\begin{equation}
1=\sum_{k=0}^{k_{\mathrm{max}}}p(k)\int\mathrm{d}u\rho_{J}(u|k)\int\{\mathrm{d}\pi\} _{k}\left\langle \left(\frac{-qu\left\langle \lambda_1 \right\rangle_{J}+\sum_{\ell=1}^k \frac{h_\ell K_\ell}{\omega_\ell}}{\lambda-\sum_{\ell=1}^k \frac{K_\ell^2}{\omega_\ell}}\right)^{2}\right\rangle_{\{K\}_{k}}\ ,
\label{eq:lambda_gen}
\end{equation}
whereas \eqref{eq:zstarcondition} gives the following condition for $q$ 
\begin{equation}
q=\sum_{k=0}^{k_{\mathrm{max}}}p(k)\int\mathrm{d}u\rho_{J}(u|k)u\int\{\mathrm{d}\pi\} _{k}\left\langle \left(\frac{-qu\left\langle \lambda_1 \right\rangle_{J}+\sum_{\ell=1}^k \frac{h_\ell K_\ell}{\omega_\ell}}{\lambda-\sum_{\ell=1}^k \frac{K_\ell^2}{\omega_\ell}}\right)\right\rangle_{\{K\}_{k}}\ .\label{eq:q_gen}
\end{equation}

The structure of the action \eqref{eq:action_pihatpi} is the same as that found in \cite{Susca2019} (see for instance Section 4.1.1 there), except for the term $S_4(z)\equiv S_4(q)$. Therefore, building on the same reasoning, the average largest eigenvalue of $\tilde{J}$, i.e. the average second largest  eigenvalue of $J$ is given by
\begin{equation}
\left\langle \tilde{\lambda_1} \right\rangle_{\tilde{J}}\equiv \left\langle \lambda_2 \right\rangle_{J}= \lambda+\left\langle \lambda_1 \right\rangle_{J}q^2\ ,
\label{eq:lambda2_gen}
\end{equation}
where $\lambda$ and $q$ are defined by \eqref{eq:lambda_gen} and \eqref{eq:q_gen}. As observed in Section \ref{sec:cavity_td_orto}, in case of full deflation we find $q=0$, hence $\left\langle \tilde{\lambda_1} \right\rangle_{\tilde{J}}\equiv \left\langle \lambda_2 \right\rangle_{J}= \lambda$.

Eq.~\eqref{eq:pi_gen}, along with the conditions  \eqref{eq:lambda_gen} and \eqref{eq:q_gen}, are typically solved by population dynamics, as shown in Section \ref{sec:pd}. They represent the generalisation in the large $N$ limit of the single-instance recursions \eqref{eq:cavity_omega} and \eqref{eq:cavity_h} along with the conditions \eqref{eq:q_cav_single} and \eqref{eq:lambda_cav_single}.

\subsection{\\Density of top eigenvector's components using replicas \label{sec:replica_evect}}
In this section, we provide the derivation for the density of components of the top eigenvector of the matrix $\tilde{J}$, in terms of $\pi$, $\lambda$ and $q$. As in the previous subsection, we consider the deflation parameter $x=\langle \lambda_1\rangle_J$, and therefore the top eigenvector of the deflated matrix $\tilde{J}$ corresponds to the \emph{second} eigenvector of the original matrix $J$. We will be following the same approach of Section 4.2 in \cite{Susca2019}. We will report here the main steps to keep this paper self-contained. In this statistical mechanics framework, the quantity
\begin{equation}
\tilde{\rho}_{\beta,\tilde{J}}\left(w\right)=\left\langle{\frac{1}{N}\sum_{i=1}^{N}\delta\left(w-v_{i}\right)}\right\rangle \label{eq:instance_density}
\end{equation}
is defined such that in the limit $\beta\rightarrow\infty$ it gives
the density of the top eigenvector components for a given $N\times N$
deflated symmetric random matrix $\tilde{J}$. The simple angle brackets $\left\langle ... \right\rangle$  stands for
thermal averaging with respect to the Gibbs-Boltzmann
distribution \eqref{eq:hard} of the system
\begin{equation}
P_{\beta,\tilde{J}}(\bm{v})=\frac{\exp\left(\frac{\beta}{2}\left(\bm{v},\tilde{J}\bm{v}\right)\right) \delta\left(\left|\bm{v}\right|^{2}-N\right)}{\int\mathrm{d}\bm{v}'\exp\left( \frac{\beta}{2}(\bm{v}',\tilde{J}\bm{v}')\right) \delta\left(\left|\bm{v}'\right|^{2}-N\right)}\ .
\end{equation}
Defining an auxiliary partition function as 

\begin{equation}
Z^{(\beta)}_\epsilon(t,\tilde{J};w)=\int\mathrm{d}\bm{v}\exp\left[\frac{\beta}{2}\left(\bm{v},\tilde{J}\bm{v}\right)+\beta t\sum_{i}\delta_\epsilon\left(w-v_{i}\right)\right] \delta\left(\left|\bm{v}\right|^{2}-N\right)\ ,
\end{equation}
where $\delta_\epsilon$ is a smooth regulariser of the delta function, the quantity \eqref{eq:instance_density} can be formally expressed as 

\begin{equation}
\tilde{\rho}_{\beta,\tilde{J}}(w)=\lim_{\epsilon\to 0^+}\frac{1}{\beta N}\frac{\partial}{\partial t}\ln Z^{(\beta)}_\epsilon(t,\tilde{J};w)\Big|_{t=0}\ .
\end{equation}

Averaging now over the matrix ensemble
\begin{equation}
\rho_{\beta,\tilde{J}}(w)=\left\langle\tilde{\rho}_{\beta,\tilde{J}}\left(w\right)\right\rangle_{\tilde{J}}
\end{equation}
and sending $\beta\to\infty$ at the very end, the density of the top eigenvector's components is eventually given by the formula 
\begin{equation}
\rho_{\tilde{J}}(w)=\lim_{\beta\to\infty}\lim_{\epsilon\to 0^+}\frac{1}{\beta N}\frac{\partial}{\partial t}\left\langle\ln Z^{(\beta)}_\epsilon(t,\tilde{J};w)\right\rangle_{\tilde{J}}\Big|_{t=0}\ ,\label{eq:remarkable}
\end{equation}
equivalent to Eq. (95) in \cite{Susca2019}.

To compute the average of the logarithm of the auxiliary partition function
$Z^{(\beta)}_\epsilon(t,\tilde{J};w)$, we employ the replica trick 
\begin{equation}
\left\langle\ln Z^{(\beta)}_\epsilon(t,\tilde{J};w)\right\rangle_{\tilde{J}}=\lim_{n\to 0}\frac{1}{n}\ln \left\langle [Z^{(\beta)}_\epsilon(t,\tilde{J};w)]^n\right\rangle_{\tilde{J}}\ .\label{replicavecZeps}
\end{equation}

The replicated partition function takes the form

\begin{equation}
 \left\langle [Z^{(\beta)}_\epsilon(t,\tilde{J};w)]^n\right\rangle_J\propto\frac{1}{\mathcal{M}} \int\mathcal{D}\psi\mathcal{D}\hat{\psi}\mathrm{d}\vec{\lambda}\mathrm{d}\vec{z}\exp\left [NS^{(\beta)}_{n}\left[\psi,\hat{\psi},\vec{\lambda},\vec{z};t,\epsilon;w\right]\right]\ ,
\end{equation}
where $\psi$ and $\hat\psi$ are functional order parameters\footnote{We use the same symbols  $\psi$ and $\hat\psi$ as in \ref{sec:replica_eigenvalue}.} . For large $N$, we employ a saddle point approximation 
\begin{equation}
 \left\langle [Z^{(\beta)}_\epsilon(t,\tilde{J};w)]^n\right\rangle_{\tilde{J}}\approx \frac{1}{\mathcal{M}} \exp\left [NS^{(\beta)}_{n}\left(\psi^\star,\hat{\psi}^\star,\vec{\lambda}^\star,\vec{z}^\star;t,\epsilon;w\right)\right]\ ,\label{actionvec}
\end{equation}
where the starred objects satisfy self-consistency equations where $t$ can be safely set to zero, since the partial derivative $\frac{\partial}{\partial t}$ in \eqref{eq:remarkable} only acts on terms containing an \emph{explicit} dependence on $t$. Again, we refer to Appendix B of \cite{Susca2019} for the evaluation of the constant $\mathcal{M}$.

The stationarity conditions defining
$\psi^\star$, $\hat{\psi}^\star$, $\lambda^\star$ and $\vec{z}^\star$ at the saddle point
for $t=0$ are identical to those found in Section \ref{sec:replica_eigenvalue}. The explicit $n$-dependence of the action $S^{(\beta)}_{n}\left(\psi^\star,\hat{\psi}^\star,\vec{\lambda}^\star,\vec{z}^\star ;t,\epsilon;w\right)$ is again extracted by representing
the order parameters $\psi^\star$ and $\hat{\psi}^\star$ as infinite superpositions of
Gaussians. The explicit $t$-dependence appears in the so-called ``single-site'' term of the action, i.e.
\begin{align}
\nonumber S_{5}(\hat{\psi}^\star,\lambda^\star,z^\star;t,\epsilon;w) &=  n\sum_{k=0}^{k_{\mathrm{max}}}p(k)\int\mathrm{d}u\rho_{J}(u|k)\int\{\mathrm{d}\pi\} _{k}~\mathrm{Log}\int\mathrm{d}v\exp\left[ -\mathrm{i}\frac{\beta}{2}\lambda^\star v^{2}
\right.\\
&+\beta t\delta_\epsilon\left(w-v\right)\left.+ \frac{\beta}{2}\{\hat{\omega}\}_k v^{2}+\beta\left(\mathrm{i}z^\star x+\{\hat{h}\}_k \right)v\right]\ .
\end{align}

By making the identifications $\mathrm{i}\lambda^\star\equiv\lambda$ and $q\equiv-\mathrm{i}z^\star$ as before, taking the $t$-derivative at $t=0$ and considering the limits $\epsilon\to0$ and $\beta\to\infty$ as prescribed by \eqref{eq:remarkable}, we eventually find

\begin{equation}
\rho_{\tilde{J}}(w)\equiv\rho_{J,2}(w)=\sum_{k=0}^{k_{\mathrm{max}}}p(k)\int\mathrm{d}u\rho_{J}(u|k)\int\{\mathrm{d}\pi\} _{k}\left\langle \delta\left(w-\frac{-qu\left\langle \lambda_1 \right\rangle_{J}+\sum_{\ell=1}^k \frac{h_\ell K_\ell}{\omega_\ell}}{\lambda-\sum_{\ell=1}^k \frac{K_\ell^2}{\omega_\ell}}\right)\right\rangle_{\{K\}_{k}}\ ,
\label{eq:rho2_gen}
\end{equation}
where we recall that $\langle\cdot\rangle_{\{K\}_{k}}$ denote averaging w.r.t. a collection of $k$ i.i.d. random variables $K$, each drawn from the bond weight distribution $p_{K}(K)$.

Eq.~\eqref{eq:rho2_gen} represents the resulting probability density function of the top eigenvector's component of the deflated matrix $\tilde{J}$ in case of full deflation,  which in turn corresponds to the distribution of the second largest eigenvector's components of $J$.  This equation is the large $N$ generalisation of the single-instance result \eqref{eq:cav_components} found by the cavity method. The set of equations \eqref{eq:pi_gen}, \eqref{eq:lambda_gen}, \eqref{eq:q_gen}, \eqref{eq:lambda2_gen} and \eqref{eq:rho2_gen} are exactly equivalent to the thermodynamic limit equations \eqref{eq:cavity_pdf_td}, \eqref{eq:cavity_norm_td}, \eqref{eq:cavity_ort_td}, \eqref{eq:cavity_evect_td} and \eqref{eq:cavity_second_eval_td} found within the cavity method in Section \ref{sec:cavity_td}.

All the observations made in Section \ref{sec:cavity_td_orto} about the fact that \eqref{eq:q_gen} in case of full deflation encodes the orthogonality condition (hence $q=0$) hold here as well. Taking into account the average orthogonality condition $q=0$, we obtain
\begin{align}
\pi(\omega,h)&=\sum_{k=1}^{k_{\mathrm{max}}}p(k)\frac{k}{c}\int\{\mathrm{d}\pi\} _{k-1}\left\langle \delta\left(\omega-\left(\lambda-\sum_{\ell=1}^{k-1}\frac{K_{\ell}^2}{\omega_{\ell}}\right)\right)\delta\left(h-\left( \sum_{\ell=1}^{k-1}\frac{h_{\ell}K_{\ell}}{\omega_{\ell}}\right)\right)\right\rangle_{\{K\}_{k-1}}\ ,\label{eq:pi_last} \\ 
1&=\sum_{k=0}^{k_{\mathrm{max}}}p(k)\int\{\mathrm{d}\pi\} _{k} \left\langle \left(\frac{\sum_{\ell=1}^k \frac{h_\ell K_\ell}{\omega_\ell}}{\lambda-\sum_{\ell=1}^k \frac{K_\ell^2}{\omega_\ell}}\right)^{2}\right\rangle_{\{K\}_{k}}\ ,\label{eq:norm_last} \\ 
0&=\sum_{k=0}^{k_{\mathrm{max}}}p(k)\int\mathrm{d}u \rho_J(u|k)u \int \{\mathrm{d}\pi\} _{k} \left\langle \left(\frac {\sum_{\ell=1}^k \frac{h_\ell K_\ell}{\omega_\ell}}{\lambda-\sum_{\ell=1}^k \frac{K_\ell^2}{\omega_\ell}}\right)\right\rangle_{\{K\}_{k}}\ ,\label{eq:orto_last} \\ 
\rho_{\tilde{J}}(w)&\equiv\rho_{J,2}(w)=\sum_{k=0}^{k_{\mathrm{max}}}p(k)\int\{\mathrm{d}\pi\} _{k}\left\langle \delta\left(w-\frac{\sum_{\ell=1}^k \frac{h_\ell K_\ell}{\omega_\ell}}{\lambda-\sum_{\ell=1}^k \frac{K_\ell^2}{\omega_\ell}}\right)\right\rangle_{\{K\}_{k}}\label{eq:evect_last}\ , \\
\left\langle \tilde{\lambda_1} \right\rangle_{\tilde{J}}&\equiv \left\langle \lambda_2 \right\rangle_J=\lambda  \label{eq:eval_last}\ .
\end{align}

Eq. \eqref{eq:pi_last}, \eqref{eq:norm_last}, \eqref{eq:orto_last},\eqref{eq:evect_last} and \eqref{eq:eval_last} provide the solution of the second largest eigenpair problem in the large $N$ limit. They are identical to eq. \eqref{eq:cav_pi_last}, \eqref{eq:cav_norm_last},\eqref{eq:cav_orto_last}, \eqref{eq:cav_evect_last} and \eqref{eq:cav_eval_last} found with the cavity method.

\section{\\Top eigenvalue evaluation in the RRG case \label{sec:RRG_appendix}}
Here we give details of the calculation of the top eigenvalue of the RRG deflated matrix in both the \emph{outer} and \emph{bulk} regimes, as anticipated in Sections \ref{sec:RRG_peaked_eval} and \ref{sec:RRG_gauss_eval}.

In the \emph{outer} regime, the top eigenvalue is found by taking into account \eqref{eq:z_peaked}, \eqref{eq:gauss-ansatz}, \eqref{eq:omega-with-x} and the identity $\bar{h}=\bar\omega-1$, which follows from \eqref{eq:z-h-omega}. The $\mathcal{O}(n)$ terms of the action
$S_{n}$ in \eqref{eq:action_pihatpi} - keeping only the leading $\beta\rightarrow\infty$ terms - are expressed
as follows

\begin{align}
\nonumber S_{1}\left[\pi,\hat{\pi}\right]  &=-nc\int\mathrm{d}\pi(\omega,h)\mathrm{d}\hat{\pi}(\hat\omega,\hat h)\ln\frac{Z_\beta(\omega-\hat{\omega},h+\hat{h})}{Z_\beta(\omega,h)} \\
 & \simeq -nc\frac{\beta}{2}\frac{\bar{h}^2}{\bar\omega}\left(\frac{2}{\bar\omega-1}\right)\ ,
\end{align}

\begin{align}
\nonumber S_{2}[\pi] & =n\frac{c}{2}\int\mathrm{d}\pi(\omega,h)\mathrm{d}\pi(\omega',h')\ln\frac{Z_\beta\left(\omega-\frac{1}{\omega'},h+\frac{h'}{\omega'}\right)}{Z_\beta(\omega,h)}\\
 &\simeq nc\frac{\beta}{2}\frac{\bar{h}^2}{\bar\omega}\left(\frac{1}{\bar\omega-1}\right)=-\frac{1}{2}S_{1}[\pi,\hat\pi]\ ,
\end{align}

\begin{align}
S_{3}\left(\lambda\right) & =\frac{\beta}{2}n\lambda\ =\frac{\beta}{2}n(c-x) ,
\end{align}

\begin{align}
S_{4}(z,x)=-n\frac{\beta}{2}x{z^{\star}}^2 =n\frac{\beta}{2}xq^2=n\frac{\beta}{2}x\ ,
\end{align}

\begin{align}
S_{5}[\hat{\pi},\lambda] & =n\int\left[\prod_{\ell=1}^{c}\mathrm{d}\hat{\pi}(\hat{\omega}_{\ell},\hat{h}_{\ell})\right]\mathrm{Log}~Z_\beta\left(\lambda-\{ \hat{\omega}\} _{c},\{ \hat{h}\} _{c}-qx\right) & {}\nonumber \\
 &\simeq n\frac{\beta}{2}\left(\lambda-\frac{c}{\bar\omega}\right)\label{S5afterbeta}\ .
\end{align}

Summing up all terms and recalling from \eqref{eq:omega-with-x} that $\lambda=c-x$, the action at the saddle point reads 

\begin{equation}
S_{n}=n\frac{\beta}{2}(c-x)\ ,
\end{equation}
which implies from \eqref{eq:formula_largest} for the average of the largest eigenvalue of $\tilde{J}$ the formula
\begin{equation}
\left\langle \tilde{\lambda}_{1}\right\rangle _{\tilde{J}}=c-x\ .
\end{equation}

In the \emph{bulk} regime, the top eigenvalue is found by taking into account  \eqref{eq:sigma-gauss} and \eqref{eq:barh-gauss} and also that $q=0$ and $\lambda=2\sqrt{c-1}$. Then the $\mathcal{O}(n)$ terms of the action
$S_{n}$ in \eqref{eq:action_pihatpi} - keeping only the leading $\beta\rightarrow\infty$ terms - are expressed
as

\begin{align}
\nonumber S_{1}\left[\pi,\hat{\pi}\right]  & =-nc\frac{\beta}{2}\frac{2\sigma^2}{\bar\omega({\bar\omega}^2-1)}\ ,\\
\nonumber S_{2}[\pi] & =nc\frac{\beta}{2}\frac{2\sigma^2}{2\bar\omega({\bar\omega}^2-1)}=-\frac{1}{2}S_{1}[\pi,\hat\pi]\ ,\\
S_{3}\left(\lambda\right) & =\frac{\beta}{2}n\lambda\ =\frac{\beta}{2}n\sqrt{c-1} ,\\
S_{4}(q,x) &=n\frac{\beta}{2}x{q}^2 =0\ ,\\
S_{5}[\hat{\pi},\lambda] &=n\frac{\beta}{2}\frac{1}{{\bar\omega}^2}\frac{c\sigma^2}{\lambda-\frac{c}{\bar\omega}}
 \label{S5afterbeta}\ .
\end{align}

Summing up all terms and exploiting the identities \eqref{eq:baromega} and \eqref{eq:barh}, the action at the saddle point reads 

\begin{equation}
S_{n}=nc\frac{\beta}{2}\frac{\sigma^2}{\bar\omega}\left[-\frac{1}{{\bar\omega^2}-1}+\frac{1}{\bar\omega}\frac{1}{\lambda-\frac{c}{\bar\omega}}\right]=n\frac{\beta}{2}\lambda=n\beta\sqrt{c-1}\ ,
\end{equation}
which implies from \eqref{eq:formula_largest} that the average of the largest eigenvalue of $\tilde{J}$ is
\begin{equation}
\left\langle \tilde{\lambda}_{1}\right\rangle _{\tilde{J}}=2\sqrt{c-1}\ ,
\end{equation}
corresponding to the upper edge of the Kesten-McKay distribution.

\section{\\Replica setup for the second largest eigenpair of sparse random Markov transition matrices \label{sec:markov_appendix}}
The partition function reads
\begin{equation}
Z=\int \mathrm{d}\bm{v} \exp\left[ \frac{\beta}{2}\sum_{i,j=1}^N v_i \frac{c_{ij}}{\sqrt{k_i k_j}} v_j-\frac{\beta}{2cN}\left( \sum_{i=1}^N v_i \sqrt{k_i}\right)^2 \right]\delta\left(\left|\bm{v}\right|^{2}-N\right)\ .\label{eq:part_funct_m}
\end{equation}
By expressing the delta function in \eqref{eq:part_funct_m} via its Fourier representation and  employing the change of variable $\tilde{v}_i\sqrt{k_i}\leftarrow v_i$, the partition function becomes
\begin{align}
\nonumber Z=&\left(\frac{\beta}{4\pi}\right)\left( \prod_{i=1}^N k_i\right)^{1/2} \int \mathrm{d}\tilde{\bm{v}}\mathrm{d}\lambda \exp\left[ \frac{\beta}{2}\sum_{i,j=1}^N \tilde{v}_i c_{ij} \tilde{v_j}-\frac{\beta}{2cN}\left( \sum_{i=1}^N \tilde{v_i} k_i\right)^2 \right]\\
&\times\exp\left[ -\mathrm{i}\frac{\beta}{2}\lambda\left(\sum_{i=1}^N {\tilde{v_i}}^2 k_i -N\right)\right]\ .\label{eq:part_funct_m_1}
\end{align}
The square in the exponent of \eqref{eq:part_funct_m_1} can be linearised by a Hubbard-Stratonovich transform as in \eqref{eq:HS}. The resulting partition function, where we rename the $\tilde{v_i}$ variables as $v_i$ to avoid cumbersome notation, reads
\begin{align}
\nonumber Z=&\left(\frac{\beta}{4\pi}\right)\left( \frac{\beta N}{2\pi c}\right)^{1/2}\left( \prod_{i=1}^N k_i\right)^{1/2} \int \mathrm{d}\bm{v}\mathrm{d}\lambda\mathrm{d}z \exp\left( \frac{\beta}{2}\sum_{i,j=1}^N v_i c_{ij} v_j\right)\\
&\times\exp\left[ -\mathrm{i}\frac{\beta}{2}\lambda\left(\sum_{i=1}^N v_i^2 k_i -N\right)\right]\exp\left(-\frac{\beta N}{2c}z^2+\mathrm{i}\frac{\beta}{c}\sum_{i=1}^N v_i k_i z\right) .\label{eq:part_funct_m_2}
\end{align}
The average w.r.t. the matrix ensemble of ${\tilde{W}}^S$ reduces to averaging over the connectivity matrix $C=\{ c_{ij}\}$. By using the replica trick, we need to compute
\begin{equation}
\left\langle \tilde{\lambda}_{1}\right\rangle _{{\tilde{W}}^S}=\lim_{\beta\rightarrow\infty}\frac{2}{\beta N}\lim_{n\rightarrow0}\frac{1}{n}\mathrm{Log}\left\langle Z^{n}\right\rangle _{C}\ .\label{formulalargest_m}
\end{equation}

Henceforth, the derivation will exactly match the steps in \ref{sec:replica_eigenvalue}.

\section*{\textemdash \textemdash \textemdash \textemdash \textemdash \textendash{}}

\end{document}